\documentclass[12pt, a4paper]{article}

\usepackage{amsmath}
\usepackage{graphicx}
\usepackage{amssymb, latexsym}
\usepackage{setspace}
\usepackage{a4wide}
\usepackage[colorlinks, breaklinks=true, bookmarks=true]{hyperref}
\usepackage{color}
\usepackage{cite}

\usepackage{tikz}
\usetikzlibrary{shapes,arrows}
\newcommand{\sech}{\mathrm{sech}}

\newcommand{\deriv}[2]{\frac{d#1}{d#2}}

\newcommand{\BigFig}[1]{\parbox{12pt}{\Huge $#1$}}
\newcommand{\BigZero}{\BigFig{0}}

\newcommand{\bvec}[1]{\vec{\mathbf{#1}}}

\newcommand{\abs}[1]{\left\vert #1 \right\vert}

\newcommand{\operator}[1]{\mathbf{\hat#1}}

\newcommand{\ie}{i.\,e.\ }
\newcommand{\eg}{e.\,g.\ }

\newcommand{\vu}{\bvec{u}}
\newcommand{\vs}{\bvec{s}}
\newcommand{\vv}{\bvec{v}}
\newcommand{\vw}{\bvec{w}}
\newcommand{\va}{\bvec{a}}
\newcommand{\vb}{\bvec{b}}
\newcommand{\vc}{\bvec{c}}
\newcommand{\ve}{\bvec{e}}
\newcommand{\vg}{\bvec{g}}
\newcommand{\vk}{\bvec{k}}
\newcommand{\vphi}{\vec{\boldsymbol{\varphi}}}
\newcommand{\vups}{\vec{\boldsymbol{\upsilon}}}
\newcommand{\vom}{\vec{\boldsymbol{\omega}}}
\newcommand{\veta}{\vec{\boldsymbol{\eta}}}
\newcommand{\vmu}{\vec{\boldsymbol{\mu}}}
\newcommand{\vbmu}{\vec{\bar{\boldsymbol{\mu}}}}

\newcommand{\tG}{\widetilde{G}}
\newcommand{\bG}{\bar{G}}
\newcommand{\sext}{\vs_{ext}}

\newcommand{\tq}{\tilde{q}}
\newcommand{\vts}{\vec{\tilde{\mathbf{s}}}}
\newcommand{\vtv}{\vec{\tilde{\mathbf{v}}}}
\newcommand{\tf}{\tilde{f}}

\newcommand{\bGamma}{\bar{\Gamma}}
\newcommand{\bUps}{\bar{\Upsilon}}
\newcommand{\nv}{\Vert \vv \Vert}
\newcommand{\tlam}{\tilde{\lambda}}

\newcommand{\Guo}{}
\newcommand{\Blanes}{}

\begin{document}
    
\title{Semi-global approach for propagation of the time-dependent Schr\"odinger equation for time-dependent and nonlinear problems}
\author{Ido Schaefer, Hillel Tal-Ezer and Ronnie Kosloff}
\date{\today}
\maketitle

\begin{abstract}
A detailed exposition of highly efficient and accurate method for the propagation of the time-dependent Schr\"odinger equation \cite{propagator_2012} is presented. 
The method is readily generalized to solve an arbitrary set of ODE's. 
The propagation is based on a global approach, in which large time-intervals are treated as a whole, replacing the local considerations of the common propagators. The new method is suitable for various classes of problems, including problems with a time-dependent Hamiltonian, nonlinear problems, non-Hermitian problems and problems with an inhomogeneous source term. In this paper, a thorough presentation of the basic principles of the propagator is given. We give also a special emphasis on the details of the numerical implementation of the method. For the first time, we present the application for a non-Hermitian problem by a numerical example of a one-dimensional atom under the influence of an intense laser field. The efficiency of the method is demonstrated by a comparison with the common Runge-Kutta approach.

\end{abstract}

\tableofcontents

\section{Introduction}

The time-dependent Schr\"odinger equation (TDSE),
\begin{equation}\label{Schr_fun}
	\deriv{\psi(t)}{t} = -\frac{i}{\hbar} \operator{H}(t)\psi(t)
\end{equation}
is a central pillar in quantum dynamics. Solution of the equation supplies insight on fundamental quantum processes. 
For the majority of problems closed form solutions do not exist. An alternative is to develop
numerical schemes able to simulate from first principle quantum processes. 
In the present paper we concentrate on the issue of time-propagation.

We assume that the formal operation \text{$\phi = \operator{H} \psi$} can be carried out in a matrix vector representation, \text{$\bvec{v} = H \bvec{u}$}. Hence, the problem of solving Eq.~\eqref{Schr_fun} becomes a special case of the problem of solving a general set of ordinary differential equations (ODE's).

Historically, the common practice in early studies was either to solve Eq.~\eqref{Schr_fun} by general methods for solving a set of ODE's, or by methods which were developed particularly for Eq.~\eqref{Schr_fun}. General solvers for a set of ODE's rely on approximations to a low order Taylor expansion of $\psi(t)$, derived from Eq.~\eqref{Schr_fun}. This leads to the necessity of a time-step propagation scheme (see Sec.~\ref{ssec:Taylor}). The most popular methods for general applications are the Runge-Kutta methods \cite{butcherrunge}. Another method, which became very popular in early quantum studies, is second order differencing \cite{FourierGrid}.

Commonly, early researchers preferred other methods, which were intended specifically for quantum applications. These methods have conservation properties of certain physical quantities, such as norm or energy. The popular methods are Crank-Nicolson implicit scheme \cite{sanz1984methods} and split operator exponentiation \cite{feit1982solution} (it is noteworthy that the second order differencing method has also special conservation properties). These methods are also equivalent to an approximation of a low order Taylor expansion of $\psi(t)$, and lead to a time-step scheme. The advantage of these methods over the general methods is questionable, since the overall quality of the obtained solution is not improved over the general methods. The convergence becomes non-uniform---the error is accumulated in the physical quantities which are not conserved by the propagation scheme. In particular, the norm conservation leads to larger accumulation of errors in phase.

All the methods that were mentioned can be classified as \emph{local methods}. They all share the common property of being equivalent to a low order Taylor expansion in time. The Taylor expansion has slow convergence properties, which limit its application for approximation purposes to low orders. This leads to the locality of the solution, and consequently, to the time-step integration scheme. The drawback of a time-step scheme is the accumulation of the errors in each time step, which limits the accuracy for realistic times. The time step $\Delta t$ is limited by the spectral range of $\operator{H}$, $\Delta E$. Typically, the propagation process is numerically stable only for time-steps which do not exceed \text{$\Delta t \sim \frac{1}{10}  \hbar/\Delta E$}. The final accuracy of a Taylor propagation method scales as \text{$O(\Delta t^n)$} where $n$ is the order of the method.

A breakthrough in solving the TDSE was the development of the global Chebyshev propagator. The global Chebyshev propagator solves Eq.~\eqref{Schr_fun} for a \emph{time-independent Hamiltonian}, \text{$\operator{H}(t)\equiv\operator{H}$}, without the necessity of a time-step scheme. The whole propagation interval is treated \emph{globally} in a single step. The development of the method was led by the insight that a local time-step integration scheme is unnecessary when integration can be performed analytically. With a time-independent Hamiltonian, Eq.~\eqref{Schr_fun} can be directly integrated to yield
\begin{equation}\label{psitTiH}
	\psi(t) = \exp\left(-\frac{i}{\hbar} \operator{H}t\right)\psi(0)
\end{equation}
The direct computation of this expression becomes highly demanding for large-scale problems (see Sec.~\ref{sssec:TiH}); the computation in the Chebyshev propagator is based on a polynomial Chebyshev expansion of the evolution operator, \text{$\operator{U} =\exp(-\frac{i}{\hbar} \operator{H}t)$} \cite{k28}. The method has uniform convergence and does not accumulate errors.
The computational effort scales as $\sim \frac{\Delta E\, t}{2 \hbar}$.
The Chebyshev scheme outperforms all other methods for problems with time independent Hamiltonian operators \cite{k71}. 
More than thirty years after it was introduced it is still the method of choice for efficient and high accuracy solution to large scale problems with a time-independent Hamiltonian \cite{zhao2016dynamics}.

The case of a non-Hermitian Hamiltonian requires a special care. The global Chebyshev propagator was developed for a Hermitian Hamiltonian. 
Without modification it is not suitable for non-Hermitian operators. \Guo{A generalization of Chebyshev are 
Faber polynomials which enable to enter the complex plane \cite{huang1994general,k144,guo1999short}. A different approach
for non-Hermitian operators led to the development of Newtonian propagators.
The Chebyshev approximation on the real axis is replaced by a Newton interpolation in the complex plane.}
The first application was the solution of the Liouville--von-Neumann equation \cite{k80}. The Newtonian scheme was later implemented also to the Schr\"odinger equation with absorbing boundary conditions \cite{k118}. \Guo{A comparison of the different schemes and the relation to other propagators
\cite{park1986unitary,zhu1994orthogonal,vijay2002polynomial,chen1998discrete} has been reviewed \cite{k104,guo2007recursive}}.

The global schemes mentioned above assume a previous knowledge on the eigenvalue domain of the Hamiltonian. Commonly, such a knowledge is missing, in particular in non-Hermitian problems. In such a case, the global approach can be implemented by the Arnoldi approach, using a restarted Arnoldi algorithm (see, for example, \cite{restartedArnoldi}). A paper on this topic has not been published to date.

The focus of the present study is solving the Schr\"odinger equation for problems in which the Hamiltonian is explicitly time-dependent. 
Such problems are common in ultrafast spectroscopy, coherent control and high harmonic generation. 
Another typical complication arises when the Hamiltonian becomes nonlinear, \ie explicitly depends on the state $\psi(t)$.
Mean field approximation typically lead to such equations. Examples are the time-dependent Gross-Pitaevskii approximation \cite{NumericalBEC}, time-dependent Hartree \cite{kulanderTDHF, meyerTDHF} and time-dependent DFT \cite{GrossTDDFT}. In general, Eq.~\eqref{Schr_fun} cannot be integrated analytically for a Hamiltonian with time-dependence or nonlinearity.
A less common complication arises when a source term is added to the Schr\"odinger equation. 
Such equations can be found in scattering theory \cite{Neuhauser_ih} and in particular problems in coherent control \cite{TDtargets,k236,k279}.

The common practice to overcome the explicit time dependence, or nonlinearity, is to resort again to a time-step scheme, which relies on Eq.~\eqref{psitTiH}. The propagation interval is divided into small time-steps, where the Hamiltonian is assumed to be stationary within the time-step. This becomes equivalent to a first order method in time. The result is either a significant increase in computational cost or low accuracy. In our opinion, this is a misuse of the global Chebyshev propagator, which is intended to overcome this very problem. 
\Blanes{A better scheme is based on the Magnus expansion \cite{magnus1954exponential} to correct for time ordering, and a low order polynomial approximation
for the exponent of an operator \cite{k79,blanes2009magnus,chuluunbaatar2008explicit,bader2016efficient}. 
This leads again to a local scheme, since the radius of convergence of the Magnus expansion is limited \cite{blanes2009magnus}. However, the scaling of the error with $\Delta t$ is improved over the common naive practice.}
\Guo{Another local approach with improved scaling is to use a high order splitting method \cite{sun2012higher}.}

Attempts have been made to implement the global approach in problems with time-dependent or nonlinear Hamiltonians. 
A very accurate scheme was developed, based on embedding in a larger space, which includes also time.
In the extended space, the problem can be formulated by global means \cite{k105}. The drawback was the very high computational cost which scaled as $\Delta E\, \Delta t\, \Delta \omega$ where $\Delta E$ is the eigenvalue range of the Hamiltonian, $\Delta t$ is the time step and $\Delta \omega$ is the bandwidth  of the explicit time-dependent function.
In addition, the method was not applicable to nonlinear problems.

Another attempt was proposed in \cite{Roi_nl}. Like in the global Chebyshev propagator, the propagation method is based on the idea of global integration of Eq.~\eqref{Schr_fun}. A direct integration leads to an integral equation:
\begin{equation}\label{Roi}
	\psi(t) = \psi(0) -\frac{i}{\hbar}\int_0^t \operator{H}\left[\psi(\tau), \tau\right]\psi(\tau)\,d\tau
\end{equation}
The integral is approximated by the expansion of the integrand in time in a truncated Chebyshev series, which can be integrated analytically. This results in a system of equations of $\psi(t)$ in multiple time-points. In the nonlinear case, the system of equations becomes also nonlinear. Seemingly, this replaces the problem of time-propagation with the even more difficult problem of optimization.
However, the system can be solved by a relatively simple iterative scheme when the time-interval is sufficiently small. This leads again to a time-step scheme. The hope was that it will be possible to use larger time-steps than other propagation schemes, thus leading to reduction of the error accumulation during propagation. Later it was found that this approach led to extremely small time-steps and a large number of iterations, thus becoming highly inefficient (from our experience, and private communications with the author). The failure of the method clearly lies in the necessity of solving a system of equations, a task which was found to be much more demanding than a local propagation scheme.

The introduction of source terms was followed by the development of a global propagation scheme for inhomogeneous problems \cite{Mamadou_ih,k247}. This led to new insight to the problem of the time-dependence or nonlinearity of the Hamiltonian. A new global scheme, based on integration in large time-steps, was first introduced in \cite{k253}. The method is based on another integrated version of Eq.~\eqref{Schr_fun}, in which the $\psi(t)$ dependence in the integral expression is minimized in comparison to \eqref{Roi}. Here again, an iterative scheme is used to solve the resulting system of equations. This scheme was a significant improvement to that introduced in \cite{Roi_nl}. However, we found that it gave inferior results in comparison to a 4'th order Runge-Kutta scheme (RK4). A drastic improvement was achieved by new insights on the propagation scheme \cite{propagator_2012}. The improved scheme was demonstrated to be significantly more efficient than the Taylor methods, particularly when high accuracy is required. Quite importantly, the scheme was generalized to solve an arbitrary set of ODE's.

The propagation approach of the new scheme, as well as Ref.~\cite{Roi_nl}), combines global and local elements. Hence, it can be classified as a \emph{semi-global propagation approach}.

The original global Chebyshev propagator \cite{k28} was easy to program. This led to fast proliferation with many applications.
The new algorithm for explicit time dependence became more involved with three user defined parameters which control
the accuracy and efficiency. However, we believe that the vast increase in efficiency is worth the effort of learning and computing the algorithm.

The present paper consolidates the numerical scheme. In addition, the application of the algorithm is extended to non-Hermitian Hamiltonians.
Our purpose is to give explicit description of all steps and considerations in the scheme to enable the potential user either to program from scratch or to be able to tailor an existing program to the problem of choice. Although the scheme is more involved than the basic Chebyshev scheme,
we hope that the explicit description will lead to proliferation of the method.

\section{Theory}

\subsection{Definition of the problem}

Let us rewrite the time-dependent Schr\"odinger equation in a matrix-vector notation:
\begin{equation}\label{Schr}
	\deriv{\bvec{u}(t)}{t} = -iH(t)\bvec{u}(t)
\end{equation}
where $\vu(t)$ represents the state, and $H(t)$ is a matrix representing the time-dependent Hamiltonian of the system. (Atomic units are used throughout, so we set $\hbar=1$.)

In our discussion, we shall consider a generalization of Eq.~\eqref{Schr}. First, we let $H$ include a dependence on the state vector, $\bvec{u}(t)$, \ie \text{$H \equiv H(\bvec{u}(t),t)$}. This results in a nonlinear equation of motion.
In addition, we include an  inhomogeneous \emph{source term} $\vs(t)$.
The time-dependent nonlinear inhomogeneous Schr\"odinger equation reads:
\begin{equation}\label{GeneralSchr}
	\deriv{\bvec{u}(t)}{t} = -iH(\bvec{u}(t),t)\bvec{u}(t) + \vs(t)
\end{equation}

Actually, Eq.~\eqref{GeneralSchr} has the form of a much more general problem. A general set of ODE's is equivalent to an equation of the following form:
\begin{equation}\label{odes0}
	\deriv{\bvec{u}(t)}{t} = \vg(\vu(t), t)
\end{equation}
where $\vg(\vu(t), t)$ is an arbitrary vector function of $\vu(t)$ and $t$. This can be always rewritten as:
\begin{equation}\label{odes}
	\deriv{\bvec{u}(t)}{t} = G(\bvec{u}(t),t)\bvec{u}(t) + \vs(t)
\end{equation}
where $G(\bvec{u}(t),t)$ is a matrix. Hence, the problem of solving Eq.~\eqref{GeneralSchr} is equivalent to the problem of solving Eq.~\eqref{odes}, by setting \text{$G(\bvec{u}(t),t)=-iH(\bvec{u}(t),t)$}. As a matter of convenience, we shall use the form of Eq.~\eqref{odes} in our discussion.

The initial condition for the vector state is:
\begin{equation}\label{initcond}
	\bvec{u}(0) = \bvec{u}_0
\end{equation}

We require the solution, $\bvec{u}(t)$, at an arbitrary time $t$.

\subsection{Local approach---Taylor methods}\label{ssec:Taylor}

The popular algorithms for solving a general set of ODE's are based on Taylor expansion considerations. In order to illustrate this approach, we will consider the \emph{Euler method} which is the simplest Taylor method.

The Euler method is based on a first order Taylor expansion for approximation of the solution at a close point. The solution at \text{$t=\Delta t$} is approximated by:
\begin{equation}\label{EulerStep}
	\vu(\Delta t) \approx \vu(0) + \Delta t\deriv{\vu(0)}{t}
\end{equation}
$\vu(0)$ is given by Eq.~\eqref{initcond}. \text{$d\vu(0)/dt$} can be computed by plugging Eq.~\eqref{initcond} into Eq.~\eqref{odes}. Using a first order approximation, the solution will be of low accuracy, unless $\Delta t$ is sufficiently small. In order to get an accurate solution far from \text{$t=0$}, we have to march in small time-steps. The solution at \text{$t=2\Delta t$} is computed in the same way, using $\vu(\Delta t)$ from Eq.~\eqref{EulerStep}, and Eq.~\eqref{odes} for obtaining \text{$d\vu(\Delta t)/dt$}. We continue by repeating this propagation technique until we reach the solution at the final time, which will be denoted as \text{$t=T$}. If desired, the accuracy of the solution can be improved by choosing a smaller time-step $\Delta t$. Of course, this requires more computational effort.

The Euler method is rarely used, because of its slow convergence properties with the decrement of $\Delta t$. The error of the solution in $T$ scales as $O(\Delta t)$. Other Taylor methods are based on higher order expansions. The error of a Taylor method of order $n$ scales as $O(\Delta t^n)$.

The most popular Taylor methods are the \emph{Runge-Kutta methods}. The idea underlying the Runge-Kutta methods is to approximate the Taylor expansion without a direct evaluation of high-order derivatives of $\vu(t)$. The Taylor expansion is approximated by first order derivative evaluations, using Eq.~\eqref{odes}. This approximation preserves the scaling of the error with $\Delta t$. For instance, we consider the Runge-Kutta method of the 4'th order (RK4). The solution at \text{$t=\Delta t$} is approximated by:
\begin{align}
	& \vu(\Delta t) \approx \vu(0) + \frac{1}{6}(\vk_1 + 2\vk_2 + 2\vk_3 + \vk_4) \label{RK4} \\
	& \vk_1 = \Delta t\,\vg(\vu(0), 0) \nonumber \\
	& \vk_2 = \Delta t\,\vg\!\left(\vu(0)+\frac{\vk_1}{2}, \frac{\Delta t}{2} \right) \nonumber \\
	& \vk_3 = \Delta t\,\vg\!\left(\vu(0)+\frac{\vk_2}{2}, \frac{\Delta t}{2} \right) \nonumber \\
	& \vk_4 = \Delta t\,\vg(\vu(0) + \vk_3, \Delta t) \nonumber
\end{align}
Eq.~\eqref{RK4} approximates a fourth order Taylor expansion. In our numerical example (Sec.~\ref{sec:example}) we shall use RK4 as a reference method.

The Taylor approach is based on local considerations---in each time-step, the solution $\vu(t)$ is approximated using our knowledge on the local behaviour of $\vu(t)$ at the previous time-point. The information on the behaviour of $\vu(t)$ is deduced from its derivatives \emph{at the time-point}. For this information to be accurate, it is essential that the time-point in which the solution is to be evaluated is close enough. Hence, it is necessary to propagate in small time-steps. The many time-step propagation scheme results in a large computational effort. Moreover, the error is accumulated during the propagation process. These drawbacks are direct consequences of the locality of the Taylor approach.

Another drawback of the Taylor approach lies in the slow convergence properties of the Taylor series. These reduce the efficiency of this approach when using high order Taylor expansions. The popular Runge-Kutta methods are based on 4'th or 5'th order expansions.

In what follows, we shall accommodate with these problems by developing a more global approach for the task of solving Eq.~\eqref{odes}.

\subsection{Global approach using closed integrated forms}

In order to approach the problem in a global manner, we seek a way to treat the whole time interval of the problem in a single stage. Indeed, Eq.~\eqref{odes} can be solved in a single step in the special cases that it can be integrated analytically. The closed integrated forms in these special cases constitute the basis for the present approach.

\subsubsection{Time-independent Hamiltonian}\label{sssec:TiH}

We start from the simplest case with a closed integrated form. The Hamiltonian is time-independent:
\begin{equation*}
	H(t) \equiv H_0
\end{equation*}
or, equivalently:
\begin{equation*}
	G(t) \equiv G_0
\end{equation*}
In addition, there is no inhomogeneous term:
\begin{equation*}
	\vs(t)\equiv 0
\end{equation*}
Eq.~\eqref{odes} becomes:
\begin{equation}\label{tiG0}
	\deriv{\bvec{u}(t)}{t} = G_0\bvec{u}(t)
\end{equation}
This equation, with the initial condition \eqref{initcond}, can be integrated directly to yield:
\begin{equation}\label{solG0}
	\vu(t) = \exp(G_0 t)\vu_0
\end{equation}
for an arbitrary $t$. In the special case of the Schr\"odinger equation, we have the well known result of the situation of stationary dynamics:
\begin{equation}\label{statSchr}
	\vu(t) = \exp(-iH_0 t)\vu_0
\end{equation}

The problem that arises is that the exponent of the matrix $G_0 t$ cannot be computed directly (unless $G_0$ is diagonal). One immediate approach is to diagonalize $G_0$ and compute the function of the matrix in the basis of the eigenvectors of $G_0$. Then we can write:
\begin{equation}\label{u_direct}
	\vu(t) = S\exp(Dt)S^{-1}\vu_0
\end{equation}
where $D$ is the diagonalized $G_0$, and $S$ is the transformation matrix from the eigenvector basis to the original basis. The problem with this approach is that when the dimension of $G_0$ is large, it becomes infeasible to diagonalize it, because of the high numerical cost of this operation---diagonalization scales as $O(N^3)$, where $N$ is the dimension of the problem.

Usually, Eq.~\eqref{statSchr} is solved by another approach, which is less demanding numerically. We expand the RHS of Eq.~\eqref{solG0} in a polynomial series in $G_0$. First, we define a function \text{$f(x) = \exp(xt)$}, where $t$ is treated as a parameter. Then we approximate it by a truncated polynomial series:
\begin{equation}\label{PolExpansion}
	f(x) \approx \sum_{n=0}^{K-1} a_n P_n(x)
\end{equation}
where $P_n(x)$ is a polynomial of degree $n$, and $a_n$ is the corresponding expansion coefficient. This requires the choice of the set of expansion polynomials $P_n(x)$, and the computation of the corresponding $a_n$'s. Then, we approximate Eq.~\eqref{solG0} as:
\begin{equation}\label{MatExpansion}
	\vu(t) \approx \sum_{n=0}^{K-1} a_n P_n(G_0)\vu_0
\end{equation}
The expansion \eqref{PolExpansion} has to be accurate in the eigenvalue domain of $G_0$ in order that the form \eqref{MatExpansion} will be useful (see Appendix~\ref{app:FMpoly}).

The RHS of Eq.~\eqref{MatExpansion} can be computed by successive matrix-vector multiplications. Matrix-vector multiplications scale just as $O(N^2)$. In many cases, the computational effort can be reduced further. The direct multiplication of $\vu_0$ by $G_0$ can be replaced by the operation of an equivalent linear operator. The operation of the linear operator can be defined by a computational procedure, which may have a lower scaling with $N$. For instance, in the Fourier grid method (see~\cite{FourierGrid}) the Hamiltonian operation scales as $O(N\ln N)$ only.

An immediate question that arises is how to choose the set of expansion polynomials $P_n(x)$. One might suggest to use the Taylor polynomials, \ie \text{$P_n(x) = x^n$}, and expand $f(x)$ in a Taylor series. However, this would be a poor choice, because of the slow convergence properties of a Taylor series. The reason for the slow convergence lies in the low quality of the Taylor polynomials as expansion functions---as $n$ increases, they are getting closer to be parallel in the function space. In order to attain a fast convergence of the polynomial series with $K$, an orthogonal set of polynomials should be used. The expansion coefficients $a_n$ are given by a scalar product of the $P_n(x)$'s with $f(x)$.

Usually, $f(x)$ is expanded in a \emph{Chebyshev polynomial} series, or equivalently, by a \emph{Newton interpolation polynomial} at the \emph{Chebyshev points} of the eigenvalue domain. When the Hamiltonian is non-Hermitian, the eigenvalue domain becomes complex, and the Chebyshev approach is not appropriate anymore. Then, the \emph{Arnoldi approach} should be used instead. In Appendix~\ref{app:PolApprox} we present the approximation methods of a function by a Newton interpolation polynomial or a Chebyshev polynomial series. In Appendix~\ref{app:FunMat} we describe the different approximation methods for the multiplication of a vector by a function of matrix, by Chebyshev or Newton series, or by the Arnoldi approach.

In the Chebyshev or Newton methods, an approximation of degree \text{$K-1$}, with \text{$K$} expansion terms, requires \text{$K-1$} matrix-vector multiplications. This is due to the recurrence relations between the expansion polynomials in both methods, as will be described in Appendix~\ref{app:FunMat}. Similarly, in the Arnoldi approach, \text{$K$} matrix-vector multiplications are required for \text{$K$} expansion terms.

Note that using Eq.~\eqref{statSchr}, the solution is given only at the chosen $t$, and not at intermediate time points. Usually, it is desirable to follow the whole physical process which leads to the result at the final time, and the intermediate times are also of interest. Actually, the solution at the intermediate time-points can be obtained with a negligible additional computational effort. Let us rewrite Eq.~\eqref{PolExpansion} for each of the time-points to be computed:
\begin{align}
	& f_j(x) = \exp(xt_j) \qquad j=1,\ldots,N_{tp} \nonumber \\
	& f_j(x) = \sum_{n=0}^{K-1} a_{n,j} P_n(x) \label{ftj}
\end{align}
where $N_{tp}$ is the number of time points, and $t_j$ is the $j$'th time-point. The only difference between the $t_j$'s is in the definition of the $f_j(x)$'s, and the corresponding expansion coefficients, $a_{n,j}$. The $P_n(x)$'s remain the same. Hence, it is sufficient to compute the $P_n(G_0)\vu_0$ just once for all the desired time points. The $a_{n,j}$'s are computed for each time-point $t_j$. The computational effort of the matrix-vector multiplications (or the equivalent linear operations) is much greater than that of the computation of the $a_{n,j}$'s, unless $N$ is very small.

\subsubsection{Addition of a source term with a polynomial time-dependence}\label{sssec:polih}

Now let us add to Eq.~\eqref{tiG0} a source term:
\begin{equation}\label{ihodes}
	\deriv{\bvec{u}(t)}{t} = G_0\bvec{u}(t) + \vs(t)
\end{equation}
\Blanes{A source term is not very common in quantum applications. Nevertheless, the treatment of the common cases of a time-dependent or nonlinear Hamiltonian relies on the results that will be derived in the present section.}

Eq.~\eqref{ihodes} can be integrated using the \emph{Duhamel principle} which relates the solution for the inhomogeneous equation to that of the corresponding homogeneous equation. Let us denote the evolution matrix for the homogeneous equation \eqref{tiG0} by:
\begin{equation}
	U_0(t) = \exp(G_0 t)
\end{equation}
The Duhamel principle states that the solution of Eq.~\eqref{tiG0} can be written by the means of $U_0(t)$ in the following way:
\begin{align}
	\vu(t) &= U_0(t)\vu_0 + \int_0^t U_0(t-\tau)\vs(\tau)\,d\tau \nonumber \\
		   &= \exp(G_0 t)\vu_0 + \int_0^t \exp[G_0 (t-\tau)]\vs(\tau)\,d\tau \nonumber \\
		   &= \exp(G_0 t)\vu_0 + \exp(G_0 t)\int_0^t \exp(-G_0\tau)\vs(\tau)\,d\tau \label{solih}
\end{align}
Eq.~\eqref{solih} assumes a closed form when the integral in the RHS of Eq.~\eqref{solih},
\begin{equation}\label{vecexpint}
	\int_0^t \exp(-G_0\tau)\vs(\tau)\,d\tau
\end{equation}
can be performed analytically.

We shall focus on a family of source terms for which \eqref{vecexpint} assumes a closed form---source terms with a polynomial time-dependence:
\begin{equation}\label{pols}
	\vs(t) = \sum_{m=0}^{M-1}\frac{t^m}{m!}\vs_m
\end{equation}
First, we need a closed expression for \eqref{vecexpint} in this particular case. For convenience, we will discuss the scalar version of \eqref{vecexpint}, without loss of generality:
\begin{equation}\label{expint}
	\int_0^t \exp(-z\tau)s(\tau)\,d\tau
\end{equation}
where $z$ is a complex variable, and $s(t)$ is a scalar function of the form
\begin{equation}\label{scalarpols}
	s(t) = \sum_{m=0}^{M-1}\frac{t^m}{m!}s_m
\end{equation}
After we obtain a closed expression, we will be able to write the RHS of Eq.~\eqref{solih} by the means of multiplication of vectors by functions of the matrix $G_0$. Finally, we will show that the solution can be written in a modified form, in which the computational effort is much reduced.

First, we discuss a source term of the form:
\begin{equation}\label{singleterm}
	s(t) = \frac{t^m}{m!}s_m
\end{equation}
Let us define:
\begin{equation}\label{Js}
	J_{m+1}(z,t) \equiv \int_0^t \exp(-z\tau)\tau^m \,d\tau, \qquad\qquad m=0,1,\ldots
\end{equation}
which are the integrals that need to be evaluated. We start with the simplest situation, when \text{$m=0$}, and $s(t)$ becomes a constant. In the case that \text{$z \neq 0$} we obtain:
\begin{equation}
	J_1(z,t) \equiv \int_0^t \exp(-z\tau)\,d\tau = \frac{1 - \exp(-zt)}{z}
\end{equation}
When \text{$z=0$}, we obtain:
\begin{equation}
	J_1(0,t) = t
\end{equation}
Now let us consider the case that $m>0$. If $z \neq 0$, we can evaluate the integral using integration by parts. A simple calculation yields:
\begin{equation}
	J_{m+1}(z,t) = -\frac{\exp(-zt)t^m}{z} + \frac{m}{z}\int_0^t \exp(-z\tau)\tau^{m-1} \,d\tau = -\frac{\exp(-zt)t^m}{z} + \frac{m}{z}J_{m}(z,t)
\end{equation}
or, equivalently:
\begin{equation}
	J_{m}(z,t) = -\frac{\exp(-zt)t^{m-1}}{z} + \frac{m-1}{z}J_{m-1}(z,t), \qquad\qquad m=2,3,\ldots
\end{equation}
Successive operations of the resulting recursion formula lead to the following expression:
\begin{equation}\label{Jsol}
	J_{m}(z,t) = \frac{(m-1)!}{z^m}\left[1 - \exp(-zt)\sum_{j=0}^{m-1} \frac{(zt)^j}{j!} \right], \qquad\qquad m=1,2,\ldots
\end{equation}
Note that Eq.~\eqref{Jsol} applies also for \text{$J_1(z,t)$}. In the case that \text{$z=0$} we have:
\begin{equation}
	J_{m}(0,t) = \frac{t^{m}}{m}, \qquad\qquad m=1,2,\ldots
\end{equation}

We proceed to the evaluation of the scalar form of Eq.~\eqref{solih}:
\begin{equation}\label{scalarsoih}
	u(t) = \exp(zt)u_0 + \exp(zt)\int_0^t \exp(-z\tau)s(\tau)\,d\tau
\end{equation}
for a source term of the form \eqref{scalarpols}. We begin with the treatment of the second term in the RHS of Eq.~\eqref{scalarsoih}. Plugging \eqref{scalarpols} into this term, we obtain:
\begin{equation}
	\exp(zt)\sum_{m=0}^{M-1}\frac{1}{m!}\int_0^t \exp(-z\tau)t^m\,d\tau\, s_m = \exp(zt)\sum_{m=0}^{M-1}\frac{1}{m!}J_{m+1}(z,t)s_m = \sum_{m=0}^{M-1} f_{m+1}(z,t)s_m
\end{equation}
where we defined:
\begin{equation}\label{fs}
	f_{m}(z,t) \equiv
	\begin{cases}		
		\frac{1}{z^m}\left[\exp(zt) - \sum_{j=0}^{m-1} \frac{(zt)^j}{j!} \right] & z \neq 0 \\
		\frac{t^m}{m!} & z = 0 
	\end{cases}
	\qquad\qquad m=1,2,\ldots
\end{equation}
The corresponding scalar form of Eq.~\eqref{solih} becomes:
\begin{equation}\label{scalarsolfs}
	u(t) = \exp(zt)u_0 + \sum_{m=0}^{M-1} f_{m+1}(z,t)s_m
\end{equation}
Let us write Eq.~\eqref{scalarsolfs} in a prettier way. First, we define the following set of constants:
\begin{equation}\label{ws}
	w_m \equiv 
	\begin{cases}
		u_0 & m=0 \\
		s_{m-1} & 0 < m \leq M
	\end{cases}
\end{equation}
Second, we note that the definition \eqref{fs} can be extended to the case of \text{$m=0$}, using the convention that
\begin{equation}\label{sumconvention}
	\sum_{j=L}^N b_j = 0, \qquad N<L
\end{equation}
for arbitrary $b_j$'s. Using this extension of definition, we have:
\begin{equation}
	f_0(z,t) = \exp(zt)
\end{equation}
Then, Eq.~\eqref{scalarsolfs} becomes:
\begin{equation}\label{scalarsolfs2}
	u(t) = \sum_{m=0}^{M} f_{m}(z,t)w_m
\end{equation}
We can use the form of Eq.~\eqref{scalarsolfs2} to write an analogous vector solution for Eq.~\eqref{ihodes}:
\begin{equation}\label{solfs}
	\vu(t) = \sum_{m=0}^{M} f_{m}(G_0,t)\bvec{w}_m
\end{equation}
where the $\bvec{w}_m$'s are defined in an analogous manner to the scalar $w_m$'s.

When we compare Eq.~\eqref{solfs} to Eq.~\eqref{solG0}, it seems that the addition of the source term is quite expensive numerically. In Eq.~\eqref{solG0} it is necessary to evaluate just one multiplication of a vector by a function of a matrix. In Eq.~\eqref{solfs}, it is necessary to perform the same kind of operation \text{$M+1$} times. Actually, the computational effort can be much reduced, if we rewrite Eq.~\eqref{solfs} in a modified form.

Let us return to the corresponding scalar equation, Eq.~\eqref{scalarsolfs2}. We are going to show that it can be rewritten using just one of the $f_m(z,t)$ functions. Observing the definition \eqref{fs}, it can be easily seen that
\begin{equation}\label{ffplus1}
	f_m(z,t) = zf_{m+1}(z,t) + \frac{t^m}{m!}
\end{equation}
Eq.~\eqref{ffplus1} implies that a function $f_m(z,t)$ can be expressed using any of the other $f_k(z,t)$ functions. If \text{$k>m$}, we need \text{$k-m$} successive operations of Eq.~\eqref{ffplus1} in order to write $f_m(z,t)$ in the terms of $f_k(z,t)$. We obtain: 
\begin{equation}\label{fmfk}
	f_m(z,t) = z^{k-m}f_k(z,t) + \sum_{j=m}^{k-1} \frac{t^j}{j!} z^{j-m}
\end{equation}
Eq.~\eqref{fmfk} can be applied also to the case of \text{$k=m$}, with the summation convention \eqref{sumconvention}.

Using Eq.~\eqref{fmfk}, we can express all the $f_m(z,t)$ functions in Eq.~\eqref{scalarsolfs2} by the function with the largest $m$, \ie $f_M(z,t)$. Eq.~\eqref{scalarsolfs2} becomes:
\begin{align}
	u(t) &= \sum_{m=0}^M \left[z^{M-m}f_{M}(z,t)w_m + \sum_{j=m}^{M-1} \frac{t^j}{j!} z^{j-m}w_m \right] \nonumber \\
		 &= f_M(z,t)\sum_{m=0}^M z^{M-m}w_m + \sum_{j=0}^{M-1} \frac{t^j}{j!} \sum_{m=0}^j z^{j-m}w_m \nonumber \\
		 &= f_M(z,t)v_M + \sum_{j=0}^{M-1} \frac{t^j}{j!} v_j \label{scalarsol}
\end{align}
where we defined:
\begin{equation}
	v_j \equiv \sum_{m=0}^j z^{j-m}w_m
\end{equation}
Returning to the vector solution of Eq.~\eqref{ihodes}, we can write an analogous expression:
\begin{equation}\label{solih2}
	\vu(t) = f_M(G_0,t)\vv_M + \sum_{j=0}^{M-1} \frac{t^j}{j!} \vv_j 
\end{equation}
where
\begin{equation}
	\vv_j \equiv \sum_{m=0}^j G_0^{j-m}\bvec{w}_m, \qquad j=0,1,\ldots
\end{equation}
Now, only one function of $G_0$ appears in the solution. However, the computation of the $\vv_j$ vectors is still an expensive operation---the computation of the \text{$M+1$} vectors involves $M$ sums, which require $O(M^2)$ matrix-vector multiplications. The computational effort can be much reduced when we notice that the $\vv_j$'s satisfy a recursion relation:
\begin{equation}\label{vrecursion}
	\vv_j = G_0\vv_{j-1} + \bvec{w}_j, \qquad\qquad j=1,2,\ldots
\end{equation}
Using Eq.~\eqref{vrecursion}, all the $\vv_j$'s can be computed by $M$ matrix-vector multiplications only, starting from
\begin{equation}\label{v0}
	\vv_0 = \bvec{w}_0 = \vu_0
\end{equation}
Taking into account also the evaluation of the first term in Eq.~\eqref{solih2}, we can conclude that the overall computational cost is reduced to \text{$M+K-1$} matrix-vector multiplications for the Chebyshev or Newton series approximation methods (see Sec.~\ref{sssec:TiH}). Similarly, \text{$M+K$} matrix-vector multiplications are required for the Arnoldi approach.

\subsection{Approximated solutions based on closed integrated forms}

\subsubsection{Source term with an arbitrary time-dependence}\label{sssec:gnenralih}

Let us consider the case of Eq.~\eqref{ihodes} with a source term $\vs(t)$ with an arbitrary time-dependence. In general, the integral \eqref{vecexpint} does not assume a closed form, so a closed solution for Eq.~\eqref{ihodes} cannot be obtained. In the present approach, we utilize the closed solution for the case of \eqref{pols} in order to approximate the solution in the general case. The idea is to approximate the general source term by a truncated polynomial series of the form of \eqref{pols}. Then, the solution is approximated by a direct application of Eq.~\eqref{solih2}.

The approximation of $\vs(t)$ by the form of \eqref{pols} requires the computation of the $\vs_m$ coefficients. The form of Eq.~\eqref{pols} might suggest that we should set \text{$\vs_m=d^m \vs(0)/dt^m$}, to yield a truncated Taylor series of $\vs(t)$. However, as has already been mentioned, a Taylor series is a poor tool for approximation purposes. Thus, this approach is not recommended.

A better approach is to approximate $\vs(t)$ by an orthogonal polynomial set at the first stage:
\begin{equation}\label{sPn}
	\vs(t) \approx \sum_{n=0}^{M-1} \bvec{b}_n P_n(t)
\end{equation}
where the $\bvec{b}_n$'s are computed by a scalar product of the $P_n(t)$'s with $\vs(t)$. The orthogonal expansion polynomials can be expressed in the terms of the Taylor polynomials:
\begin{equation}\label{PnTaylor}
	P_n(t) = \sum_{m=0}^n q_{n,m} \frac{t^m}{m!}
\end{equation}
Plugging Eq.~\eqref{PnTaylor} into Eq.~\eqref{sPn} we obtain:
\begin{equation}
	\vs(t) \approx \sum_{n=0}^{M-1} \sum_{m=0}^{n} q_{n,m}\bvec{b}_n \frac{t^m}{m!} = \sum_{m=0}^{M-1} \left(\sum_{n=m}^{M-1} q_{n,m}\bvec{b}_n \right) \frac{t^m}{m!}
\end{equation}
Then, the result is equated to the Taylor form, 
\begin{equation}
	\sum_{m=0}^{M-1} \left(\sum_{n=m}^{M-1} q_{n,m}\bvec{b}_n \right) \frac{t^m}{m!} = \sum_{m=0}^{M-1}\frac{t^m}{m!}\vs_m
\end{equation}
to yield the $\vs_m$ Taylor polynomial coefficients:
\begin{equation}\label{q2s}
	\vs_m = \sum_{n=m}^{M-1} q_{n,m}\bvec{b}_n
\end{equation}
These are in general different from the Taylor expansion coefficients. In this way, we preserve the Taylor polynomial form of Eq.~\eqref{pols}, but with the advantage of the fast convergence of an orthogonal polynomial set.

It is recommended to use the Chebyshev polynomials as the $P_n(t)$ set. An equivalent option is to use a Newton interpolation expansion in the Chebyshev points.

We still need a procedure for a systematic computation of the $q_{n,m}$ coefficients. Recursive algorithms can be derived from recursive definitions of different polynomial sets. In appendix~\ref{app:pol2Taylor} we develop recursive conversion algorithms from Chebyshev and Newton expansions to a Taylor form.

Of course, the expansion \eqref{sPn} should not be confused with the similar expansion \eqref{PolExpansion}. The first approximates the function $\vs(t)$ in \emph{time}, within the time interval of the solution, while the second approximates a function of the matrix $G_0$ in the \emph{eigenvalue domain of $G_0$}.

\subsubsection{Time-dependent Hamiltonian}\label{sssec:TDH}

Now we shall consider the case of a time-dependent Hamiltonian, \text{$H=H(t)$}. For the sake of generality, a source term is included in the equation. We have:
\begin{equation}\label{TDH}
	\deriv{\bvec{u}(t)}{t} = -iH(t)\bvec{u}(t) + \vs(t)
\end{equation}
or,
\begin{equation}\label{TDodes}
	\deriv{\bvec{u}(t)}{t} = G(t)\bvec{u}(t) + \vs(t)
\end{equation}
In this case, the Duhamel principle cannot be applied directly for obtaining a closed form solution, as in the previous cases (see Sec.~\ref{sssec:polih}). However, we shall see that the results from the previous cases can be utilized for obtaining a procedure which approximates the solution in the present case.

First, it is always possible to split $G(t)$ into a sum of time-dependent and time-independent parts:
\begin{equation}\label{splitG}
	G(t) = \tG + \bG(t)
\end{equation}
where $\tG$ is arbitrary, and
\begin{equation}
	\bG(t) \equiv G(t) - \tG
\end{equation}
Let us define:
\begin{equation}\label{sext1}
	\sext(\vu(t), t) = \vs(t) + \bG(t)\vu(t)
\end{equation}
$\sext(\vu(t), t)$ is a new, extended ``source term''. Now, Eq.~\eqref{TDodes} can be written as
\begin{equation}\label{TDodes1}
	\deriv{\bvec{u}(t)}{t} = \tG\bvec{u}(t) + \sext(\vu(t), t)
\end{equation}
which resembles the form of Eq.~\eqref{ihodes}. The Duhamel principle can be applied to yield:
\begin{equation}\label{DuhamelTD}
	\vu(t) = \exp(\tG t)\vu_0 + \exp(\tG t)\int_0^t \exp(-\tG\tau)\sext(\vu(\tau), \tau)\,d\tau
\end{equation}
As in the case of Sec.~\ref{sssec:gnenralih}, we can write an approximation of this equality in the form of Eq.~\eqref{solih2}:
\begin{equation}
	\vu(t) \approx f_M(\tG,t)\vv_M + \sum_{j=0}^{M-1} \frac{t^j}{j!} \vv_j 
\end{equation}
The $\vv_j$ vectors are computed by expanding $\sext$ in time in the form of \eqref{pols}, and using the resulting coefficients, as in Sec.~\ref{sssec:gnenralih}.

Apparently, this gives nothing---Eq.~\eqref{DuhamelTD} is an integral equation, and the RHS includes a dependence on $\vu(t)$ itself, which is still unknown. Consequently, the $\vv_j$'s also depend on $\vu(t)$. However, it is possible to utilize this form for obtaining a solution by an \emph{iterative procedure}. First, we \emph{guess} a solution $\vu_g(t)$ in the desired time interval. Then, we use $\vu_g(t)$ for the computation of the RHS of the equation. We obtain for the LHS a new approximated solution. It seems reasonable that it should be closer to the actual solution than $\vu_g(t)$. We can use the improved solution for obtaining a better one by inserting it into the RHS, and so on. This procedure can be continued until the solution converges with a desired accuracy.

This iterative scheme sounds reasonable. However, we have not given a rigorous justification to it. Thus, one might suspect if it should actually work. Experience shows that this iterative process does converge to the solution, given that \emph{the time-interval is sufficiently small}. Thus, the iterative procedure has a \emph{convergence radius}. The size of the convergence radius is problem dependent. When the time interval is larger than the convergence radius, the solution diverges, \ie it tends to infinity with the number of iterations.



A more rigorous justification to the iterative procedure can be obtained by a convergence analysis. This topic is left for a future paper.

In the case that the time of the desired solution is outside the convergence interval of the algorithm, this procedure cannot be used directly. Instead, we can use a \emph{time-step algorithm}, in a similar manner to the Taylor approach. We divide the time-interval into smaller time-steps, in which the iterative procedure converges. In each time-step we solve the sub-problem of obtaining the solution within the time-step. We use the solution obtained in order to compute $\vu_g(t)$ for the next time-step, as will be described later.

We see that at the end of the day we still need a time-step propagation, as in the Taylor approach. The advantage of the present approach is that we can use much larger time-steps, which means that the accumulation of errors and the computational effort can be much reduced. The approach for the computation of each time-step is global, replacing the local considerations of the Taylor approach. However, the algorithm still contains an obvious local element, in the sense that the solution is computed separately in each local time-step. Hence, we can call this approach a \emph{``semi-global approach''}. 

Note that the definition of the $\vs_j$'s, the $\vw_j$'s and corresponding $\vv_j$'s is different for each time-step. Let us denote the $k$'th time-point by $t_k$. The time-interval of the $k$'th time-step is \text{$[t_k, t_{k+1}]$}. The $\vs_j$'s, the $\vw_j$'s and the $\vv_j$'s in the $k$'th time-step will be denoted by $\vs_{k,j}$, $\vw_{k,j}$ and $\vv_{k,j}$, accordingly. The $\vs_{k,j}$'s are computed using the expansion \eqref{sPn} of $\sext(t)$ \emph{within the $k$'th time-interval}. The $\vw_{k,j}$'s are defined as:
\begin{equation}\label{vws_tk}
	\vw_{k,j} \equiv 
	\begin{cases}
		\vu(t_k) & j=0 \\
		\vs_{k,j-1} & 0 < j \leq M
	\end{cases}
\end{equation}
The $\vv_{k,j}$'s are computed accordingly by the recursion
\begin{align}
	& \vv_{k,0} = \vu(t_k) \nonumber \\
	& \vv_{k,j} = \tG\vv_{k,j-1} + \vw_{k,j}, \qquad\qquad j=1,2,\ldots \label{vrecursion_tk}
\end{align}
The solution within the $k$'th time-step is:
\begin{equation}\label{sol_tk}
	\vu(t) = f_M(\tG, t-t_k)\vv_{k,M} + \sum_{j=0}^{M-1} \frac{(t-t_k)^j}{j!} \vv_{k,j}, \qquad\qquad t \in [t_k, t_{k+1}] 
\end{equation}

Two questions remained open: How $G(t)$ should be split (see Eq.~\eqref{splitG}), and how the guess solution $\vu_g(t)$ should be chosen. We begin with the first question. From a physical point of view, it seems that a natural choice in many problems is to split the Hamiltonian in the following way:
\begin{equation}\label{splitH0}
	H(t) = H_0 + V(t)
\end{equation}
where $H_0$ is the unperturbed Hamiltonian and $V(t)$ is a time-dependent perturbation. If, in addition, $\vs(t)\equiv 0$, Eq.~\eqref{DuhamelTD} becomes
\begin{equation}\label{DuhamelH0V}
	\vu(t) = \exp(-iH_0 t)\vu_0 -i\int_0^t \exp[-iH_0(t-\tau)] V(\tau)\vu(\tau)\,d\tau
\end{equation}
which has a striking resemblance to the well-known expression of the first-order time-dependent perturbation theory (see, for example, \cite[Chapter~XIII]{CohenTan}). Indeed, the expressions become identical by the replacement of $\vu(\tau)$ in the integral by $\vu_0$. However, although this option is appealing in the sense of the directness of the physical interpretation, it needn't be the best option from a numerical point of view. The result may converge faster with other choices of splitting.

A more educated choice of splitting comes to us when we realize that the weak point of the algorithm lies in the point where we ``cheat''. This point is the treatment of the $\vu(t)$ dependent ``source term'', $\sext(\vu(t), t)$, as an inhomogeneous, $\vu(t)$ independent term. We should choose the splitting in a way that minimizes the size of the $\vu(t)$ dependence in $\sext(\vu(t), t)$. Hence, $\bG(t)$ should be as small as possible. Consider the $k$'th time-step, in the time-interval \text{$[t_k, t_{k+1}]$}. For the sake of generality, we consider also a non-equidistant time-grid. Let us denote: \text{$\Delta t_k = t_{k+1}-t_k$}. Usually, $\Delta t_k$ can be assumed to be small, by the requirements of the convergence radius of the algorithm. Hence, we can assume that $G(t)$ does not change much during the time-interval. Then it becomes reasonable to choose the following splitting:
\begin{align}
	& \tG = G\left(t_k + \frac{\Delta t_k}{2} \right) \label{tG}\\
	& \bG(t) \equiv G(t) - G\left(t_k + \frac{\Delta t_k}{2} \right) & t \in [t_k, t_{k+1}] \label{bG}
\end{align}
Obviously, the splitting of Eqs.~\eqref{tG}-\eqref{bG} is time-step dependent, unlike the splitting of Eq.~\eqref{splitH0}.

The choice of the guess solution $\vu_g(t)$ is lead by two contradicting considerations: On one hand, we need a sufficiently accurate starting point for the iterative process. On the other hand, we require that it can be obtained with minimal amount of extra numerical effort. One obvious choice, which requires no extra numerical effort, is the zero'th order approximation---within the $k$'th time-step interval, \text{$[t_k, t_{k+1}]$}, the guess solution is
\begin{equation}\label{guess0}
	\vu_g(t) \equiv \vu(t_k) \qquad\qquad  t \in [t_k, t_{k+1}]
\end{equation}
However, this approximation is of low accuracy. Actually, a very accurate approximation can be obtained from an \emph{extrapolation} of the solution in the previous time-step. All we need to do is to use Eq.~\eqref{sol_tk} for the previous time-step to compute the solution in the current time-step. We obtain the following approximated guess solution:
\begin{equation}\label{guess_tk}
	\vu_g(t) = f_M(\tG,t-t_{k-1})\vv_{k-1,M} + \sum_{j=0}^{M-1} \frac{(t-t_{k-1})^j}{j!} \vv_{k-1,j}, \qquad\qquad t \in [t_k, t_{k+1}] 
\end{equation}
This solution approximates the solution in the interval \text{$[t_k, t_{k+1}]$}, using information from the previous interval \text{$[t_{k-1}, t_k]$}. Note that the second argument of the function $f_M(z,t)$ in Eq.~\eqref{guess_tk} represents a different time interval from that of Eq.~\eqref{sol_tk}. As in Eq.~\eqref{ftj}, the function is expanded in $z$, and $t$ serves as a parameter. Hence, the functions to be computed are different in the two equations, and new expansion coefficients have to be computed in the new interval. The numerical effort of this operation is negligible in comparison to the matrix-vector multiplications (or the equivalent linear operations), unless the dimension $N$ of the problem is very small. Thus, by Eq.~\eqref{guess_tk} we obtain an accurate guess with a relatively low computational cost.

In the first time-step, $\vu_g(t)$ can be computed by Eq.~\eqref{guess0}. More iterations will be required in comparison with the other time-steps, but the overall additional computational effort in a many time-point grid is negligible. 

\subsubsection{Nonlinear Hamiltonian}

Let the Hamiltonian include also a dependence on $\vu(t)$, \ie \text{$H \equiv H(\vu(t),t)$}. Now we get to the general case of Eq.~\eqref{GeneralSchr}, or equivalently, Eq.~\eqref{odes}. The treatment of this case is completely analogous to that of the linear time-dependent Hamiltonian.

Let us split $G(\vu(t), t)$ in the following way:
\begin{equation}\label{splitGu}
	G(\vu(t), t) = \tG + \bG(\vu(t), t)
\end{equation}
where $\tG$ is linear and time-independent. Let us define:
\begin{equation}\label{sext2}
	\sext(\vu(t), t) = \vs(t) + \bG(\vu(t), t)\vu(t)
\end{equation}
The rest of the algorithm is identical to that of Sec.~\ref{sssec:TDH}, for the same considerations.
 
In an analogous manner to the time-dependent linear case, it is recommended to choose the following splitting in the time-step algorithm:
\begin{align}
	& \tG = G\left[\vu \left(t_k + \frac{\Delta t_k}{2}\right), t_k + \frac{\Delta t_k}{2} \right] \label{tGnl}\\
	& \bG(\vu(t), t) \equiv G(\vu(t), t) - G\left[\vu\left(t_k + \frac{\Delta t_k}{2} \right), t_k + \frac{\Delta t_k}{2} \right] & t \in [t_k, t_{k+1}] \label{bGnl}
\end{align}

\section{Implementation}\label{sec:implementation}

\subsection{The propagation time-grid}\label{ssec:tgrid}

In order to obtain the $\vs_m$ Taylor like coefficients, we have to know the total time-dependence of $\sext(\vu(t), t)$ (see Sec.~\ref{sssec:gnenralih}). Of course, we have no explicit expression for the total time-dependence of $\sext(\vu(t), t)$, because of its dependence on $\vu(t)$. Hence, the time-dependence of $\sext(\vu(t), t)$ has to be approximated from several \emph{sampling points} within the time interval. In each sampling point $t_l$, we need the values of $\vu(t_l)$, $G(\vu(t_l), t_l)$ and $\vs(t_l)$ in order to compute $\sext(\vu(t_l), t_l)$. The $\vb_n$'s from Eq.~\eqref{sPn} are obtained from the samplings of $\sext(\vu(t), t)$ within the time-interval. Then, the $\vs_m$'s can be computed from the $\vb_n$'s as described in Sec.~\ref{sssec:gnenralih}.

It is recommended to choose the \emph{Chebyshev points} within the time-interval as the sampling points. Then, $\sext(\vu(t), t)$ can be expanded in time either by a Chebyshev polynomial expansion, or by a Newton interpolation at the Chebyshev points (see Appendix~\ref{app:PolApprox}). When a Chebyshev polynomial expansion is used, the $\vb_n$'s from Eq.~\eqref{sPn} correspond to the Chebyshev coefficients, that will be denoted by $\vc_n$, and the polynomials are the Chebyshev polynomials. When a Newton interpolation is used, the $\vb_n$'s are the divided differences, that will be denoted by $\va_n$, and the polynomials are the Newton basis polynomials. In appendix~\ref{app:PolApprox} we describe how the $\va_n$'s or the $\vc_n$'s can be obtained from the samplings at the Chebyshev points.

In the time-step algorithm, the time interval of each time-step is sampled at the Chebyshev points of the interval. Thus, the structure of the time-grid necessary for the propagation is complex; it consists of adjacent time-intervals, each with an internal Chebyshev grid. In order to refer also to the internal grid of each interval, we shall replace the single index notation of the time-grid from Sec.~\ref{sssec:TDH}, $t_k$, by a double index notation, $t_{k,l}$. The first index $k$ refers to the $k$'th time-interval, where \text{$k = 1,2,\ldots,N_t$}. The second index $l$ indexes the points in the internal Chebyshev grid of each interval, as will be readily seen.

The length of the $k$'th time-interval is denoted by $\Delta t_k$, as in Sec.~\ref{sssec:TDH}. For the sake of generality, we will consider also the possibility that the number of Chebyshev points is different for each time-step. The number of Chebyshev points in the $k$'th interval is denoted by \text{$M_k$}. $M_k$ is also the number of expansion terms for the approximation of $\sext(\vu(t), t)$ (see Eq.~\eqref{sPn}).

We use the set of boundary including Chebyshev points. In the Chebyshev domain, \text{$[-1, 1]$}, the Chebyshev grid is defined as follows (Cf.~Appendix~\ref{App:NewtonCheb}, Eq.~\eqref{chebpbrev}; note that the equations of Appendix~\ref{app:PolApprox} are formulated in the terms of the order of the polynomial approximation, which is equivalent to \text{$M_k - 1$}):
\begin{equation}\label{chebdgrid}
	y_{k,l} \equiv -\cos\left(\frac{l\pi}{M_k - 1} \right), \qquad l=0,1,\ldots,M_k-1
\end{equation}
In the $k$'th time-step domain, the Chebyshev points become (Cf.~Eq.~\eqref{y2x}):
\begin{equation}\label{tchebgrid}
	t_{k,l} = t_{k,0} + \frac{\Delta t_k}{2}(1 + y_{k,l})
\end{equation}
Note that we have:
\begin{equation}\label{multip}
	t_{k,M_k-1} = t_{k,0} + \Delta t_k =  t_{k+1,0}
\end{equation}
Eqs.~\eqref{tchebgrid}, \eqref{multip} define together the entire time grid, where $t_{1,0}$ and the $\Delta t_k$'s are given.

\subsection{Algorithm}\label{ssec:algorithm}

We assume that the initial condition, {$\vu(t_{1,0})$}, is given. In addition, it is assumed that the structure of the propagation time-grid (\ie $\Delta t_k$ and $M_k$ for each time-step) is known in advance. Alternatively, it is possible to choose it adaptively during the propagation, by an internal procedure. The number of expansion terms for the approximation of \text{$f_{M_k}(\tG, t-t_{k,0})\vv_{M_k}$} (see Eq.~\eqref{sol_tk}) is also supplied by the user. It may depend on $k$.

The accuracy of the solution is determined by a tolerance parameter, which will be denoted by $\epsilon$. It represents the order of the accepted relative error of the solution. The tolerance parameter is supplied by the user.

The scheme of propagation goes as follows:

\begin{enumerate}
	\item Set the guess solution of the first time-step in the internal Chebyshev grid (Cf. Eq.~\eqref{guess0}): 
	\[
		\vu(t_{1,l}) = \vu(t_{1,0}), \qquad l=0,1,\ldots,M_1 - 1
	\]
	\item for $k=1$ to $N_t$
	\begin{enumerate}
		\item Set the middle point of the internal grid, \text{$t_{mid} = t_{k, M_k\backslash 2}$}, where $\backslash$ denotes integer division.
		\item (\text{$l=0,1,\ldots,M_k - 1, \qquad n=0,1,\ldots,M_k - 1, \qquad j=0,1,\ldots,M_k$})
		\item do \label{pr:do}
		\begin{enumerate}
			\item Set \text{$\sext^l = \vs(t_{k,l}) + [G(\vu(t_{k,l}),t_{k,l}) - G(\vu(t_{mid}),t_{mid})]\vu(t_{k,l})$} (Cf.~Eqs.~\eqref{sext2}, \eqref{bGnl}). \label{pr:loop_beginning}
			\item Use the $\sext^l$'s to compute the expansion coefficients of \text{$\sext(\vu(t), t)$} in the time-step.
			\begin{itemize}
				\item \emph{For a Newton interpolation}: Compute the divided differences \text{$\va_n$} recursively, as described in Appendix~\ref{app:NewtonApprox} (relevant equations: \eqref{dvd0}, \eqref{dvd},\eqref{an}). Use \text{$4t_{k,l}/\Delta t_k$} as the sampling points (see Sec.~\ref{app:Newton4}), and the corresponding $\sext^l$'s as the function values.
				\item \emph{For a Chebyshev expansion}: Compute the Chebyshev coefficients \text{$\vc_n$}, as described in Appendix~\ref{app:ChebApprox} (relevant equation: \eqref{chebc_arbxb_rev}). Use the $\sext^l$'s as the function values.
			\end{itemize} \label{pr:sext2b}
			\item Compute the $\vs_n$ Taylor-like coefficients recursively from the $\va_n$'s or the $\vc_n$'s, using the conversion schemes described in Appendix~\ref{app:pol2Taylor} (relevant equations: \eqref{Nrecursion_qn04}-\eqref{Nrecursion_qnn4}, \eqref{q2s_dvd} for the $\va_n$'s, \eqref{Crecursion_qn0}-\eqref{q2s_cheb} for the $\vc_n$'s). \label{pr:b2s}
			\item Compute the $\vv_j$ vectors recursively from $\vu_{k,0}$ and the $\vs_n$'s (see Eqs.~\eqref{vrecursion_tk}, \eqref{vws_tk}), where \text{$\tG=G(\vu(t_{mid}), t_{mid})$} (Cf.~Eq.~\eqref{tGnl}). \label{pr:s2v}
			\item Store the current solution at the time-step edge for convergence check, \text{$\vu_{old}=\vu(t_{k, M_k - 1})$}.
			\item Compute a new solution from the $\vv_j$'s by the expression (Cf.~Eq.~\eqref{sol_tk}):
			\[			
				\vu(t_{k,l}) = f_{M_k}(\tG, t_{k,l}-t_{k,0})\vv_{M_k} + \sum_{j=0}^{M_k-1}\frac{(t_{k,l}-t_{k,0})^j}{j!}\vv_j
			\]
			$f_{M_k}(\tG, t_{k,l}-t_{k,0})\vv_{M_k}$ is approximated by one of the methods described in Appendix~\ref{app:FunMat} (note that the expansion vectors required for the approximation are computed just once for all time points---see Sec.~\ref{sssec:TiH}). \label{pr:uts}
			\item Repeat from step \ref{pr:loop_beginning} while
			\[
				\frac{\Vert\vu(t_{k, M_k - 1}) - \vu_{old}\Vert}{\Vert \vu_{old}\Vert}>\epsilon
			\] \label{pr:conv}
		\end{enumerate}
		\item end do
		\item Compute the solution at any desired point \text{$t_p\in[t_{k,0}, t_{k,M_k-1}]$} by
		\[			
			\vu(t_p) = f_{M_k}(\tG, t_p-t_{k,0})\vv_{M_k} + \sum_{j=0}^{M_k-1}\frac{(t_p-t_{k,0})^j}{j!}\vv_j
		\] \label{pr:up}
		\item Set the guess solution for the next time-step; by definition: \text{$\vu(t_{k+1,0}) = \vu(t_{k,M_k-1})$} (see Eq.~\eqref{multip}). The guess solution at the rest of the Chebyshev internal points is computed by (Cf.~Eq.~\eqref{guess_tk}):
		\begin{align*}
			& \vu(t_{k+1,m}) = f_{M_k}(\tG, t_{k+1,m}-t_{k,0})\vv_{M_k} + \sum_{j=0}^{M_k-1}\frac{(t_{k+1,m}-t_{k,0})^j}{j!}\vv_j,\\
			& m = 1,\ldots,M_{k+1}-1		
		\end{align*} \label{pr:ug}
	\end{enumerate}
	\item end for
\end{enumerate}

The algorithm is sketched schematically in Fig.~\ref{fig:flowchart}.

\tikzstyle{decision} = [diamond, draw, fill=yellow!30, 
    text width=1.5cm, text badly centered, node distance=3cm, inner sep=0pt]
\tikzstyle{block} = [rectangle, draw, fill=green!30, 
    text width=6.7cm, text centered, node distance=1.6cm, minimum height=3em]
\tikzstyle{narrowblock} = [rectangle, draw, fill=green!30, 
    text width=6.7cm, text centered, node distance=1.1cm, minimum height=2em]
\tikzstyle{smallblock} = [rectangle, draw, fill=green!30, 
    text width=2cm, text centered, node distance=1.1cm, minimum height=1.5em]
\tikzstyle{startstop} = [rectangle, rounded corners, draw, fill=blue!30, 
    text width=2cm, text centered, node distance=1.1cm, minimum height=1.5em]
\tikzstyle{arrow} = [thick,->,>=stealth]
\tikzstyle{connector} = [draw, circle, fill=red!20, node distance=1.1cm, minimum height=2em]

\begin{figure}

{\fontsize{10pt}{12pt}


\begin{center}

\vspace{-1cm}

\begin{tikzpicture}[node distance = 2cm, auto]
    \node [startstop] (start) {Start};
    \node [narrowblock, below of = start] (guess0) {Set $\vu(t_{1,l})\equiv \vu_0$ for all $l$'s};
    \node [smallblock, below of = guess0] (k1) {$k=1$};
    \node [connector, below of = k1] (connector1) {};
    \node [connector, below of = connector1] (connector2) {};
    \node [narrowblock, below of = connector2, yshift = -0.1cm] (sext) {Compute $\sext(\vu(t_{k,l}),t_{k,l})$ for all $l$'s};
    \node [block, below of = sext, yshift = 0.1cm] (cheb) {Approximate $\sext(\vu(t),t)$ in the $k$'th time-interval by a Chebyshev approximation};
    \node [block, below of = cheb] (taylor) {Convert the Chebyshev approximation of $\sext(\vu(t),t)$ to a Taylor form};
    \node [narrowblock, below of = taylor, yshift = -0.3cm] (vj) {Compute the $\vv_j$'s recursively};
    \node [narrowblock, below of = vj, yshift = -0.1cm] (sol) {Compute the solution $\vu(t_{k,l})$ for all $l$'s};
    \node [decision, below of = sol, yshift = 1cm] (conv) {$\vu(t)$ converged?};
    \node [block, below of = conv, yshift = -0.6cm] (sol2) {Compute $\vu(t)$ at any desired point in the $k'th$ time-interval};
    \node [block, below of = sol2] (guess) {Compute $\vu(t_{k+1,l})$ for all $l$'s by extrapolating the solution into the next time-interval};
    \node [smallblock, below of = guess, yshift = -0.2cm] (kupdate) {$k=k+1$};
    \node [decision, below of = kupdate, yshift = 1.3cm] (klimit) {$k\leq N_t$?};
    \node [startstop, below of = klimit, yshift = -0.5cm] (end) {End};

    \draw [arrow] (start) -- (guess0);
    \draw [arrow] (guess0) -- (k1);
    \draw [arrow] (k1) -- (connector1);
    \draw [arrow] (connector1) -- (connector2);
    \draw [arrow] (connector2) -- (sext);
    \draw [arrow] (sext) -- (cheb);
    \draw [arrow] (cheb) -- (taylor);
    \draw [arrow] (taylor) -- (vj);
    \draw [arrow] (vj) -- (sol);
    \draw [arrow] (sol) -- (conv);
    \draw [arrow] (conv) -- node [xshift = 0.1cm] {yes} (sol2);
	\draw [arrow] (conv) -- node [yshift = -0.1cm] {no} ++ (-4.3cm, 0) |- (connector2);
	\draw [arrow] (sol2) -- (guess);
    \draw [arrow] (guess) -- (kupdate);
	\draw [arrow] (kupdate) -- (klimit);
	\draw [arrow] (klimit) -- node [xshift = 0.1cm] {no} (end);
	\draw [arrow] (klimit) -- node [yshift = -0.5cm] {yes} ++ (4.3cm, 0) |- (connector1);

\end{tikzpicture}

\end{center}

\vspace{-0.5cm}

\caption{The semi-global propagator algorithm}\label{fig:flowchart}   

}
\end{figure}
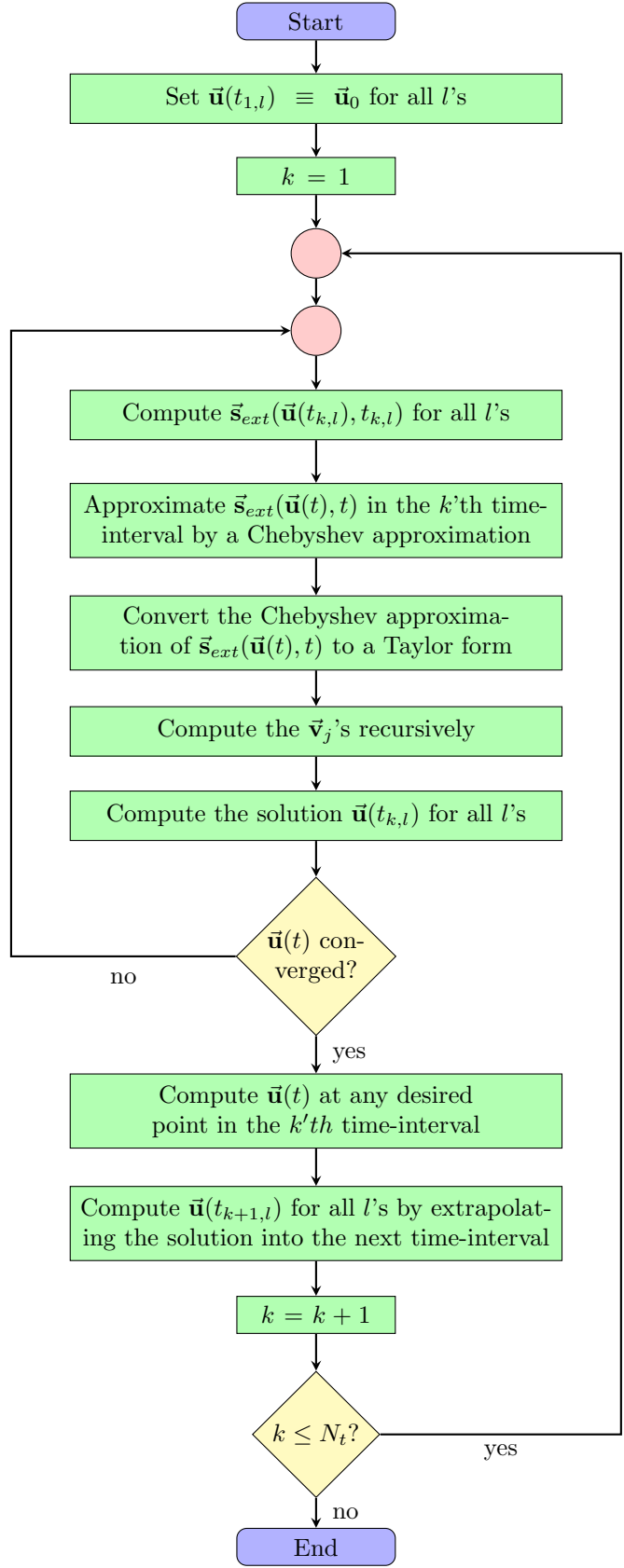

\subsection{Programming}\label{ssec:programming}

\subsubsection{Numerical stability of the time polynomial expansion}\label{sssec:coefficients}

In Eqs.~\eqref{pols}, \eqref{PnTaylor}, the coefficients of the $t$ polynomials are defined as the coefficients of $t^m/m!$, in analogy to the Taylor expansion form. Accordingly, we obtained in Eq.~\eqref{solih2} the $\vv_m$'s as the coefficients of $t^m/m!$\ . The $1/m!$ factor decreases very fast as $m$ grows. Consequently, the $\vs_m$'s, the $q_{n,m}$'s and the $\vv_m$'s tend to attain huge values. This may lead to numerical instability.

The problem can be solved by the definition of alternative polynomial expansions, in which the coefficients absorb the $1/m!$ factor. The alternative expansions will lead to expressions which are more stable numerically. The source term $\vs(t)$ is expanded as
\begin{equation}\label{polst}
	\vs(t) \approx \sum_{m=0}^{M-1}t^m\vts_m
\end{equation}
where \text{$\vts_m=\vs_m/m!$}\ . Accordingly, Eq.~\eqref{PnTaylor} is replaced by
\begin{equation}\label{PnTaylort}
	P_n(t) = \sum_{m=0}^n \tq_{n,m} t^m
\end{equation}
where \text{$\tq_{n,m}=q_{n,m}/m!$}\ .

First we derive the solution equation for a time-independent Hamiltonian. We plug the expansion \eqref{polst} into Eq.~\eqref{solih}. The derivation of the solution is similar to that of Sec.~\ref{sssec:polih}, but less appealing from an aesthetic point of view. We end with the following equation:
\begin{equation}\label{soliht}
	\vu(t) = \tf_M(G_0,t)\vtv_M + \sum_{j=0}^{M-1} t^j \vtv_j 
\end{equation}
where we defined
\begin{equation}\label{tfs}
	\tf_{m}(z,t) \equiv
	\begin{cases}		
		\frac{m!}{z^m}\left[\exp(zt) - \sum_{j=0}^{m-1} \frac{(zt)^j}{j!} \right] & z \neq 0 \\
		t^m & z = 0 
	\end{cases}
	\qquad\qquad m=0,1,\ldots
\end{equation}
The $\vtv_j$'s are defined recursively in the following way:
\begin{align}
	& \vtv_0 = \vu_0 \nonumber \\
	& \vtv_j = \frac{G_0\vtv_{j-1} + \vts_{j-1}}{j}, \qquad\qquad j=1,2,\ldots \label{tvrecursion}
\end{align}
Note that we have:
\begin{align}
	& \tf_{m}(z,t) = m!f_{m}(z,t) \\
	& \vtv_j = \frac{\vv_j}{j!}
\end{align}
Thus, it can be easily verified that Eq.~\eqref{soliht} is equivalent to Eq.~\eqref{solih2}.

In the case of a time-dependent nonlinear Hamiltonian, we expand $\sext(t)$ as in Eq.~\eqref{polst}, and replace $G_0$ in Eqs.~\eqref{soliht}, \eqref{tvrecursion}, by $\tG$.

The computation of the $\tq_{n,m}$'s is completely analogous to that of the $q_{n,m}$'s. The procedure is given in Appendix~\ref{app:pol2Taylor}.

In summary, in order to improve the stability of the program, the following changes should be made in the algorithm:
\begin{itemize}
	\item In step \ref{pr:b2s}, the $\vts_n$'s are computed instead of the $\vs_n$'s, via the computation of the $\tq_{n,m}$'s (relevant equations: \eqref{Nrecursion_tqn04}-\eqref{Nrecursion_tqnn4}, \eqref{tq2s_dvd} for the $\va_n$'s, \eqref{Crecursion_tqn0}-\eqref{tq2s_cheb} for the $\vc_n$'s).
	\item In step \ref{pr:s2v}, the $\vtv_j$'s are computed instead of the $\vv_j$'s, by the recursion
	\begin{align*}
		& \vtv_0 = \vu_0 \nonumber \\
		& \vtv_j = \frac{\tG\vtv_{j-1} + \vts_{j-1}}{j}, \qquad j=1,2,\ldots
	\end{align*}
	where \text{$\tG = G(\vu(t_{mid}), t_{mid})$}.
	\item In steps \ref{pr:uts}, \ref{pr:up} and \ref{pr:ug}, the solution at the relevant time-points is computed by
	\begin{equation}\label{tsol}
		\vu(t) = \tf_M(\tG ,t-t_{k,0})\vtv_M + \sum_{j=0}^{M-1} (t-t_{k,0})^j \vtv_j
	\end{equation}
\end{itemize}

\subsubsection{The computation of $\tf_m(z,t)$}

The function $f_m(z,t)$, and its variant, $\tf_m(z,t)$, include the following expression (see Eqs. \eqref{fs}, \eqref{tfs}):
\begin{equation}\label{exp_diff}
	\exp(zt) - \sum_{j=0}^{m-1} \frac{(zt)^j}{j!}
\end{equation}
The sum is just a truncated Taylor expansion of $\exp(zt)$. If $zt$ is small, the difference between $\exp(zt)$ and its truncated expansion becomes extremely small. Often, this results in roundoff errors.

The problem can be solved by an alternative computation of $\tf_m(z,t)$ for small $zt$ values. Let us expand $\exp(zt)$ from \eqref{exp_diff} by a Taylor expansion. \eqref{exp_diff} can be expressed as a ``tail'' of the Taylor expansion in the following way:
\begin{equation}
	\exp(zt) - \sum_{j=0}^{m-1} \frac{(zt)^j}{j!} = \sum_{j=0}^\infty \frac{(zt)^j}{j!} - \sum_{j=0}^{m-1} \frac{(zt)^j}{j!} = \sum_{j=m}^\infty \frac{(zt)^j}{j!} 
\end{equation}
The expression for $\tf_m(z,t)$ becomes:
\begin{equation}\label{tfTaylor}
	\tf_m(z,t) = m!\,t^m\sum_{j=0}^\infty \frac{(zt)^j}{(j+m)!}
\end{equation}
$\tf_m(z,t)$ can be computed by truncating the sum in \eqref{tfTaylor}. The Taylor expansion converges very slowly. Hence, the expansion should be truncated only after achieving the machine accuracy in the summation procedure. The form \eqref{tfTaylor} should not be used when $zt$ is large enough to be computed directly.

\subsubsection{Efficiency of the computation of $\sext(\vu(t), t)$}

The number of required matrix-vector multiplications for the computation of the integrated form \eqref{solih2} is \text{$M+K-1$} for a polynomial approximation of the function of matrix, and $M+K$ for the Arnoldi approach, Cf.~Sec.~\ref{sssec:polih}. The computation of $\sext(\vu(t_{k,l}), t_{k,l})$ in the general case (step~\ref{pr:loop_beginning} in the algorithm) seems to cost considerable amount of additional computational effort. It can be readily seen from step~\ref{pr:loop_beginning} that a direct computation of each of the $\sext^l$'s requires a subtraction of two matrices, and a matrix-vector multiplication. Subtraction of matrices has the same scaling as a matrix-vector multiplication, $O(N^2)$. An alternative, which is less time-consuming, is to perform the computation as \text{$\sext^l = \vs(t_{k,l}) + G(\vu(t_{k,l}),t_{k,l})\vu(t_{k,l}) - G(\vu(t_{mid}),t_{mid})\vu(t_{k,l})$}. This requires two matrix-vector multiplications for each $l$. This is with the exception of \text{$l=M_k\backslash 2$}, which indexes the middle internal time-point, $t_{mid}$; the extended part of the inhomogeneous term vanishes, and we are left with \text{$\sext^{M_k\backslash 2} = \vs(t_{mid})$}. Thus, the computation in the middle point does not involve additional expensive operations. The overall additional cost of step~\ref{pr:loop_beginning} is $2(M_k - 1)$ matrix-vector multiplications. This roughly doubles the computational effort.

Most frequently, the computational effort can be considerably reduced by a proper formulation of the calculation. First, in many problems, the operator represented by $G(\vu(t_{k,l}),t_{k,l}) - G(\vu(t_{mid}),t_{mid})$ is diagonal in the basis of representation. Thus, the scaling of its operation on $\vu(t_{k,l})$ becomes linear, $O(N)$. The computational cost of this operation is negligible. A common example is a Hamiltonian which is composed from a stationary part and a time-dependent nonlinear potential,
\begin{equation}\label{H0V}
	H(\vu(t), t) = H_0 + V(\vu(t), t)
\end{equation}
In this case we have:
\begin{equation}
	[G(\vu(t_{k,l}),t_{k,l}) - G(\vu(t_{mid}),t_{mid})]\vu(t_{k,l}) = -i[V(\vu(t_{k,l}),t_{k,l}) - V(\vu(t_{mid}),t_{mid})]\vu(t_{k,l})
\end{equation}
When the problem is represented in the spatial basis, the potential becomes diagonal. Thus, the computation of $\sext(\vu(t_{k,l}), t_{k,l})$ does not require any additional expensive operation.

Moreover, in many situations, the dependence of $G(\vu(t), t)$ on $\vu(t)$ and $t$ is determined by a small number of parameters. This may reduce the required computational cost. For example, let us consider a Hamiltonian of the form of \eqref{H0V} with a potential
\begin{equation}
	V(\vu(t), t) = \zeta(\vu(t), t)\mu
\end{equation}
where $\zeta(\vu(t), t)$ is a scalar parameter and $\mu$ is a matrix. $\zeta(\vu(t), t)$ may represent an electric or magnetic field (up to a sign), and $\mu$ may represent the electric or magnetic moment operator, respectively (state dependent electric or magnetic fields occur, \eg, in coherent control problems, when the Krotov algorithm is employed; see \cite{QOCT} for a review). We have:
\begin{equation}
	[G(\vu(t_{k,l}),t_{k,l}) - G(\vu(t_{mid}),t_{mid})]\vu(t_{k,l}) = -i[\zeta(\vu(t_{k,l}),t_{k,l}) - \zeta(\vu(t_{mid}),t_{mid})]\,\mu\,\vu(t_{k,l})
\end{equation}
We see that the computation of $\sext(\vu(t_{k,l}), t_{k,l})$ requires just one matrix-vector multiplication (when $\mu$ is a non-diagonal matrix). Thus, the additional computational cost of the $\sext^l$'s is just $M_k - 1$ matrix-vector multiplications.

More generally, let us consider $G(\vu(t), t)$ which can be represented in the following form:
\begin{equation}
	G(\vu(t) ,t) = G_0 + \sum_{j=1}^L \xi_j(\vu(t),t)G_j
\end{equation}
where the $\xi_j(\vu(t), t)$'s are scalar parameters, the $G_j$'s are matrices, and $L\ll N^2$. We have:
\begin{equation}
	[G(\vu(t_{k,l}),t_{k,l}) - G(\vu(t_{mid}),t_{mid})]\vu(t_{k,l}) = \sum_{j=1}^L [\xi_j(\vu(t_{k,l}),t_{k,l}) - \xi_j(\vu(t_{mid}),t_{mid})]G_j\vu(t_{k,l})
\end{equation}
The cost of this operation is less than a single multiplication of a vector by $G(\vu(t), t)$. Since $L\ll N^2$, the computational cost of the expressions \text{$\xi_j(\vu(t_{k,l}),t_{k,l}) - \xi_j(\vu(t_{mid}),t_{mid})$} becomes negligible in comparison to a matrix-vector multiplication (note that the condition for $L$ becomes different when the multiplication by the matrix is represented by an equivalent linear operation procedure with lower scaling than $O(N^2)$).

\subsection{Parameter choice}\label{ssec:parmeters}

Several free parameters are involved in the propagation algorithm. They need to be supplied by the user in each problem. The choice of the parameters determines the efficiency and the accuracy of the algorithm. With an inappropriate choice, the algorithm might become inefficient, inaccurate, or even completely fail. Of course, the parameters which yield good results are problem dependent.

There are two main criteria for a successful choice of the parameters:
\begin{enumerate}
	\item The accuracy of the results;
	\item The efficiency of the algorithm.
\end{enumerate}
The treatment of the first criterion is closely related to the ability to estimate the error of the different approximations involved in the algorithm. This important topic is left to Appendix~\ref{app:error}, due to its length. The present discussion mainly focuses on the second criterion.

In general, a successful choice of parameters can be achieved by trial and error. The choice may improve with experience with the algorithm, and after the treatment of similar problems. Here we give several recommendations which are based on our experience with the algorithm.

The choice of the $\epsilon$ tolerance parameter (see step~\ref{pr:conv}) is obvious: It should be determined by the desired accuracy of the solution. Note that the tolerance parameter is defined for a single time-step. The error of the final solution is expected to accumulate during the propagation, roughly as the sum of the errors of each time-step.

In addition, there are three free parameters which need to be specified in each time-step:
\begin{enumerate}
	\item The length of the time-step interval, $\Delta t_k$;
	\item The number of expansion terms for the approximation of $\sext(\vu(t), t)$, $M_k$;
	\item The number of expansion terms for the approximation of $\tf_{M_k}(\tG, t-t_{k,0})\vtv_{M_k}$, $K_k$.
\end{enumerate}

Our experience shows that the parameters should be chosen such that \emph{a single iteration is required in each time-step}. In other words, the steps of the loop in stage \ref{pr:do} of the algorithm are performed just once. This is with the exception of the first time-step, in which the guess solution is of low accuracy (see Eq.~\eqref{guess0}), and usually two or more iterations are required for a sufficient accuracy. The observation that the algorithm becomes most efficient with a minimal number of iterations is consistent with the reasoning that lead us to the choice of the $G(\vu(t),t)$ splitting (see Sec.~\ref{sssec:TDH})---\emph{the weak point of the algorithm lies in the iterative process}. Hence, the part that the iterative process takes in the computation of an accurate solution should be minimized.

Typically, the values of $M_k$ and $K_k$ should be chosen to lie in the range \text{$5-13$}. For higher values, even though $\Delta t_k$ can be increased, the algorithm usually becomes less efficient. The guess solution for the next time-step begins to become less accurate for large $M$. 
This is due to the high sensitivity of a high order extrapolation to roundoff errors. A high order $K$ is usually simply unnecessary. It should be noted that for high $M$ orders, the algorithm may become numerically unstable. The source of the instability lies in the fact that both \text{$(t-t_{k,0})^j$} and $\tf_j(\tG,t-t_{k,0})$ in Eq.~\eqref{tsol} typically become exceedingly small for high $j$'s. Accordingly, the $\vtv_j$'s attain very large values, and the computational process for obtaining them becomes unstable.

In the future, we plan to develop a version of the algorithm which is parameter free. In such an algorithm, the parameters are specified adaptively by the procedure during the propagation process, according to the accuracy requirements.


\section{Numerical example}\label{sec:example}

In the present section we test the efficiency of the semi-global propagator in solving a numerical problem of physical importance. Several numerical examples of relatively simple problems are already presented in Ref.~\cite{propagator_2012}. In this paper, we test the performance of the propagator in a more realistic physical model problem. Moreover, for the first time, we demonstrate the capability of the algorithm to solve a problem with a \emph{complex spectrum}, \ie a problem in which the eigenvalue spectrum of $G$ is distributed on the \emph{complex plane}. This requires the use of the Arnoldi approach (see Sec.~\ref{app:Arnoldi}) for the computation of \text{$f_M(\tG, t-t_k)\vv_{k,M}$} (see the procedure in Sec.~\ref{ssec:algorithm}, step~\ref{pr:uts}).

We choose a physical problem which is known to be challenging numerically---an atom subject to an intense laser field. This physical situation is characterized by extreme conditions for which an accurate numerical calculation becomes difficult. Under the influence of the intense field, a partial ionization of the atom occurs. The ionized part of the electron wave-function has the characteristics of an unbound particle, thus is spread to large spatial distances from the parent atom. Its dynamics is characterized by a strongly accelerated motion under the influence of the intense field. The central potential of the parent atom has a Coulomb character, which is steep in nature. These characteristics of the problem contribute to the difficulty of an accurate computation of the dynamics.

In order to give a reliable description of the problem, \emph{absorbing boundary conditions} have to be employed. These are implemented here by a \emph{complex absorbing potential}. This is the origin of the complex spectrum in our problem. The topic will be discussed in more detail in Sec.~\ref{ssec:num_imp}.

In Sec.~\ref{ssec:details} we present the physical details of the model problem. In Sec.~\ref{ssec:num_imp} we present the details of the numerical implementation of the physical problem. In Sec.~\ref{ssec:results} the results are presented, and compared to a reference method.

\subsection{The details of the physical problem}\label{ssec:details}

\emph{Remark}: Atomic units are used throughout.

\begin{figure}
	\centering \includegraphics[width=3in]{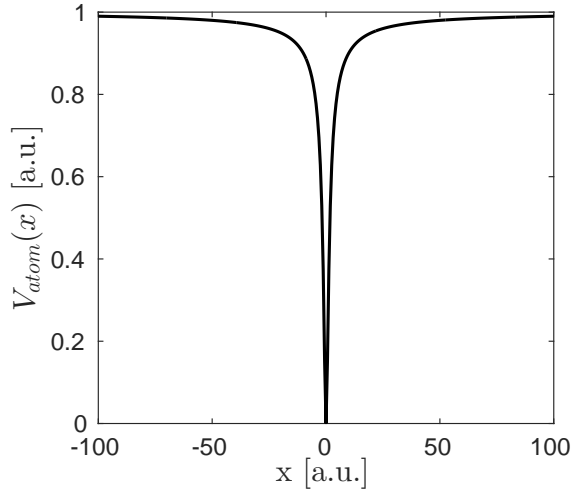}
	\caption{The central atom model potential}\label{fig:Vatom}	
\end{figure}

We use a one-dimensional model for the problem. The central potential of the parent atom is represented by a truncated Coulomb potential (see Fig.~\ref{fig:Vatom}):
\begin{equation}\label{eq:coulV}
	V_{atom}(x) = 1 - \frac{1}{\sqrt{x^2 + 1}}
\end{equation}
In this model, the singularity of the Coulomb potential at \text{$x=0$} is removed. The model has been extensively studied in the context of intense laser atomic physics. The fundamental energy difference, \text{$\Delta E_1 =0.395\,a.u.$}, is similar to that of the hydrogen atom ($0.375\,a.u.$).

\begin{figure}
	\centering \includegraphics[width=3in]{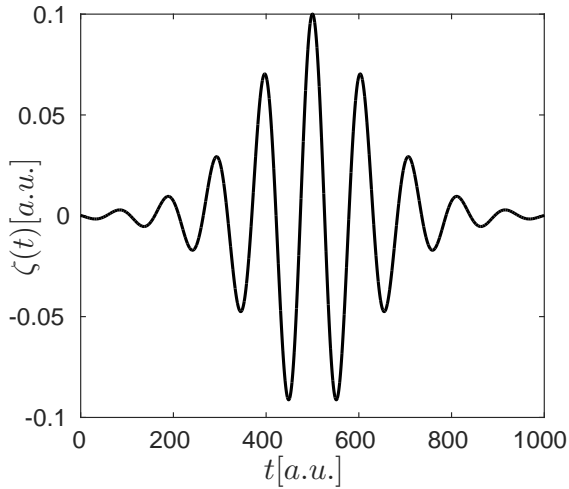}
	\caption{The electric field}\label{fig:field}	
\end{figure}

The laser pulse electric field has the following form (see Fig.~\ref{fig:field}):
\begin{equation}
	\zeta(t) = 0.1\,\sech^2\left(\frac{t-500}{170}\right)\cos[0.06(t-500)]
\end{equation}
The central frequency of the pulse is \text{$\omega = 0.06\,a.u.$}, which corresponds to a wavelength \text{$\lambda = 760_{nm}$}---similar to the central wavelength of the common Titanium-Sapphire laser. The envelope $\sech^2$ form is known to be similar to the actual form of laser pulses. The peak amplitude of the field, \text{$\zeta_{max} = 0.1\,a.u.$}, corresponds to an intensity of \text{$I_{max} = 3.52\times 10^{14}\,{W/cm^2}$}. The final time is \text{$T=1000\,a.u.$}, which corresponds to a pulse duration of $24.2\,{fs}$.

The dipole approximation is employed for the potential induced by the laser field. The total potential of the \emph{physical model} is
\begin{equation}
	V_{phys}(x, t) = V_{atom}(x) + V_{field}(x, t) = 1 - \frac{1}{\sqrt{x^2 + 1}} - x\zeta(t)
\end{equation}
However, the potential of the \emph{numerical problem} must be modified in order to obtain a reliable description of the physical situation, as will be explained in Sec.~\ref{ssec:num_imp}.

The mass of the electron is \text{$m=1\,a.u.$}, and the kinetic energy becomes $p^2/2$. The total time-dependent \emph{physical Hamiltonian} is
\begin{equation}
	H(t) = \frac{p^2}{2} + 1 - \frac{1}{\sqrt{x^2 + 1}} - x\zeta(t)
\end{equation}

The dynamics is governed by the time-dependent Schr\"odinger equation, Eq.~\eqref{Schr}.

\subsection{Numerical implementation of the problem}\label{ssec:num_imp}

The Fourier grid method \cite{FourierGrid} is employed for the Hamiltonian operation.

The $x$ domain is \text{$[-240, 240)$}. We use an equidistant grid, with 768 points. The distance between adjacent grid points becomes $0.625\,a.u.$.

The present physical situation, which involves a partial ionization of the electron, requires a special numerical treatment, in order to prevent the appearance of spurious effects. The reason is that the ionized part of the electron behaves as a free particle, and is spread to very large spatial distances from the parent atom. Hence, the problem cannot be described as is in a finite spatial grid of a reasonable length. The description of the problem by a finite grid involves spurious effects of wraparound or reflection (depending on the computational method) of the wave function at the boundaries of the grid.

Usually, this problem is overcome by the employment of absorbing boundary conditions. The absorbing boundaries are implemented here by the addition of a complex absorbing potential at both boundaries of the grid (for a thorough review see \cite{CAP}). The part of the wave-function which incomes into the complex absorbing potential decays gradually under the influence of the potential, until it becomes practically zero at the edge of the grid. Thus, the spurious effects are prevented. With the addition of the complex potential, the Hamiltonian becomes non-Hermitian, and consequently, the eigenvalue spectrum becomes complex.

Different absorbing potentials vary in their absorption capabilities. The part of the amplitude which is not absorbed by the potential is either reflected by the potential or transmitted. Thus, the efficiency of the absorbing boundaries in the prevention of spurious effects depends on the choice of the absorbing potential. The question of the choice of the absorbing potential becomes important when there is an interest in small amplitude effects. One of the major applications of the present physical situation is in the generation of high-harmonic spectrum, which is a small amplitude effect. It has already been recognized in the former high-harmonic generation simulations (see~\cite{kulander92}) that reflection from the absorbing boundaries is responsible to large spurious effects in the calculation of the high-harmonic spectrum. An appropriate absorbing potential for this calculation could not be found by inspection.

\begin{figure}
	\centering \includegraphics[width=3in]{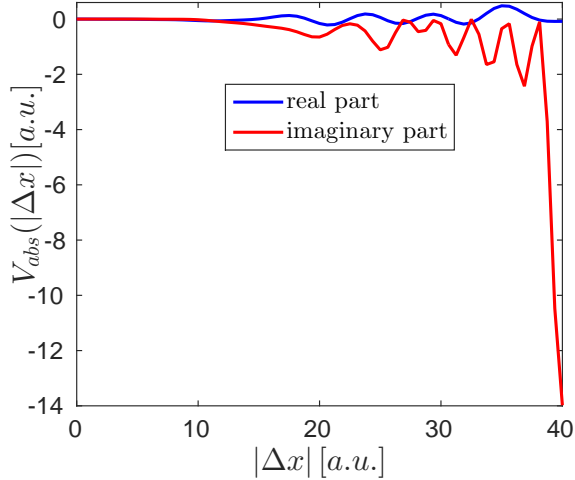}
	\caption{The real and imaginary parts of the absorbing potential, as a function of the absolute distance $\vert\Delta x\vert$ from the beginning of the absorbing boundary at \text{$x=\pm200\,a.u.$}.}\label{fig:Vabs}	
\end{figure}

In our simulation, we use an absorbing potential which is optimized numerically to maximize the absorption. The procedure basically relies on the principles presented in~\cite{squareBarriers}, but with several necessary modifications. The real part of the absorbing potential is constructed from a finite cosine series. The imaginary part is constructed from another finite cosine series, where the imaginary potential is given by squaring the cosine series and adding a minus sign. The optimization parameters are the cosine coefficients. This topic will be hopefully presented elsewhere. We choose the length of the absorbing boundary to be $40\,a.u.$. The obtained potential has a large imaginary part, which induces a significant shift of the Hamiltonian eigenvalues from the real axis into the fourth quarter of complex plane. The real and imaginary parts of the potential are plotted in Fig.~\ref{fig:Vabs} as a function of the absolute distance from the beginning of the absorbing boundary at \text{$x=\pm200\,a.u.$}. The absorbing potential added at the left boundary is the mirror image of that added at the right boundary. The values of the absorbing potential are available in the complementary material.

It was verified that the results do not change significantly if the grid length is doubled, and the form of the high-harmonic spectrum is very well preserved (a test which failed in Ref.~\cite{kulander92} for high intensity field, even for very large grid). The peak error of $\abs{u_i}^2$ from the doubled grid does not exceed the order of $10^{-5}$, where $u_i$ is the $i$'th component of $\vu$ (of course, in the physical region, \text{$\abs{x}\leq 200$}). Thus, the boundary effects are reduced to a reasonable magnitude.

\begin{figure}
	\centering \includegraphics[width=3in]{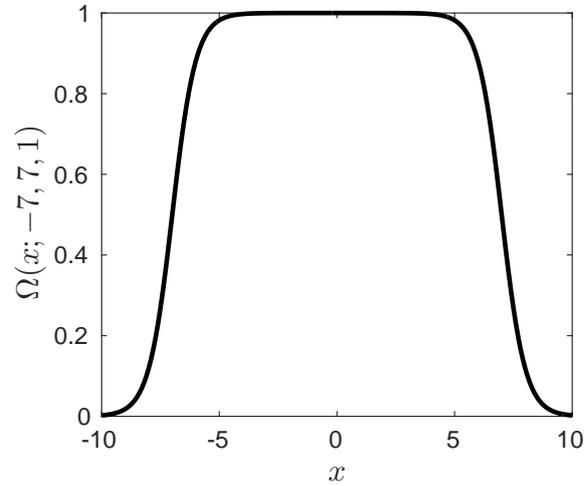}
	\caption{The soft rectangular function, with parameters $a=-7$, $b=7$, $\alpha=1$.}\label{fig:softrect}	
\end{figure}

Since the absorbing potential is optimized for an ideal absorption, the influence of the physical potential at the boundaries should be ``turned off'', in order that the absorption will not be damaged. This is achieved by a modification of the physical potential to a potential which is constant at the absorbing boundaries. In order to avoid discontinuities in the potential derivatives, it is desirable to ``turn off'' the physical potential in a continuous manner. For that purpose, we use the following practice. Let us define a \emph{soft rectangular function}:
\begin{equation}
	\Omega(x; a,b,\alpha) = \frac{1}{2}\left\lbrace\tanh\left[\alpha(x-a)\right] - \tanh\left[\alpha(x-b)\right] \right\rbrace
\end{equation}
The function is plotted in Fig.~\ref{fig:softrect} for arbitrary parameters. The modified potential, which is constant at the absorbing boundaries, will be denoted as $V_{mod}(x)$. It is defined to satisfy the following conditions:
\begin{align}
	& V_{mod}(0) = V_{phys}(0) \\
	& V_{mod}'(x) = V_{phys}'(x)\Omega(x; a,b,\alpha)
\end{align}
$V_{mod}(x)$ is obtained by integration in the following way:
\begin{align}
	V_{mod}(x) &= V_{phys}(0) + \int_0^x V_{phys}'(\xi)\Omega(\xi; a,b,\alpha)\,d\xi \nonumber \\
	&= V_{phys}(0) + V_{phys}(x)\Omega(x; a,b,\alpha) - V_{phys}(0)\Omega(0; a,b,\alpha) - \int_0^x V_{phys}(\xi)\Omega'(\xi; a,b,\alpha)\,d\xi \nonumber \\
	&\approx V_{phys}(x)\Omega(x; a,b,\alpha) - \int_0^x V_{phys}(\xi)\Omega'(\xi; a,b,\alpha)\,d\xi \label{Vmod}
\end{align}
where we utilized the fact that \text{$V_{phys}(0)\Omega(0; a,b,\alpha)\approx V_{phys}(0)$}. We have:
\begin{equation}
	\Omega'(x; a,b,\alpha) = \frac{1}{2}\alpha\left\lbrace\sech^2\left[\alpha(x-a)\right] - \sech^2\left[\alpha(x-b)\right] \right\rbrace
\end{equation}
The integration in Eq.~\eqref{Vmod} is performed numerically. 

In our problem, we choose the following parameters: \text{$a=-197.5\,a.u.$}, \text{$b=197.5\,a.u.$}, \text{$\alpha=1$}.

\begin{figure}
	\centering \includegraphics[width=3in]{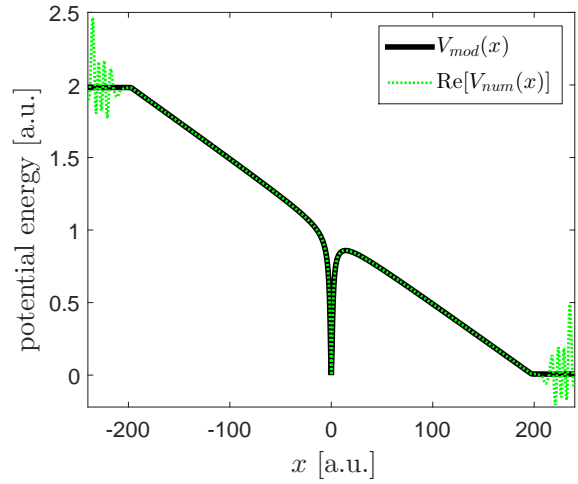}
	\caption{$V_{mod}(x)$ and the real part of $V_{num}(x)$ for \text{$V_{phys}(x) = V_{atom}(x) - 0.005x$}.}\label{fig:Vnum}	
\end{figure}

The numerical potential is given by
\begin{equation}
	V_{num}(x) \equiv V_{mod}(x) +
	\begin{cases}
		0 & \abs{x} < 200 \\
		V_{abs}(\abs{x} - 200) & \abs{x}\geq 200
	\end{cases}
\end{equation}
In Fig.~\ref{fig:Vnum}, we plot both $V_{mod}(x)$ and the real part of $V_{num}(x)$ for a relatively small field, $\zeta=0.005\,a.u.$.

\subsection{Results}\label{ssec:results}

The problem was solved by the semi-global propagator for different choices of $M$ and $K$ values. For each $M$ and $K$ choice, the problem was solved several times with different values of $\Delta t$ (the time-step is constant throughout the propagation, as well as $M$ and $K$). We compute the magnitude of the relative error of the final solution for each parameter choice. The efficiency of the propagator is demonstrated by a comparison of the resulting errors with those obtained by Runge-Kutta of the 4'th order (RK4, see Sec.~\ref{ssec:Taylor}). We compare also between the results of the semi-global propagator for the different $M$ and $K$ values.

In order to compare between different methods and parameter choices, we should compare the computational effort required for a similar accuracy. This may be done by choosing several specific examples. However, a fuller and a more reliable comparison is obtained by investigating the \emph{behaviour} of the error decay with the computational effort. This is done by plotting the \emph{error decay curve} for each method and parameter choice. In the error decay curve, the error is plotted Vs.\ the computational effort for several choices of $\Delta t$. A log-log plot is used. The computational effort is measured here by the number of Hamiltonian operations, which constitute the majority of the computational effort. 

In order to obtain a consistent behaviour of the error decay for the semi-global propagator, we use a slightly different version of the algorithm from that presented in Sec.~\ref{ssec:algorithm}; we restrict the number of iterations to a single iteration, \ie the steps of the loop in stage \ref{pr:do} of the algorithm are performed just once. This is with the exception of the first time-step, in which the solution is computed without a limitation on the number of iterations, where the parameter $\epsilon$ (see Sec.~\ref{ssec:algorithm}) represents the machine accuracy of the double-precision. This version of the algorithm restricts the inner freedom in the algorithm, thus ensures the consistency of the error decay curve. 

The relative error should be computed from a highly accurate solution of the problem. A highly accurate solution cannot be obtained from RK4 with double-precision, even with an extremely small time-step. This is due to accumulation of the machine errors. Hence, we use a reference solution obtained by the semi-global propagator, with the following parameters: \text{$M=9$}, \text{$K=13$}, \text{$\Delta t = 1/30$}. No limitation is imposed on the number of iterations, and $\epsilon$ represents the machine accuracy of the double precision. It was verified that the estimated errors, computed by the tests for the different sources of the error in Appendix~\ref{app:error}, do not exceed the order of the machine accuracy. The high accuracy of the obtained solution is evident from the shapes of the error decay curves.

\begin{figure}
	\centering \includegraphics[width=3in]{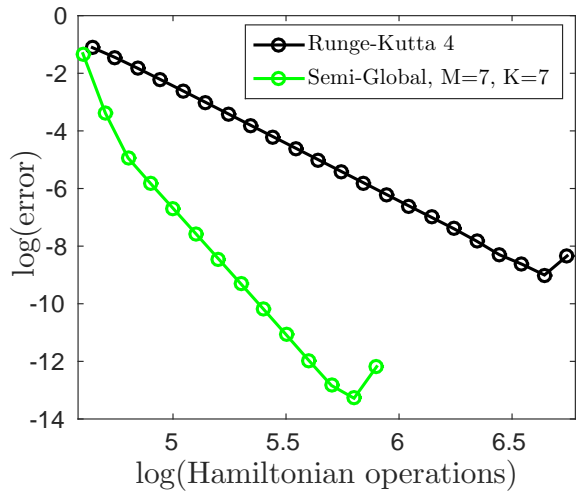}
	\caption{The error decay curves of the RK4 method, and the semi-global propagator with parameters \text{$M=K=7$}. The $\log_{10}$ of the relative error is plotted Vs.\ the $\log_{10}$ of the number of Hamiltonian operations. The last sampling point in each curve represents the limit in which the effects of roundoff errors become important, and the error ceases to decay. The RK4 curve shows a linear behaviour with a slope very close to $-4$. There is also a seemingly linear region in the semi-global curve, with a slope close to $-9$.}\label{fig:RK_SG}	
\end{figure}

First, we shall compare the results of the semi-global propagator with the parameters \text{$M=K=7$}, with those of RK4. The error decay curves are plotted in Fig.~\ref{fig:RK_SG}. The sampling points represent gradually decreasing values of $\Delta t$, for which the computational effort gradually increases. In each curve, $\Delta t$ is decreased until the error stops to decay, due to the effects of roundoff errors.

The linear behaviour of the RK4 curve is apparent. This is in consistence with theory---the error of the RK4 method is of $O(\Delta t^4)$ (see Sec.~\ref{ssec:Taylor}). Since \text{$\Delta t = T/N_t$}, the error decays as $N_t^{-4}$. The number of Hamiltonian operations is linear with $N_t$. Thus, the log-log plot yields a $-4$ slope. The slope obtained by a linear fit of the linear region of the curve agrees very well with theory (the obtained slope is $-3.99$). The semi-global curve has also a seemingly linear region. The slope obtained by a linear fit of the linear region is close to $-9$ (the precise value obtained is $-8.77$). The advantage of the semi-global propagator can be clearly seen.

Another advantage of the semi-global propagator is the maximal accuracy which can be obtained with the same machine accuracy. The minimal relative error which was obtained by RK4 is $9.96\times 10^{-10}$. Thus, in this problem, the RK4 method with double-precision is limited to an accuracy of about $10^{-9}$. The minimal error obtained for the semi-global propagator with this choice of parameters is $5.25\times 10^{-14}$.

Relying on the linear fit of both curves, one can estimate the number of Hamiltonian operations needed for a requested accuracy for each method. Regular accuracy requirements for most physical applications are of the order of $10^{-5}$. One finds that for an accuracy of $10^{-5}$, the RK4 method requires $6.8$ times the number of Hamiltonian operations required for the semi-global propagator. The advantage of the semi-global propagator becomes more apparent for applications which require a high accuracy solution. For an accuracy of $10^{-9}$, the RK4 method requires $24$ times the number of Hamiltonian operations required for the semi-global propagator.

\begin{figure}
	\centering \includegraphics[width=3in]{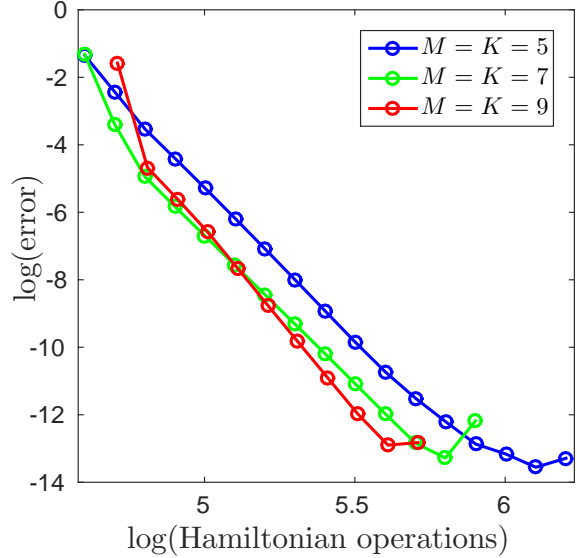}
	\caption{The error decay curves of the semi-global propagator with different choices of $M$ and $K$. The $\log_{10}$ of the relative error is plotted Vs.\ the $\log_{10}$ of the number of Hamiltonian operations. All curves include a region with a seemingly linear behaviour.}\label{fig:SG_KM}	
\end{figure}

We proceed with a comparison between different choices of $K$ and $M$. We shall compare between the following choices: \text{$M=K=5$}, \text{$M=K=7$}, \text{$M=K=9$} (the results of \text{$M=K=7$} have already been presented in Fig.~\ref{fig:RK_SG}). The error decay curves are shown in Fig.~\ref{fig:SG_KM}. The choice of \text{$M=K=5$} is shown to give inferior results in comparison to the higher orders. The \text{$M=K=7$} choice is slightly advantageous over the \text{$M=K=9$} for regular accuracy requirements. The \text{$M=K=9$} choice becomes superior for high accuracy requirements. These findings are by no means general; the ideal parameter choice for a required accuracy is problem dependent. 

All curves include a region with a seemingly linear behaviour. The slope of the linear region for \text{$M=K=5$} is very close to $-9$ (the precise value is $-8.98$). The slope for \text{$M=K=9$} is $-10.6$. As has already been mentioned, the slope of \text{$M=K=7$} is $-8.77$. 

The explanation of these results requires a detailed error analysis. One can show that the error resulting from each of the three error sources, mentioned in Appendix~\ref{app:error}, behaves polynomially with $\Delta t$, under certain approximation assumptions. This is the origin of the seemingly linear behaviour in the error decay curves. However, the order of each error source with $\Delta t$ is different. Since the overall error depends on several error sources, its behaviour with $\Delta t$ is considerably more complicated than that of RK4, in which there is only one error source. A detailed error analysis is beyond the scope of the present paper, and is left for a future publication.

Nevertheless, the error decay rate in each curve is shown to be much advantageous over the common Taylor methods. The error decay rate is higher for each parameter choice than a Taylor method of the same order (the order of approximation for each of the three parameter choices is \text{$M-1$}, and the slope predicted for a corresponding Taylor method is \text{$-(M-1)$}). Particularly, the \text{$M=K=5$} choice has the same approximation order as RK4, and the decay rate is much higher.

We can summarize that the semi-global propagator has two advantages over the Taylor approach, which lead to a higher error decay rate:
\begin{enumerate}
	\item In general, approximation by higher orders leads to higher error decay rate. The use of high order expansion is not recommended in Taylor methods, because of the inefficiency of a Taylor series as an approximation tool in higher orders (see Sec.~\ref{ssec:Taylor}). The semi-global approach, being free of Taylor considerations, allows to use higher order expansions than the Taylor approach;
	\item The error decay rate of the semi-global propagator is higher even for the same expansion order as the Taylor method.
\end{enumerate}

\Blanes{In this context, it is interesting to compare between the current approach and a class of propagators, known as \emph{exponential integrators} (see, \eg, \cite{exp_integ_rev2005, Hochbruck1999, Hochbruck2005, Hochbruck2006, Caliari2009, Suhov2014}). The propagation technique of the exponential integrators is also based on the $f_m(z,t)$ functions (defined in Eq.~\eqref{fs}). Certain elements of the current approach can be found to have been  employed in exponential integrators (see, in particular, \cite{Friesner1989}). However, there is a fundamental difference between this class of propagators and the current approach: The exponential integrator studies always employ local Taylorian considerations for constructing the propagation, while the propagation technique of the present approach is Taylor free. Hence, the error decay rate of the exponential integrators is limited by the order of the Taylor approximation employed, with the slope predicted for a simple Taylor method of the same order (see, for example, \cite{Suhov2014}). In contrast, the error decay rates in the present algorithm significantly exceed those of a Taylor method with the same expansion order. This fundamental difference between the approaches also allows to use higher expansion orders and larger time steps in the present approach in comparison to the exponential integrators.}

\section{Conclusion}

The solution of the time-dependent Schr\"odinger equation is one of the most important tasks in quantum physics. 
It can be solved by global means when the Hamiltonian is stationary.  This approach leads to vast improvement in accuracy and efficiency. In the present paper we presented a generalization of the global approach to the general case of a time-dependent, nonlinear Hamiltonian, with the additional inclusion of an inhomogeneous source term. The global approach can be implemented for the inhomogeneous Schr\"odinger equation with a stationary Hamiltonian. The solution method in this case constitutes the basis for the present approach for the general case of time-dependence or nonlinearity of the Hamiltonian. The general case can be treated by a semi-global approach, which combines global and local elements. The semi-global approach is characterized by propagation in relatively large time-steps, each of which is treated by global means.

The semi-global approach was shown to be significantly more efficient than the common local approach, which is based on Taylor considerations. Since the propagation method is Taylor free, it has the advantage of being able to use higher order approximations. The error decay rates were shown to be much higher in the new approach, thus enabling to achieve highly accurate solutions with a vast decrease in computational effort.

The semi-global algorithm applies also to the solution of a general set of ODE's, a fundamental problem in numerical analysis. The solution of the Schr\"odinger equation is an application in a special case of the general problem.

It was demonstrated that the semi-global algorithm is applicable also to non-Hermitian operators by the use of the Arnoldi approach. Thus, it applies also to problems in non-Hermitian quantum mechanics or to the solution of the Liouville von-Neumann equation.

We asserted that the success of the present approach is mainly attributed to the global element of the method, in which large intervals are treated as a whole in a unified process. This significantly reduces the main problem in the regular local schemes, in which the step-by-step propagation leads to a large numerical effort and error accumulation.

This advantage of the global approach over the local one reveals a more fundamental difference. The local approach relies (explicitly or implicitly) on a Taylor approximation. The Taylor considerations are based on the derivative concept, which is local in nature. The ability to deduce global information from local information is limited. Hence, it is not surprising that the Taylor expansion has poor convergence properties, thus becomes an inefficient tool for approximation purposes. In our opinion, the extreme importance of the Taylor expansion for analysis led, unjustly, to its wide spread as an approximation tool. In contrary, the present approach relies on orthogonal polynomial expansions, which have fast convergence properties. The approximation by an orthogonal set is intimately related to the fundamental concept of \emph{interpolation} (see Appendix~\ref{app:ChebApprox}). The interpolation concept is global in nature, where the information is deduced from samplings which are distributed all over the approximation interval. Thus, it becomes significantly advantageous over the Taylor expansion as an approximation tool.

It should be noted that even the local element in the semi-global approach is advantageous over the local Taylor considerations. The initial information for the propagation into the next time-step is obtained by \emph{extrapolation}, rather than local derivative information. The extrapolation concept is just an extension of the interpolation concept. A global interpolation approximation inside the interpolation interval can supply relatively accurate information for extrapolation outside the interval.

The main disadvantage of the present version of the algorithm is the necessity of specifying three parameters by the user. A successful choice of the parameters requires a trial and error process and experience. In order to accommodate with this problem, a parameter free version of the algorithm should be developed. The parameters will be determined adaptively by the procedure during the propagation process, to achieve maximal efficiency for the required accuracy. Such a version of the algorithm is expected also to significantly enhance the efficiency. The Chebyshev approximation should be replaced by a Leja approximation \cite{reichel1990newton} for the flexibility of the parameter determination process. The efficiency of the procedure requires an accurate error estimation. One of the advantages of interpolation approximations is the relative ease of the error analysis and estimation. Thus, the adaptive semi-global scheme is expected to be more successful than the available adaptive schemes for Taylor propagators.

We believe that a proliferation of the semi-global algorithm will lead to a significant improvement of accuracy and efficiency in quantum applications, as well as in the vast variety of problems which require the solution of a large set of ODE's.

\subsection*{Acknowledgements}
We want to thank Christiane Koch, Lutz Marder and Erik Torrontegui for their interest and help in this project.
Calculations were carried out on high performance computers purchased with the help of the Wolfson Foundation.
Work supported by Army Research Office (ARO) contract number W911NF-15-10250.
\appendix

\section{Polynomial approximations}\label{app:PolApprox}

\subsection{Approximation by a Newton interpolation}\label{app:NewtonApprox}

\subsubsection{The Newton interpolation}

The Newton interpolation is a polynomial interpolation of a function $f(x)$. The function is sampled at specific points distributed in the domain of approximation. The Newton interpolation polynomial approximates $f(x)$ within the domain from the sampling points. The approximation becomes exact at the sampling points.

Let us denote the sampling points by
\begin{equation}\label{samplingp}
	x_j, \qquad j=0,1,\ldots,N
\end{equation}
The value of $f(x_j)$ is given for all sampling points. The Newton interpolation approximates $f(x)$ using the following form:
\begin{align}
	f(x) &\approx a_0 + a_1(x - x_0) + a_2(x - x_0)(x - x_1) + \ldots + a_N(x-x_0)(x-x_1)\cdots(x - x_{N-1}) \nonumber \\
	&= \sum_{n=0}^{N} a_n R_n(x) \label{Newtonform}
\end{align}
where the $a_n$'s are coefficients, and the $R_n(x)$'s are defined by
\begin{align}
	& R_0(x) = 1 \nonumber \\
	& R_n(x) = \prod_{j=0}^{n-1} (x - x_j) & n>0 \label{Rdef}
\end{align}
The $R_n(x)$'s are called ``Newton basis polynomials''. In order to find the $a_n$'s, we first have to become familiar with the concept of \emph{divided difference}, which will be presented below.

Let us consider a function $f(x)$, and a set of points, as in Eq.~\eqref{samplingp}. The divided differences have a recursive definition. We will give the definition, and examples will follow immediately. The divided difference for a single point $x_0$ is defined as
\begin{equation}\label{dvd0}
	f[x_0] \equiv f(x_0)
\end{equation}
The divided difference of more than a single point is defined as
\begin{equation}\label{dvd}
	f[x_0,x_1,\ldots,x_n] = \frac{f[x_1,x_2,\ldots,x_n] - f[x_0, x_1, \ldots, x_{n-1}]}{x_n - x_0}
\end{equation}
For instance, the divided difference of two points is
\begin{equation}
	f[x_0, x_1] \equiv \frac{f[x_1]-f[x_0]}{x_1 - x_0} = \frac{f(x_1)-f(x_0)}{x_1 - x_0}
\end{equation}
The divided difference of three points is
\begin{equation}
	f[x_0, x_1, x_2] \equiv \frac{f[x_1, x_2]-f[x_0, x_1]}{x_2 - x_0} = \frac{\frac{f(x_2)-f(x_1)}{x_2 - x_1} - \frac{f(x_1)-f(x_0)}{x_1 - x_0}}{x_2 - x_0}
\end{equation}

The $a_n$ coefficients are given by the divided differences as follows:
\begin{equation}\label{an}
	a_n = f[x_0,x_1,\ldots,x_n]
\end{equation}

\subsubsection{Interpolation at Chebyshev points}\label{App:NewtonCheb}

When using a high order polynomial interpolation at equally spaced points, we may encounter a phenomenon which prevents it from being useful. The interpolation polynomial does not converge to the function $f(x)$ at the edges of the domain of approximation (even though it may converge very well at the middle of the domain). Instead, we might observe very large oscillations at the edges. The problem becomes more severe as $N$ grows. This phenomenon is known as \emph{Runge phenomenon}.

\begin{figure}
	\centering \includegraphics[width=3in]{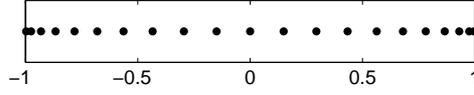}
	\caption{The Chebyshev points (Eq.~\eqref{chebp}) for \text{$N=20$}.}\label{fig:chebp}	
\end{figure}

The Runge phenomenon disappears when we choose the sampling points of $f(x)$ appropriately. There is more than one appropriate choice of a set of sampling points. The most commonly used set is given by the so called \emph{Chebyshev points} of the domain of approximation. The Chebyshev points become denser at the edges of the domain (see Fig.~\ref{fig:chebp}; this property is common to all sets of points which solve the Runge phenomenon). The Chebyshev points are originally defined for the domain \text{$[-1, 1]$}, but they can be easily transformed to another domain by a simple linear transformation, as will be described later.

The Chebyshev points for the domain \text{$[-1, 1]$} are:
\begin{equation}\label{chebp}
	y_j \equiv \cos\left[\frac{(2j+1)\pi}{2(N+1)} \right], \qquad j=0,1,\ldots,N
\end{equation}
Note that \text{$y_0>y_1>\ldots>y_N$}. Frequently, it is more convenient to index the points in an increasing order. We can reverse the order of the points, by defining them in the following way:
\begin{equation}\label{chebprev}
	y_j \equiv -\cos\left[\frac{(2j+1)\pi}{2(N+1)} \right], \qquad j=0,1,\ldots,N
\end{equation}
There is an alternative set of Chebyshev points, with similar characteristics:
\begin{equation}\label{chebpb}
	y_j \equiv \cos\left(\frac{j\pi}{N} \right), \qquad j=0,1,\ldots,N
\end{equation}
or, with a reversed order:
\begin{equation}\label{chebpbrev}
	y_j \equiv -\cos\left(\frac{j\pi}{N} \right), \qquad j=0,1,\ldots,N
\end{equation}
In this set of points, the function is sampled at the boundaries of the domain, unlike in the set \eqref{chebp}. This can be advantageous in certain circumstances (for example, in the context of the semi-global propagation scheme, in which adjacent time-steps share a common point; see Sec.~\ref{ssec:tgrid}).

The Chebyshev points can be transformed to an arbitrary domain on the real axis, $[x_{min}, x_{max}]$. First, they are stretched or compressed to match the size \text{$\Delta x = x_{max} - x_{min}$} of the domain:
\begin{equation*}
	y_j \longrightarrow \frac{\Delta x}{2}y_j
\end{equation*}
Then, they are shifted to the middle of the domain by adding the middle point,
\begin{equation*}
	\frac{x_{min} + x_{max}}{2}
\end{equation*}
The sampling points are finally obtained by the following linear transformation:
\begin{equation}\label{y2x}
	x_j \equiv \frac{1}{2}(y_j\,\Delta x + x_{min} + x_{max})
\end{equation}

There are other sets of points which solve the Runge phenomenon. The set of \emph{Leja points} \cite{reichel1990newton} can be advantageous when the required degree of approximation $N$ is difficult to be estimated in advance. This topic is beyond the scope of this paper.

\subsubsection{Numerical stability of the Newton interpolation}\label{app:Newton4}

The Newton interpolation usually becomes numerically unstable when $N$ grows. The main problem is that the $a_n$'s of high $n$ tend to become very large, and the corresponding adjacent polynomials become very small, or vice versa. This problem does not exist when the domain of approximation, \text{$[x_{min}, x_{max}]$}, is of length $4$ \cite{Newton_stability}. Hence, for a domain defined on the real axis, the problem can be overcome by transforming the problem to a length $4$ domain.

First, we transform the sampling points to a length $4$ domain. The points in the new domain are defined as
\begin{equation}
	\bar{x}_j \equiv \frac{4}{\Delta x}x_j
\end{equation}
In general, we can define a transformed \emph{variable}:
\begin{equation}
	\bar{x} = \frac{4}{\Delta x}x
\end{equation}
We also define a new function $\bar{f}(x)$, such that:
\begin{equation}
	\bar{f}(\bar{x}) = f(x)
\end{equation}
Then, we can approximate $f(x)$ at an arbitrary $x$ value in the original domain, by using the Newton interpolation for the function $\bar{f}(x)$ at the sampling points $\bar{x}_j$. The approximation to $f(x)$ at an arbitrary $x$ is given by the interpolation polynomial value at $\bar{x}$. 

\subsection{Chebyshev approximation}\label{app:ChebApprox}

In the Chebyshev approximation, a function $f(x)$ is approximated by a truncated series of orthogonal polynomials, in the form of Eq.~\eqref{PolExpansion}. The basis of expansion consists of the \emph{Chebyshev polynomials}. They are defined as follows:
\begin{equation}
	T_n(x) = \cos(n\cos^{-1}x), \qquad x\in[-1,1], \qquad n=0,1,\ldots 
\end{equation}
In order to clarify the meaning of this weird definition, let us define a variable $\theta$ such that
\begin{equation}\label{theta2x}
	x = \cos\theta
\end{equation}
Then we obtain the following equivalent definition of the Chebyshev polynomials:
\begin{equation}\label{Tcos}
	T_n(\cos\theta) = \cos(n\theta), \qquad \theta\in[0, \pi], \qquad n=0,1,\ldots 
\end{equation}
For instance,
\begin{align}
	& T_0(x) = \cos(0) = 1 \label{T0x}\\
	& T_1(x) = \cos\theta = x \label{T1x}\\
	& T_2(x) = \cos(2\theta) = 2\cos^2\theta - 1 = 2x^2 - 1
\end{align}

Note that the functions $\cos(n\theta)$ from the RHS of Eq.~\eqref{Tcos} span the function space in the domain \text{$\theta\in[0, \pi]$}. In addition, they are \emph{orthogonal} in this domain:
\begin{equation}\label{cos_ortho}
	\int_0^\pi \cos(m\theta)\cos(n\theta)\,d\theta = \frac{\pi \alpha_n}{2}\delta_{mn}, \qquad 
	\alpha_n \equiv
	\begin{cases}
		2 & n=0 \\
		1 & n>0
	\end{cases}
\end{equation}
The Chebyshev polynomials are obtained from the cosine basis by mapping $\theta$ into $x$ with the nonlinear transformation \eqref{theta2x}. The resulting functions are also orthogonal in the $x$ space, but with respect to a \emph{weight function} in the new space. This can be seen by changing the integration variable of Eq.~\eqref{cos_ortho} from $\theta$ to $x$. We obtain:
\begin{equation}\label{T_ortho}
	\int_0^\pi \cos(m\theta)\cos(n\theta)\,d\theta = -\int_{1}^{-1} \frac{T_m(x)T_n(x)}{\sin\theta}\,dx = \int_{-1}^1 \frac{T_m(x)T_n(x)}{\sqrt{1-x^2}}\,dx = \frac{\pi \alpha_n}{2}\delta_{mn}
\end{equation}
We have obtained the orthogonality relation of the Chebyshev polynomials under the weight function
\begin{equation}\label{chebweight}
	w(x) = \frac{1}{\sqrt{1-x^2}}
\end{equation}

The Chebyshev polynomials satisfy the following recurrence relation:
\begin{equation}\label{Trecursion}
	T_{n+1}(x) = 2xT_n(x) - T_{n-1}(x), \qquad n \geq 1
\end{equation}
All Chebyshev polynomials can be obtained recursively from $T_0(x)$ and $T_1(x)$ (see Eqs.~\eqref{T0x}, \eqref{T1x}), using Eq.~\eqref{Trecursion}.

Now suppose we want to approximate a function $f(x)$ in a given domain \text{$[x_{min}, x_{max}]$}, from several samplings of the function in the domain. Suppose we can choose the sampling points as we wish. The function can be approximated from the sampling points by a Chebyshev series, as will be described below. For simplicity, let us first assume that the domain of approximation is \text{$[-1, 1]$}.

$f(x)$ can be spanned in the following form:
\begin{equation}\label{ChebExpansion}
	f(x) \approx \sum_{n=0}^{N} c_n T_n(x)
\end{equation}
The $c_n$'s are called the \emph{Chebyshev coefficients} of $f(x)$. They can be obtained by projecting $f(x)$ onto each of the $T_n(x)$ basis functions. Using the orthogonality relation \eqref{T_ortho}, the Chebyshev coefficients are given by the following scalar product expression:
\begin{equation}\label{chebcx}
	c_n = \frac{2}{\pi \alpha_n} \int_{-1}^1 f(x)T_n(x)w(x)\,dx
\end{equation}
Equivalently, we can perform the scalar product in the $\theta$ space:
\begin{equation}\label{chebctheta}
	c_n = \frac{2}{\pi \alpha_n} \int_0^\pi f(\cos \theta)\cos(n\theta)\,d\theta
\end{equation}

Suppose that the form of $f(x)$ is unknown, or that the integral cannot be performed analytically. We have to compute the Chebyshev coefficients \emph{numerically} from a finite number of samplings of $f(x)$ within the domain. Fortunately, the problem of computing the Chebyshev coefficients can be reformulated by discrete means, without reduction of the quality of the approximation.

We utilize the fact that there exist discrete versions of the orthogonality relations between the cosine basis functions. The orthogonality relations are given by a finite sum over samplings of the cosine functions in equally spaced points in $\theta$. For instance, we have the following orthogonality relation:
\begin{equation}\label{cos_ortho_d}
	\sum_{j=0}^K \cos(k\theta_j)\cos(l\theta_j) = \frac{(K+1)\alpha_k}{2}\delta_{kl}, \qquad \theta_j\equiv\frac{(2j+1)\pi}{2(K+1)}
\end{equation}
where $\alpha_k$ is defined in Eq.~\eqref{cos_ortho}, and \text{$K\geq 0$}. Note that in the $x$ space, the points \text{$x_j=\cos\theta_j$} are just the Chebyshev points defined in \eqref{chebp}. The orthogonality relation \eqref{cos_ortho_d} can be utilized in order to find the Chebyshev coefficients from the sampling of $f(x)$ at the Chebyshev points, as we shall see. Let us define:
\begin{equation}
	g(\theta) \equiv f(\cos \theta)
\end{equation}
Eq.~\eqref{ChebExpansion} can be rewritten as
\begin{equation}\label{ChebExpansion_cos}
	g(\theta) \approx \sum_{m=0}^N c_m \cos(m\theta)
\end{equation}
Let us multiply $g(\theta)$ by the basis function $\cos(n\theta)$, and sum over the set of $N+1$ Chebyshev points:
\begin{equation*}
	\sum_{j=0}^N g(\theta_j)\cos(n\theta_j), \qquad \theta_j=\frac{(2j+1)\pi}{2(N+1)}
\end{equation*}
We substitute $g(\theta)$ with the approximation \eqref{ChebExpansion_cos}, and obtain:
\begin{equation}\label{dprojection}
	\sum_{j=0}^N g(\theta_j)\cos(n\theta_j) \approx \sum_{j=0}^N \sum_{m=0}^N c_m \cos(m\theta_j)\cos(n\theta_j) = \frac{(N+1)\alpha_n}{2}c_n
\end{equation}
where we applied the orthogonality relation \eqref{cos_ortho_d}. The Chebyshev coefficients are finally given by
\begin{equation}\label{chebcd}
	c_n = \frac{2}{(N+1)\alpha_n} \sum_{j=0}^N g(\theta_j)\cos(n\theta_j) = \frac{2}{(N+1)\alpha_n} \sum_{j=0}^N f\left[\cos\left(\frac{(2j+1)\pi}{2(N+1)}\right)\right]\cos\left[\frac{n(2j+1)\pi}{2(N+1)}\right]
\end{equation}
Note that the Chebyshev coefficients defined by Eq.~\eqref{chebcd} are not identical with those defined by the integral version of Eq.~\eqref{chebctheta}. However, the inaccuracy in \eqref{chebcd} is originated in the truncation error of Eq.~\eqref{ChebExpansion} itself (see Eq.~\eqref{dprojection}). Thus, the total error will be of the same order of magnitude as the truncation error, and the quality of the approximation will be similar.

Actually, the set of $N+1$ equations defined by \eqref{chebcd} is a \emph{linear transformation} of the function value vector, \text{$[g(\theta_0), g(\theta_1),\ldots, g(\theta_N)]^T$}, into the coefficient vector, \text{$[c_0, c_1, \ldots, c_N]^T$}. This transformation is called a \emph{discrete cosine transform} (DCT)\@. It has an apparent similarity to the discrete Fourier transform (DFT)\@. There are several kinds of DCT's. The transformation defined in Eq.~\eqref{chebcd} is sometimes referred as a \emph{DCT of the second kind}.

The DCT's are reversible transformations. The inverse transformation of Eq.~\eqref{chebcd} is actually defined by Eq.~\eqref{ChebExpansion_cos} for the set of $\theta_j$'s. Hence, the approximation is \emph{exact} for the Chebyshev sampling points. In that sense, Eq.~\eqref{ChebExpansion} with the $c_n$'s computed by Eq.~\eqref{chebcd} defines an \emph{interpolation} of $f(x)$ in the set of $N+1$ Chebyshev points. A fundamental theorem of interpolation theory states that the interpolation polynomial for a given set of sampling points is \emph{unique}. Thus, the approximation presented here is equivalent to the approximation by a Newton interpolation at the Chebyshev points, described in Sec.~\eqref{App:NewtonCheb}.

The DCT transformations can be computed very efficiently by algorithms derived from the fast Fourier transform (FFT) algorithm. Hence, the scaling of the computational effort is of $O(N\ln N)$. There are available programs for computation of the DCT\@. When using them, a care should be taken on the exact definition of the transformation---usually, there are slight differences in the definition of the transformation coefficients.

It is also possible to approximate the Chebyshev coefficients by sampling $f(x)$ at the set of boundary including Chebyshev points (see Eq.~\eqref{chebpb}). We use the following discrete orthogonality relation:
\begin{equation}\label{cos_ortho_db}
	\sum_{j=0}^K \frac{1}{\beta_j}\cos(k\theta_j)\cos(l\theta_j) = \frac{K\beta_k}{2}\delta_{kl}, \qquad \theta_j \equiv \frac{j\pi}{K}, \qquad
	\beta_j \equiv
	\begin{cases}
		2 & j=0,K \\
		1 & 1 \leq j \leq K-1
	\end{cases}
\end{equation}
where \text{$K\geq 1$}. Following similar steps as above, we obtain:
\begin{equation}\label{chebcdb}
	c_n = \frac{2}{N\beta_n} \sum_{j=0}^N \frac{1}{\beta_j} g(\theta_j)\cos(n\theta_j) =
	\frac{2}{N\beta_n} \sum_{j=0}^N \frac{1}{\beta_j} f\left[\cos\left(\frac{j\pi}{N} \right)\right] \cos\left(\frac{nj\pi}{N}\right)
\end{equation}
Here again, the set of $N+1$ equations defined by Eq.~\eqref{chebcdb} is a linear transformation of the function value vector into the coefficient vector. It is another kind of a DCT, sometimes referred as a \emph{DCT of the first kind}. Its inverse is defined by Eq.~\eqref{ChebExpansion_cos} for this set of points. The boundary including points are preferable when the exact value of $f(x)$ at the boundaries is important.

In the case that the domain of approximation of $f(x)$ is different from the Chebyshev domain, \text{$[-1, 1]$}, it is required to shift the problem to the Chebyshev domain. The treatment of the problem is similar to that of Sec.~\ref{App:NewtonCheb}. For instance, let us discuss the approximation by the boundary including Chebyshev points. We denote the domain of approximation by \text{$x\in[x_{min}, x_{max}]$}. Let us denote the set of Chebyshev points by $y_j$:
\begin{equation}
	y_j \equiv \cos\left(\frac{j\pi}{N} \right), \qquad j=0,1,\ldots,N
\end{equation}
The Chebyshev points in the domain \text{$[x_{min}, x_{max}]$} are defined by the linear transformation \eqref{y2x}. We can also define a \emph{variable} \text{$y\in [-1, 1]$}, which is given by the inverse linear transformation of $x$:
\begin{equation}\label{x2y}
	y \equiv \frac{2x - x_{min} - x_{max}}{\Delta x}
\end{equation}
We define a function $\bar f(x)$ such that
\begin{equation}
	\bar f(y) = f(x)
\end{equation}
The approximation to $f(x)$ is given by
\begin{equation}
	f(x) = \bar{f}(y) \approx \sum_{n=0}^N c_n T_n(y)
\end{equation}
where
\begin{equation}\label{chebc_arbxb}
	c_n = \frac{2}{N\beta_n} \sum_{j=0}^N \frac{1}{\beta_j} \bar{f}(y_j) \cos(n\theta_j) = \frac{2}{N\beta_n} \sum_{j=0}^N \frac{1}{\beta_j} f(x_j) \cos\left(\frac{nj\pi}{N}\right)
\end{equation}
We see that the Chebyshev coefficients are simply given by a discrete cosine transform of $f(x)$, sampled at the Chebyshev points of the domain \text{$[x_{min}, x_{max}]$}.

The treatment in the case of the points of \eqref{chebp} is completely identical. We obtain:
\begin{equation}\label{chebc_arbx}
	c_n =  \frac{2}{(N+1)\alpha_n} \sum_{j=0}^N f(x_j) \cos\left[\frac{n(2j+1)\pi}{2(N+1)}\right]
\end{equation}
where the $x_j$'s are given by Eq.~\eqref{y2x}, with the $y_j$'s of Eq.~\eqref{chebp}.

As was mentioned in Sec.~\ref{App:NewtonCheb}, it is often more convenient to reverse the order of the Chebyshev points, in order to obtain increasing values of $x$ with the point index. This can be done by defining them as in Eqs.~\eqref{chebprev}, \eqref{chebpbrev}. However, care should be taken to preserve the original form of Eqs.~\eqref{chebc_arbxb}, \eqref{chebc_arbx}, in which each of the $f(x_j)$'s is multiplied by the cosine of the corresponding angle. Hence, the order of the angles should also be reversed. This is equivalent to the addition of a minus sign to the RHS of the two equations. Eq.~\eqref{chebc_arbxb} is replaced by
\begin{equation}\label{chebc_arbxb_rev}
	c_n = -\frac{2}{N\beta_n} \sum_{j=0}^N \frac{1}{\beta_j} f(x_j) \cos\left(\frac{nj\pi}{N}\right)
\end{equation}
and Eq.~\eqref{chebc_arbx} is replaced by
\begin{equation}\label{chebc_arbx_rev}
	c_n =  -\frac{2}{(N+1)\alpha_n} \sum_{j=0}^N f(x_j) \cos\left[\frac{n(2j+1)\pi}{2(N+1)}\right]
\end{equation}

\section{Approximation methods for the multiplication of a vector by a function of matrix}\label{app:FunMat}

Here we discuss several methods for the computation of the following vector:
\begin{equation}\label{fAv}
	\vu = f(A)\vv
\end{equation}
where $\vv$ is an arbitrary vector, $A$ is a matrix, and $f(x)$ is a function. All approximation methods are based on the following realization: When the multiplication of $f(A)$ with $\vv$ is all what required, we can avoid the direct computation of $f(A)$, which is highly demanding numerically. Instead, we use successive multiplications of vectors by the matrix $A$. As has already been mentioned in Sec.~\ref{sssec:TiH}, in certain cases we can replace the direct multiplication of the vector by the matrix $A$, by a computational procedure which is less demanding numerically.

\subsection{Polynomial series approximations}\label{app:FMpoly}

The first two methods presented here are based on approximation of $f(x)$ by a polynomial $Q_L(x)$ of degree $L$. $Q_L(x)$ is a truncated polynomial series of $f(x)$ (Cf.~Eq.~\eqref{PolExpansion}):
\begin{equation}\label{PolExpansion_app}
	f(x) \approx Q_L(x) \equiv \sum_{n=0}^{L} b_n P_n(x)
\end{equation}
where the $P_n(x)$'s are polynomials of degree $n$, and the $b_n$'s are the corresponding expansion coefficients. We can approximate $\vu$ by the following expression (Cf.~Eq.\eqref{MatExpansion}):
\begin{equation}\label{MatExpansion_app}
	\vu \approx Q_L(A)\vv = \sum_{n=0}^{L} b_n P_n(A)\vv
\end{equation}
The idea is to compute the expressions $P_n(A)\vv$ by successive multiplications of vectors by $A$, instead of computing $P_n(A)$ and multiplying $\vv$ by the resulting matrix.

As we have seen in Appendix~\ref{app:PolApprox}, when using a polynomial expansion for the approximation of a function, we first have to define the approximation domain, \text{$[x_{min}, x_{max}]$}. The approximation is expected to be accurate only inside the approximation domain. In our problem, the approximation should be accurate in the \emph{eigenvalue domain} of $A$. This can be readily seen by considering the decomposition of $\vv$ into the eigenvectors of $A$:
\begin{equation}\label{v_decomp}
	\vv = \sum_{j=0}^{N-1} v_j\vphi_j
\end{equation}
where the $\vphi_j$'s are the eigenvectors of $A$, the $v_j$'s are the components of $\vv$ in the eigenvector basis, and $N$ is the dimension of $A$. Plugging \eqref{v_decomp} into \eqref{fAv}, we obtain:
\begin{equation}\label{u_decomp}
	\vu = \sum_{j=0}^{N-1} v_j f(\lambda_j)\vphi_j
\end{equation}
where $\lambda_j$ is the eigenvalue of $\vphi_j$. When using the approximation of Eq.~\eqref{PolExpansion_app}, we actually replace the accurate expression of Eq.~\eqref{u_decomp} by the following approximated expression:
\begin{equation}
	\vu \approx \sum_{j=0}^{N-1} v_j Q_L(\lambda_j)\vphi_j
\end{equation}
It is clear that $Q_L(\lambda_j)$ should be accurate for each of the $\lambda_j$'s. Hence, the approximation domain has to cover the whole eigenvalue domain of $A$.


Frequently, the eigenvalue domain of $A$ is unknown, and we have to estimate it. Note that an overestimation of the eigenvalue domain size costs additional numerical effort, because more terms are required to approximate $f(x)$. In cases that the eigenvalue domain cannot be estimated, or when the eigenvalue domain is complex, the Arnoldi approach (see Sec.~\ref{app:Arnoldi}) should be used instead of the polynomial expansion methods.

\subsubsection{Newton interpolation}\label{app:fMnewton}

One approach for approximation of $\vu$ by a polynomial series is by using a Newton interpolation of $f(x)$ at the Chebyshev points of the eigenvalue domain, defined by Eq.~\eqref{y2x} (see Appendix~\ref{App:NewtonCheb}) \cite{Newton_stability}. The Newton interpolation polynomial is (Cf.~Eq.~\eqref{Newtonform}):
\begin{equation}
	Q_L(x) = \sum_{n=0}^L a_n R_n(x)
\end{equation}
The $x_j$ sampling points which define the $a_n$'s and the $R_n(x)$'s are given by Eq.~\eqref{y2x}. The $R_n(x)$'s satisfy the following recurrence relation:
\begin{equation}\label{Rrecursionx}
	R_{n+1}(x) = (x-x_n)R_n(x)
\end{equation}
$\vu$ is approximated by
\begin{equation}\label{MatNewton_app}
	\vu \approx \sum_{n=0}^L a_n R_n(A)\vv
\end{equation}
The recurrence relation \eqref{Rrecursionx} can be utilized in order to compute the expressions $R_n(A)\vv$ successively.

Let us index the sampling points by \text{$j=0,1,\ldots,L$}. The algorithm for the computation of $\vu$ is described below:
\begin{enumerate}
	\item Compute the Chebyshev points $y_j$ by Eq.~\eqref{chebprev} or by Eq.~\eqref{chebpbrev}.
	\item Compute the Chebyshev points $x_j$ in the eigenvalue domain \text{$[x_{min}, x_{max}]$} from the $y_j$'s by Eq.~\eqref{y2x}.
	\item Compute the function values \text{$f_j = f(x_j)$}.\label{alg:fn}
	\item Compute the divided differences $a_n$, \text{$n=0,1,\ldots,L$}, recursively from $f_j$ and $x_j$, using Eqs.~\eqref{an}, \eqref{dvd0}, \eqref{dvd}. \label{alg:dvd}
	\item $\vw=\vv$
	\item $\vu=a_0\vw$
	\item for $i=1$ to $L$
	\begin{enumerate}
		\item $\vw = A\vw - x_{i-1}\vw$ \label{alg:neww}
		\item $\vu=\vu + a_i\vw$
	\end{enumerate}
	\item end for
\end{enumerate}

In practice, the Newton interpolation problem should be transferred to a domain of length $4$, for numerical stability (see Appendix~\ref{app:Newton4}). This amounts of two slight changes. We have to add to the recursion relation of Eq.~\eqref{Rrecursionx} an additional conversion factor, in order to transform $x$ to the domain of length $4$:
\begin{equation}\label{Rrecursionx4}
	R_{n+1}(x) = \frac{4}{\Delta x}(x-x_n)R_n(x)
\end{equation}
where \text{$\Delta x = x_{max} - x_{min}$}. In addition, the sampling points $x_j$ that appear in the denominator of the divided difference formula \eqref{dvd} are replaced by \text{$\bar{x}_j = 4x_j/\Delta x$}. Accordingly, we insert the following changes into the algorithm:
\begin{enumerate}
	\item The $x_j$'s in stage \ref{alg:dvd} are replaced by $4x_j/\Delta x$ (note that the $f_j$'s from stage \ref{alg:fn} remain the same).
	\item Stage \ref{alg:neww} is replaced by $\vw = (A\vw - x_{i-1}\vw)4/\Delta x$
\end{enumerate}

$\vu$ can be computed with other sets of sampling points which solve the Runge phenomenon (for instance, the Leja points; see Appendix~\ref{App:NewtonCheb}). We use the same recursion formula and algorithm, where the $x_j$'s are the desired set of points.

\subsubsection{Chebyshev expansion}\label{app:fMcheb}

Another approach for approximating $\vu$ is by the expansion of $f(A)$ in a Chebyshev series \cite{tal1989polynomial}. The approximation polynomial $Q_L(x)$ is given by
\begin{equation}
	Q_L(x) = \sum_{n=0}^L c_n T_n(y)
\end{equation}
where
\begin{equation}\label{x2y_2}
	y \equiv \frac{2x - x_{min} - x_{max}}{\Delta x}
\end{equation}
The $c_n$'s are given by Eq.~\eqref{chebc_arbx} or Eq.~\eqref{chebc_arbxb}, where the $x_j$'s are given by Eq.~\eqref{y2x}, together with Eq.~\eqref{chebp} or Eq.~\eqref{chebpb}, respectively (see Appendix~\ref{app:ChebApprox}).

$\vu$ is approximated by
\begin{equation}
	\vu \approx Q_L(A)\vv = \sum_{n=0}^L c_n T_n(\bar{A})\vv
\end{equation}
where
\begin{equation}
	\bar{A} = \frac{2A - x_{min} - x_{max}}{\Delta x}
\end{equation}
The Chebyshev polynomials satisfy the recurrence relation
\begin{equation}\label{Trecursion_2}
	T_{n+1}(y) = 2yT_n(y) - T_{n-1}(y), \qquad n \geq 1
\end{equation}
where
\begin{align}
	& T_0(y) = 1 \label{T0_2}\\
	& T_1(y) = y \label{T1_2}
\end{align}
We can compute the expressions $T_n(\bar{A})\vv$ successively by utilizing the recurrence relation, where $y$ in Eqs.~\eqref{Trecursion_2}, \eqref{T1_2}, is substituted by $\bar{A}$.

The algorithm for the computation of $\vu$ is described below:

\begin{enumerate}
	\item Compute the Chebyshev points $y_j$ by Eq.~\eqref{chebp} or by Eq.~\eqref{chebpb}.
	\item Compute the Chebyshev points $x_j$ in the eigenvalue domain \text{$[x_{min}, x_{max}]$} from the $y_j$'s by Eq.~\eqref{y2x}.
	\item Compute the function values \text{$f_j = f(x_j)$}.
	\item Compute the Chebyshev coefficients $c_n$ from the $f_j$'s by Eq.~\eqref{chebc_arbx} or by Eq.~\eqref{chebc_arbxb}.
	\item $\vw_1 = \vv$
	\item $\vw_2 = [2A\vv - (x_{min} + x_{max})\vv]/\Delta x$
	\item $\vu = c_0\vw_1 + c_1\vw_2$
	\item for $i=2$ to $L$
	\begin{enumerate}
		\item $\vw_3 = 2[2A\vw_2 - (x_{min} + x_{max})\vw_2]/\Delta x - \vw_1$
		\item $\vu= \vu + c_i\vw_3$
		\item $\vw_1 = \vw_2$
		\item $\vw_2 = \vw_3$
	\end{enumerate}
	\item end for	
\end{enumerate}

\subsection{Arnoldi approach}\label{app:Arnoldi}

The approximations of Sec.~\ref{app:FMpoly} are based on the assumption that the eigenvalue domain is known, or can be estimated. They cannot be applied when it is impossible to estimate the eigenvalue domain. The difficulty in the estimation of the eigenvalue domain becomes severe when the eigenvalues of $A$ are distributed on the complex plane, and not only on the real or on the imaginary axis.

Moreover, the concept of Chebyshev sampling is defined for a one dimensional \emph{axis}, which may be the real or imaginary axis. Hence, a Chebyshev approximation can be applied for functions of a real variable, or a purely imaginary variable. However, a Chebyshev approximation is not suitable for functions of \emph{complex} variables, which are distributed on the two dimensional \emph{complex plane}. Thus, the methods of Sec.~\ref{app:FMpoly} are not applicable when the eigenvalue spectrum of $A$ is complex.

In the present section, we introduce the Arnoldi approach for approximation of \eqref{fAv} \cite{restartedArnoldi}. In the Arnoldi approach, the eigenvalue domain needn't be known, and the sampling is chosen by different considerations. Thus, it becomes suitable also for the treatment of matrices with a complex spectrum.

The Arnoldi approximation is intimately related to the polynomial approximations, but the approach to the problem is different. We can view our basic problem as the problem of reduction of large-scale matrix calculations into simplified approximations. The polynomial approximations reduce the \emph{calculation} of the matrix into a \emph{simplified calculation} of the \emph{same matrix}, in which a function is reduced into a low degree polynomial. In the Arnoldi approach, the \emph{matrix itself} is reduced into a small-scale matrix. This is done by the choice of a ``good'' set of a small number of vectors, which can be representative in the framework of the specific problem. The matrix $A$ is represented in the reduced subspace spanned by the vectors.

The approximation is based on a reduction of the problem to the subspace spanned by the following vectors:
\begin{equation}\label{Krylov}
	\vv, A\vv, A^2\vv,\ldots, A^L\vv
\end{equation}
The subspace is spanned by multiplications of the vector $\vv$ by powers of $A$, from degree $0$ to $L$. A vector space of this kind is called a \emph{Krylov space}. The space spanned by \eqref{Krylov} is a \emph{Krylov space of dimension \text{$L+1$}}. The Krylov space is a ``good'' subspace in our problem for two reasons, which are interrelated:
\begin{enumerate}
	\item Any polynomial approximation of $\vu$ of degree $L$ in the form of \eqref{MatExpansion_app} can be spanned by the Krylov space. This is a direct consequence of the fact that any polynomial can be expressed in the terms of the Taylor polynomials. Thus, the Krylov space is actually the characteristic subspace of polynomial approximations of degree $L$.
	\item Suppose $\vv$ can be spanned by a subspace in the eigenvector spectrum of $A$. The subspace is invariant to multiplication by $A$ or $f(A)$, which are diagonal in the eigenvector basis (this can be readily seen by expanding $\vv$ in the eigenvector space, as in Eq.~\eqref{u_decomp}). Thus, $\vu$ remains in the same eigenvector subspace as $\vv$. The Krylov space also remains in the eigenvector subspace, for the same reason. Hence, it is expected to be effective for the representation of $\vu$. Note that if $\vv$ is spanned by an eigenvector space of dimension up to \text{$L+1$}, $\vu$ can be fully represented by the Krylov space. Even when $\vv$ is spanned by the whole eigenvector space, frequently a small number of eigenvectors dominate its composition. Thus, the Krylov space may still be effective in approximating it.
\end{enumerate}

The vectors of Eq.~\eqref{Krylov} are in general non-orthogonal. Moreover, the vectors are getting closer to be parallel with the degree of $A$. When using them as a basis set, this might be a source of numerical instability. We should work with an orthonormal basis set for spanning the Krylov space, for the sake of numerical stability and the simplicity of the calculations.

In order to obtain an orthonormal basis, we use the \emph{Gram-Schmidt process}. The idea underlying the Gram-Schmidt orthonormalization goes as follows: The orthonormal basis vectors are computed successively from the original non-orthogonal set of vectors. We subtract from each vector from the original set its projection on the subspace spanned by the already computed orthonormal vectors. We are left with a vector which is orthogonal to the subspace spanned by the previous vectors. Then, we simply normalize it, and obtain an orthonormal set which is enlarged by one dimension. Then, we continue to the next vector from the original set, and so on.

In practice we use the \emph{Modified-Gram-Schmidt} (MGS) algorithm which is equivalent, mathematically, to Gram-Schmidt but less sensitive to roundoff errors. The orthonormalization in the context of the Krylov space can be implemented by an iterative process, as will be seen. The iterative algorithm is referred as \emph{Arnoldi iteration}.

Let us denote the orthonormal set by
\[
	\vups_0, \vups_1,\ldots,\vups_{L}
\]
As will be seen, the vector $\vups_n$ belongs to the Krylov space of dimension \text{$n+1$}. In the algorithm, we compute an additional orthonormal vector, $\vups_{L+1}$, which belongs to the Krylov space of dimension \text{$L+2$}. The necessity of its computation will be clarified in what follows. We denote the scalar product of two vectors, $\bvec{r}$ and $\bvec{s}$, by \text{$\left\langle\bvec{r}, \bvec{s}\right\rangle$}.

The algorithm is described below. During the algorithm, we also store in the memory a set of constants, which participate in the computation. The necessity of this will be clarified in what follows.

\begin{enumerate}
	\item Compute the first vector in the orthonormal set as the normalized $\vv$:
	\[
		\vups_0 = \frac{\vv}{\Vert \vv  \Vert}
	\]
	\item for $j=0$ to $L$
	\begin{enumerate}
		\item Compute a non-orthonormalized new vector by setting: \text{$\vups_{j+1} = A\vups_{j}$}. \label{pr:newups}
		\item for $i=0$ to $j$ \label{pr:arnoldi_loop2}
		\begin{enumerate}
			\item Set: \text{$\Gamma_{i,j}=\left\langle\vups_i,\vups_{j+1}\right\rangle$}. \label{pr:gamma_ij}
			\item Subtract from $\vups_{j+1}$ its projection on $\vups_i$: \text{$\vups_{j+1} = \vups_{j+1} - \Gamma_{i,j}\vups_{i}$}.
		\end{enumerate}
		\item end for
		\item Set: \text{$\Gamma_{j+1,j}=\Vert \vups_{j+1} \Vert$}. \label{pr:gamma_jp1j}
		\item Normalize $\vups_{j+1}$ by setting:
		\[
			\vups_{j+1} = \frac{\vups_{j+1}}{\Gamma_{j+1,j}}
		\] \label{pr:ups_norm}
	\end{enumerate}
	\item end for
\end{enumerate}

Clearly, the $\vups_{j+1}$ computed in stage \ref{pr:newups} belongs to the Krylov space of dimension \text{$j+2$}, since it is composed from some linear combination of the first \text{$j+2$} Krylov vectors. Thus, in each iteration, the Krylov space is enlarged by one dimension.

At the end of the algorithm, we are left with an orthonormal set of dimension \text{$L+2$}. We also computed, seemingly by the way, the $\Gamma_{i,j}$'s. These have a special significance. Actually, $\Gamma_{i,j}$ is the \emph{matrix element of A in the orthonormal basis}, \ie, it is equivalent to \text{$A_{i,j} = \left\langle \vups_i, A\vups_j\right\rangle$}. This can be readily seen from the algorithm. In stage \ref{pr:gamma_ij}, the vector $\vups_{j+1}$ is given by
\begin{equation*}
	A\vups_{j} - \sum_{k=0}^{i-1}\Gamma_{k,j}\vups_{k}
\end{equation*}
with the summation convention of Eq.~\eqref{sumconvention}. Using the orthonormality relations between the vectors, we obtain immediately:
\begin{equation}
	\Gamma_{i,j} = \left\langle\vups_i,A\vups_{j} - \sum_{k=0}^{i-1}\Gamma_{k,j}\vups_{k}\right\rangle = \left\langle \vups_i, A\vups_j\right\rangle = A_{i,j}, \qquad i\leq j
\end{equation}
It can also be shown that $A_{j+1,j}$ is equivalent to $\Gamma_{j+1,j}$, computed in stage \ref{pr:gamma_jp1j}. In analogy to stage \ref{pr:gamma_ij}, the matrix element $A_{j+1,j}$ is given by the projection of $\vups_{j+1}$ in stage \ref{pr:gamma_jp1j}, on the final $\vups_{j+1}$, obtained in stage \ref{pr:ups_norm}. In stage \ref{pr:gamma_jp1j}, $\vups_{j+1}$ is in its final direction, but is still unnormalized. The projection of a vector on a unit vector in the same direction gives its norm. This clarifies the definition of $\Gamma_{j+1,j}$ in stage \ref{pr:gamma_ij}.

The matrix elements $\Gamma_{i,j}$ with \text{$i>j+1$} are not computed in the algorithm. It can be easily seen that they vanish. The expression $A\vups_{j}$ belongs to the Krylov space of dimension \text{$j+2$}. The vectors $\vups_i$ with \text{$i>j+1$} are orthogonal to this space, and hence, the scalar product vanishes.

Now we are able to construct the matrix representation of $A$ in the reduced orthonormalized Krylov basis. It will be denoted by $\Gamma$. The matrix is a square matrix of dimension \text{$L+1$}, with the following structure:
\begin{equation}\label{Hessenberg}
	\Gamma \equiv
	\begin{bmatrix}
		\Gamma_{0,0} & \Gamma_{0,1} & \Gamma_{0,2} & \cdots & \Gamma_{0,L-1} & \Gamma_{0,L} \\
		\Gamma_{1,0} & \Gamma_{1,1}	& \Gamma_{1,2} & \cdots & \Gamma_{1,L-1} & \Gamma_{1,L} \\
		 			 & \Gamma_{2,1}	& \Gamma_{2,2} & \cdots & \Gamma_{2,L-1} & \Gamma_{2,L} \\
					 &  			& \Gamma_{3,2} & \cdots & \Gamma_{3,L-1} & \Gamma_{3,L} \\
					 & \BigZero		&  			   & \ddots & \vdots		 & \vdots \\
					 &  	 		&  		  	   &   		& \Gamma_{L,L-1} & \Gamma_{L,L}
	\end{bmatrix}
\end{equation}
The matrix structure is similar to an upper triangular matrix, with zero elements below the first subdiagonal. A matrix of this structure is called an \emph{upper Hessenberg matrix}. In the context of the Arnoldi process, $\Gamma$ is referred as the \emph{Hessenberg matrix of $A$}. Note that in the algorithm, we compute also $\Gamma_{L+1,L}$, which is not included in the definition of $\Gamma$. It will be useful to define also an extended matrix $\bar{\Gamma}$, of dimension \text{$(L+2)\times (L+1)$}. $\bar{\Gamma}$ is given by the extension of $\Gamma$ by one row, in the following way:
\begin{equation}
	\bGamma \equiv
	\begin{bmatrix}
		\\
		  &	  & \BigFig{\Gamma} &   & \\
		\\
		0 & 0 & \cdots 			& 0 & \Gamma_{L+1,L}
	\end{bmatrix}
\end{equation}

Note that the last column of $\bGamma$ is computed in the algorithm during the process of obtaining $\vups_{L+1}$. We see that this process is necessary for the computation of the Hessenberg matrix, even though $\vups_{L+1}$ does not participate in the approximation itself, which takes place in a Krylov space of dimension \text{$L+1$} only. The computation of an additional vector costs an additional matrix-vector multiplication. Thus, in the Arnoldi process, we need \text{$L+1$} matrix-vector multiplications, in comparison to \text{$L$} for a polynomial approximation of the same order. Nevertheless, the extension of the Krylov space by one dimension is useful for the estimation of the error of the approximation, as will be seen.

The Arnoldi process can be summarized by \text{$L+1$} vector equations in the following way:
\begin{equation}
	\vups_{n+1} = \frac{A\vups_n - \sum_{k=0}^{n}\Gamma_{k,n}\vups_{k}}{\Gamma_{n+1,n}}, \qquad 0\leq n\leq L
\end{equation}
or,
\begin{equation}
	A\vups_n = \sum_{k=0}^{n+1}\Gamma_{k,n}\vups_{k}, \qquad 0\leq n\leq L
\end{equation}
These equations can be written compactly in a matrix form:
\begin{equation}\label{ArnoldiM}
	A\Upsilon = \bUps\bGamma
\end{equation}
where $\Upsilon$ is an \text{$N\times(L+1)$} matrix ($N$ denotes the dimension of $A$) which its columns are the $\vups_n$'s:
\begin{equation}\label{Upsilon}
	\Upsilon \equiv
	\begin{bmatrix}
		\vups_0, & \vups_1, & \cdots & \vups_L
	\end{bmatrix}
\end{equation}
and $\bUps$ is an extension of $\Upsilon$ by one column, which contains $\vups_{L+1}$:
\begin{equation}\label{bUpsilon}
	\bUps \equiv
	\begin{bmatrix}
		\vups_0, & \vups_1, & \cdots & \vups_L, & \vups_{L+1}
	\end{bmatrix}
\end{equation}
We can write Eq.~\eqref{ArnoldiM} in an alternative form, which does not involve the extended matrices. We utilize the fact that the last row of $\bGamma$ contains only one non-zero element, which affects only the last column of the resulting \text{$N\times (L+1)$} matrix. It can be easily seen that we have:
\begin{equation}\label{ArnoldiM2}
	A\Upsilon = \Upsilon\Gamma + \Gamma_{L+1,L}\,\vups_{L+1}\,\ve_{L+1}^T
\end{equation}
where $\ve_n$ denotes a unit vector of dimension \text{$L+1$}, which its $j$'th element is given by $\delta_{j,n}$. 

Now we are about to use the reduced basis, obtained by the Arnoldi iteration, for writing an approximation of $\vu$. The $\vups_n$'s in the orthonormalized Krylov basis representation are given by $\ve_{n+1}$. Thus, we can define the vector which represents $\vv$ in the reduced basis in the following way:
\begin{equation}\label{omega}
	\vom \equiv \nv\, \ve_1
\end{equation}
$\Upsilon$ has an important significance---it is the transformation matrix from the reduced Krylov basis representation to the original $N$ dimensional representation of $A$ and $\vv$:
\begin{equation}
	\vups_n = \Upsilon\ve_{n+1}
\end{equation}
Following the reasoning that was introduced in the beginning of this section, we can approximate $\vu$ by performing the calculation in the reduced basis representation, and transforming the result to the original basis. Let us define the corresponding vector of $f(A)\vv$ in the reduced representation:
\begin{equation}\label{eta}
	\veta \equiv f(\Gamma)\vom
\end{equation}
$\veta$ is transformed to the original representation, to yield the following approximation of $\vu$:
\begin{equation}\label{u_Arnoldi}
	\vu \approx \Upsilon \veta
\end{equation}
As was mentioned in Sec.~\ref{sssec:TiH}, the direct computation of a function of matrix via diagonalization, in the form of Eq.~\eqref{u_direct}, is very expensive numerically for a large scale matrix like $A$. However, it is not an expensive operation to diagonalize $\Gamma$, which is a small scale matrix, in order to calculate $\veta$. Thus, $\veta$ can be computed as
\begin{equation}\label{eta_diag}
	\veta = S f(D) S^{-1} \vom
\end{equation}
where $D$ is the diagonal matrix which represents $\Gamma$ in the basis of the $\Gamma$'s eigenvectors, and $S$ is the transformation matrix from the diagonalized basis representation to the $\vups_n$ basis representation.

In the approximation obtained, we do not need any previous knowledge about the eigenvalue domain of $A$. Moreover, we do not rely on any assumption on the form of the eigenvalue domain. Hence, it can be applied also for $A$ with a complex spectrum.

The justification that was given to the approximation in the form of Eq.~\eqref{u_Arnoldi} is intuitive rather than rigorous. Hence, it becomes unclear what is the quality of the approximation, and how to estimate the resulting error. In what follows, we shall give a fuller justification to the Arnoldi approach approximation. It will be shown that Eq.~\eqref{u_Arnoldi} is actually equivalent to a \emph{polynomial approximation} of $\vu$. This will enable us to estimate the error of the approximation.

Let us return, for the moment, to the polynomial approximation methods. We shall present a new approach for the choice of the approximation polynomial $Q_L(z)$ (see Eqs. \eqref{PolExpansion_app}, \eqref{MatExpansion_app}), which is led by different considerations from the polynomial approximations of Sec.~\ref{app:FMpoly}.

Until now, we used the concept of interpolation as an approximation tool for a function. Now we shall see that there exists a more intimate relation between a function of matrix and the interpolation concept. Let $g(z)$ be a function which interpolates $f(z)$ in the eigenvalues of $A$, \ie:
\begin{equation}\label{evi_condition}
	g(\lambda_j) = f(\lambda_j), \qquad j=0,1,\ldots,N-1	
\end{equation}
It can be easily shown that when the set of eigenvectors spans the $N$ dimensional space, then $g(A)$ is completely equivalent to $f(A)$:
\begin{equation}\label{intMeq}
	g(A) = f(A).
\end{equation}
This can be proved by showing that
\begin{equation*}
	[f(A) - g(A)]\vw = \bvec{0}
\end{equation*}
for an arbitrary vector $\vw$. Let us expand $\vw$ in the eigenvectors of $A$:
\begin{equation}\label{w_decomp}
	\vw = \sum_{j=0}^{N-1} w_j\vphi_j
\end{equation}
This yields:
\begin{equation}\label{proofintMeq}
	[f(A) - g(A)]\vw = \sum_{j=0}^{N-1} w_j[f(\lambda_j) - g(\lambda_j)]\vphi_j = \bvec{0}
\end{equation}
The condition \eqref{evi_condition} on $g(z)$ is necessary for the equivalence of the \emph{matrices} (Eq.~\eqref{intMeq}). Note that the condition for the \emph{vector} equivalence \text{$g(A)\vv=f(A)\vv$} may be even weaker. This happens when $\vv$ is spanned by a subspace of several eigenvectors. Then, it is sufficient to choose $g(z)$ which interpolates $f(z)$ in the eigenvalues of these eigenvectors only, as is apparent from \eqref{proofintMeq}.

We see that in order to obtain $f(A)$, we needn't know the behaviour of $f(z)$ everywhere, but only its values at the $\lambda_j$'s. The origin of this somewhat surprising finding lies in the fact that the spectrum of $A$ is \emph{discrete}. The discrete spectrum originates in the fact that $A$ itself is a discrete entity. Thus, $f(A)$ itself is also a discrete entity, and its approximation requires discrete knowledge on $f(z)$.

This leads to a different approach for the choice of $Q_L(z)$: Instead of trying to approximate the full behaviour of $f(z)$ by the ideal approximation polynomial of $f(z)$ in the domain, we can choose a polynomial which well represents the behaviour of $f(z)$ in several representative eigenvalues of $A$. This feature is satisfied by an \emph{interpolation polynomial of $f(z)$} in these eigenvalues. The question that remains is: How can we know the eigenvalues of $A$, without diagonalizing it?

Here the Arnoldi approach enters: We state that the eigenvalues of the reduced matrix $\Gamma$ provide estimation of several representative eigenvalues of $A$ itself. These can be used as interpolation points of $f(z)$. The eigenvalues of $\Gamma$ tend to be distributed in the eigenvalue domain of $A$ in the same way as the whole eigenvalue spectrum of $A$. In the case that several eigenvectors dominate the composition of $\vv$, the spectrum of $\Gamma$ usually contains estimation of most of them. This is because of the second feature of the Krylov space mentioned in the beginning of this section---the Krylov space remains in the same eigenvector subspace as $\vv$, and reflects its eigenvector composition, at least to some extent. This characteristic of the $\Gamma$ spectrum is particularly useful for the specific approximation of $f(A)\vv$, since the interpolation polynomial becomes adjusted to represent the behaviour of $f(z)$ in the dominant eigenvectors in $\vv$.

Let us denote the eigenvalues of $\Gamma$ by $\tlam_j$. Our approximation polynomial is a Newton interpolation polynomial,
\begin{equation}
	Q_L(z) = \sum_{n=0}^L a_n R_n(z)
\end{equation}
with
\begin{align}
	& R_0(z) \equiv 1 \nonumber\\
	& R_n(z) \equiv \prod_{j=0}^{n-1}(z-\tlam_j), \qquad n>0 \label{Rnz}
\end{align}
$\vu$ is approximated in the form of Eq.~\eqref{MatNewton_app}. If we follow the scheme of Sec.~\eqref{app:fMnewton}, we need $L$ matrix-vector multiplications for this operation. This is in addition to the \text{$L+1$} matrix-vector multiplications required for the Arnoldi iteration process. It seems that the overall cost of this approximation is more than twice the cost of the approximations of Sec.~\eqref{app:FMpoly}. Actually, we can avoid any additional large-scale matrix-vector multiplication, as will be seen. Thus, the overall number of the required large-scale matrix-vector multiplications will remain \text{$L+1$}.

In order to show that, we shall use the first feature of the Krylov space, mentioned in the beginning of this section. Let us consider again Eq.~\eqref{MatExpansion_app}, with an arbitrary $Q_L(z)$. Both $\vv$ and the approximated $\vu$ lie in the Krylov space of dimension \text{$L+1$}, since $Q_L(A)\vv$ can be decomposed into the Taylor polynomial vectors \eqref{Krylov}. Thus, the operation of the matrix $Q_L(A)$ on $\vv$ takes place in the reduced Krylov space, as well as the operation of any of the $Q_L(A)$'s polynomial components. Hence, the whole calculation can take place in the reduced space, where $A$ is replaced by its reduced representation, $\Gamma$, and $\vv$ is replaced by $\vom$. The result is transferred back to the full $N$ dimensional space, to yield the same $\vu$ as in Eq.~\eqref{MatExpansion_app}, without any further approximation. Thus, the following equality is \emph{exact} for \emph{any approximation polynomial} $Q_L(z)$:
\begin{equation}\label{pol_reduced}
	Q_L(A)\vv = \Upsilon Q_L(\Gamma)\vom
\end{equation}
In our particular case, $\vu$ is approximated as
\begin{equation}\label{u_polKr}
	\vu \approx \Upsilon \sum_{n=0}^{L} a_n R_n(\Gamma) \vom
\end{equation}
Now we note that our approximation polynomial, which interpolates $f(z)$ in the entire spectrum of $\Gamma$, is actually equivalent to $f(\Gamma)$, by Eq.~\eqref{intMeq}. Thus, we have (Cf.~\eqref{eta}):
\begin{equation}\label{eta_pol}
	\veta = \sum_{n=0}^{L} a_n R_n(\Gamma) \vom 
\end{equation}
Then, Eq.~\eqref{u_polKr} becomes equivalent to Eq.~\eqref{u_Arnoldi}. This concludes the justification to the approximation of Eq.~\eqref{u_Arnoldi}, which is shown to be equivalent to a specific polynomial approximation.

The new interpretation of the Arnoldi approximation enables us to estimate the resulting error. Assuming fast convergence of the polynomial expansion with $n$, we can estimate the truncation error of Eq.~\eqref{u_polKr} by the magnitude of the next term in the sum,
\begin{equation}\label{EArnoldi}
	E = \vert a_{L+1} \vert\, \Vert R_{L+1}(A)\vv \Vert
\end{equation}
In order to specify the next term, we first have to choose an additional sampling point. The additional point should be in the eigenvalue domain of $A$, which is unknown. It is reasonable to choose the average point of the estimated eigenvalues:
\begin{equation}\label{zavg}
	z_{L+1} = \frac{\sum_{j=0}^L \tlam_j}{L+1}
\end{equation}
It should be verified that $z_{L+1}$ is not equal to any of the $\tlam_j$'s. $R_{L+1}(z)$ is independent of $z_{L+1}$, and is given by Eq.~\eqref{Rnz} with \text{$n=L+1$}. $z_{L+1}$ determines $a_{L+1}$ only. 

The error should be computed without additional large-scale matrix-vector multiplications. Hence, it is desirable to express $R_{L+1}(A)\vv$ in the terms of the reduced Krylov space, which involves only small-scale calculations. Here we encounter the problem that the expression $R_{L+1}(A)\vv$ belongs to a Krylov space of dimension \text{$L+2$}. Hence, $A$ cannot be simply replaced by $\Gamma$, which contains information on the Krylov space of dimension \text{$L+1$} only. However, the computation of the reduced basis vectors involved also the additional vector, $\vups_{L+1}$. Thus, it is still possible to obtain a reduced calculation of \eqref{EArnoldi} by the information we already have. First, we write:
\begin{equation}\label{RLplus1}
	R_{L+1}(A)\vv = (A-\tlam_L)R_L(A)\vv
\end{equation}
$R_L(A)\vv$ lies in the Krylov space of dimension \text{$L+1$}. Hence, we can express it in the terms of the \text{$L+1$} dimensional space:
\begin{equation}
	R_L(A)\vv = \Upsilon R_L(\Gamma)\vom = \Upsilon \vmu
\end{equation}
where we defined:
\begin{equation}\label{vmu}
	\vmu \equiv R_L(\Gamma)\vom
\end{equation}
Eq.~\eqref{RLplus1} becomes:
\begin{equation}
	R_{L+1}(A)\vv = A\Upsilon\vmu - \tlam_k\Upsilon\vmu
\end{equation}
Then, we apply Eq.~\eqref{ArnoldiM2} to yield:
\begin{align}
	R_{L+1}(A)\vv &= \left(\Upsilon\Gamma + \Gamma_{L+1,L}\,\vups_{L+1}\,\ve_{L+1}^T\right)\vmu - \tlam_k\Upsilon\vmu \nonumber \\
				  &= \Upsilon\left(\Gamma - \tlam_k\right)\vmu + \Gamma_{L+1,L}\mu_{L+1}\,\vups_{L+1} \nonumber \\
				  &= \Upsilon R_{L+1}(\Gamma)\vom + \Gamma_{L+1,L}\mu_{L+1}\,\vups_{L+1}
\end{align}
where $\mu_{L+1}$ is the $(L+1)$'th element of $\vmu$. We see that the resulting expression is spanned by the extended orthonormalized Krylov set, which includes $\vups_{L+1}$. We can use an extended representation of the extended set, of dimension \text{$L+2$}, in order to represent $R_{L+1}(A)\vv$. In the extended representation, $R_{L+1}(A)\vv$ is defined by the following vector:
\begin{equation}
	\vbmu =
	\begin{bmatrix}
		R_{L+1}(\Gamma)\vom \\
		\Gamma_{L+1,L}\mu_{L+1}
	\end{bmatrix}
\end{equation}
In principle, $\vbmu$ can be transferred back to the original representation by the extended transformation matrix, $\bUps$, in the following way:
\begin{equation}
	R_{L+1}(A)\vv = \bUps \vbmu
\end{equation}
However, this operation is unnecessary, as long as we are interested only in the \emph{norm} of this expression (see \eqref{EArnoldi}). The vectors in $\bUps$ are \emph{unit vectors}, and the norm remains unaltered by the change of representation: 
\begin{equation}
	\Vert \bUps \vbmu \Vert = \Vert \vbmu \Vert
\end{equation}
The error is finally given by
\begin{equation}
	E = \vert a_{L+1} \vert \, \Vert \vbmu \Vert
\end{equation}
The relative error is given by
\begin{equation}
	E_{rel} = \frac{E}{\left\Vert \vu \right\Vert} \approx \frac{E}{\left\Vert \Upsilon\veta \right\Vert} = \frac{E}{\left\Vert \veta \right\Vert}
\end{equation}

If we compute the error, Eq.~\eqref{eta_diag} becomes unnecessary; $\veta$ should be computed by the Newton interpolation form, Eq.~\eqref{eta_pol}, utilizing the operations needed for the computation of the error. The $R_n(\Gamma)\vom$ terms and the $a_n$'s are computed iteratively, as explained in Appendix~\ref{app:NewtonApprox} and Sec.~\ref{app:fMnewton}. $R_{L+1}(\Gamma)\vom$ and $a_{L+1}$ for the error estimation are obtained by continuing the iterative process to the next order.

In order to increase the numerical stability of the Newton interpolation, the size of the approximation domain should be changed, as in the case of interpolation on a one-dimensional axis (see Appendix~\ref{app:Newton4}). In the case of a two-dimensional domain, defined on the complex plain, the problem should be transformed to a domain which its size is divided by the \emph{capacity of the domain} \cite{tal1989polynomial}. We give only an expression for the estimation of the capacity, and not an exact definition. The capacity is estimated by choosing a point in the domain, $z_p$, and computing the following expression:
\begin{equation}\label{rho}
	\rho = \left( \abs{z_p-z_0}\abs{z_p-z_1}\cdots\abs{z_p-z_N} \right)^{\frac{1}{N+1}}
\end{equation}
where the $z_n$'s (\text{$n=0,1,\ldots,N$}) are the sampling points, and \text{$N+1$} is the number of sampling points. $z_p$ can be chosen as the average point of the sampling points, as in Eq.~\eqref{zavg}. In the case of interpolation on an axis, in a domain of length $4$ (see Appendix~\ref{app:Newton4}), the capacity is $1$. This can be observed intuitively from \eqref{rho}, by choosing $z_p$ in the middle of the domain. If the capacity of the domain is different from $1$, we should perform a transformation similar to that described in Appendix~\ref{app:Newton4}, where the conversion factor of $4/\Delta x$ is replaced by $1/\rho$. In practice, the instructions at the end of Sec.~\eqref{app:fMnewton} should be followed, with the appropriate conversion factor.

The interpolation polynomial interpretation of the Arnoldi approach reveals an important advantage of the Arnoldi algorithm over the polynomial approximations. This advantage exists in certain circumstances, even for $A$ with eigenvalues distributed on a one-dimensional axis. If most of the eigenvalues are concentrated in a portion of the eigenvalue domain, with an additional small number of spread eigenvalues in the whole domain, the Arnoldi approach is expected to be more efficient. The reason is that the $\tlam$'s are distributed in the eigenvalue domain in a similar way to the $\lambda$'s. Thus, the important region in which most of the eigenvalues are concentrated is better represented than the less important regions. In contrary, the Chebyshev sampling is uniformly distributed in the domain.

In the application of the time-dependent Hamiltonian propagator, we apply a time-step scheme. The short time intervals can be treated by a relatively small Krylov space, typically---up to \text{$L=15$}. When the Hamiltonian is time-independent, the whole time-interval is treated in a single step, and the required approximation space is large. The Arnoldi approach usually becomes problematic in a large space. The main reasons are:
\begin{enumerate}
	\item In the Arnoldi process, we store \text{$L+2$} vectors in the memory. For very large $N$ and large $L$, this might be impermissible.
	\item During the Arnoldi process, we perform \text{$(L+1)^2/2$} scalar products, each of which scales as $N$. This becomes quite demanding for large $L$.
\end{enumerate}
When a large dimension approximation is required, a \emph{restarted Arnoldi algorithm} should be used (see, for example, \cite{restartedArnoldi}). This topic is beyond the scope of this paper.

\section{Conversion schemes of polynomial expansions to a Taylor form}\label{app:pol2Taylor}

\subsection{Conversion scheme for a Newton expansion}

\subsubsection{Conversion scheme for the $q_{n,m}$ coefficients}

Let us write the Newton expansion form for $\vs(t)$ (Cf.~Eq.~\eqref{Newtonform}):
\begin{equation} \label{sRn}
	\vs(t) \approx \sum_{n=0}^{M-1} \va_n R_n(t)
\end{equation}
The $R_n(t)$'s satisfy the following recursion formula (see Eq.~\eqref{Rdef}):
\begin{equation}\label{Rrecursion}
	R_{n+1}(t) = (t-t_n)R_n(t)
\end{equation}
where
\begin{equation}\label{R0}
	R_0(t)=1
\end{equation}
and the $t_n$'s are the sampling points.

We need the conversion coefficients of the $R_n(t)$'s to a Taylor form (Cf.~Eq.~\eqref{PnTaylor}):
\begin{equation}\label{R2Taylor}
	R_n(t) = \sum_{m=0}^n q_{n,m} \frac{t^m}{m!}
\end{equation}
It can be immediately observed from Eq.~\eqref{R0} that:
\begin{equation}\label{q00}
	q_{0,0} = 1
\end{equation}
The rest of the $q_{n,m}$'s can be computed from Eq.~\eqref{q00} by the derivation of recurrence relations. Plugging Eq.~\eqref{R2Taylor} into \eqref{Rrecursion} we obtain:
\begin{equation}\label{Rnplus1a}
	R_{n+1}(t) = \sum_{m=0}^n q_{n,m} \frac{t^{m+1}}{m!} - t_n\sum_{m=0}^n q_{n,m} \frac{t^m}{m!} =
	 \sum_{m=1}^{n+1} q_{n,m-1} \frac{t^{m}}{(m-1)!} - t_n\sum_{m=0}^n q_{n,m} \frac{t^m}{m!}
\end{equation}
On the other hand, we have from \eqref{R2Taylor}:
\begin{equation}\label{Rnplus1b}
	R_{n+1}(t) = \sum_{m=0}^{n+1} q_{n+1,m} \frac{t^m}{m!}
\end{equation}
Equating the RHS of Eqs.~\eqref{Rnplus1a} and \eqref{Rnplus1b} we obtain:
\begin{equation}
	\sum_{m=0}^{n+1} q_{n+1,m} \frac{t^m}{m!} = \sum_{m=1}^{n+1} q_{n,m-1} \frac{t^{m}}{(m-1)!} - t_n\sum_{m=0}^n q_{n,m} \frac{t^m}{m!}
\end{equation}
Equating the coefficients of similar powers of $t$, we obtain the following recurrence relations for the $q_{n,m}$'s:
\begin{align}
	& q_{n+1,0} = -t_n q_{n,0} \label{Nrecursion_qn0} \\
	& q_{n+1,m} = m q_{n, m-1} - t_n q_{n,m} & 1\leq m\leq n \label{Nrecursion_qnm}\\
	& q_{n+1, n+1} = (n+1)q_{n,n} \label{Nrecursion_qnn}
\end{align}
Starting from Eq.~\eqref{q00}, all $q_{m,n}$'s can be computed in a recursive manner, using Eqs.~\eqref{Nrecursion_qn0}-\eqref{Nrecursion_qnn}.

Once the $q_{n,m}$'s are obtained, the $\vs_m$ Taylor coefficients can be computed by (Cf. Eq.~\eqref{q2s}):
\begin{equation}\label{q2s_dvd}
	\vs_m = \sum_{n=m}^{M-1} q_{n,m}\va_n
\end{equation}

\subsubsection{Conversion scheme for the $\tq_{n,m}$ coefficients}

As was mentioned in Sec.~\ref{ssec:programming}, for numerical stability, it is recommended to absorb the $1/m!$ factor in the $\vs_m$ Taylor coefficients from Eq.~\eqref{pols}. Accordingly, Eq.~\eqref{R2Taylor} is replaced by:
\begin{equation}\label{R2Taylor2}
	R_n(t) = \sum_{m=0}^n \tilde q_{n,m} t^m
\end{equation}
The recurrence relations for the $\tq_{n,m}$'s are slightly different from those of the $q_{n,m}$'s. Following the same steps as above, we obtain the recursion formulas:
\begin{align}
	& \tq_{n+1,0} = -t_n \tq_{n,0} \label{Nrecursion_tqn0} \\
	& \tq_{n+1,m} = \tq_{n, m-1} - t_n \tq_{n,m} & 1\leq m\leq n \label{Nrecursion_tqnm}\\
	& \tq_{n+1, n+1} = \tq_{n,n} \label{Nrecursion_tqnn}
\end{align}
In addition, we have from \eqref{R0}:
\begin{equation}
	\tq_{0,0} = 1
\end{equation}
which completes the required information for computing the $\tq_{n,m}$'s recursively.

The $\vts_m$'s are computed by:
\begin{equation}\label{tq2s_dvd}
	\vts_m = \sum_{n=m}^{M-1} \tq_{n,m}\va_n
\end{equation}

\subsubsection{Conversion scheme for a length $4$ domain}

We mentioned in Sec.~\ref{app:NewtonApprox} that it is recommended to use a domain of length $4$ in a Newton expansion, for numerical stability. When transferring the original $t$ domain to a length $4$ domain, the recursion formula for the $R_n(t)$'s gains an additional conversion factor (Cf.~Eq.~\eqref{Rrecursion}):
\begin{equation}\label{Rrecursion4}
	R_{n+1}(t) = \frac{4}{\Delta t}(t-t_n)R_n(t)
\end{equation}
where $\Delta t$ is the length of the original domain. Accordingly, the RHS of the recurrence relations \eqref{Nrecursion_qn0}-\eqref{Nrecursion_qnn} and \eqref{Nrecursion_tqn0}-\eqref{Nrecursion_tqnn} is multiplied by the same factor, to yield:
\begin{align}
	& q_{n+1,0} = -\frac{4}{\Delta t}t_n q_{n,0} \label{Nrecursion_qn04}\\
	& q_{n+1,m} = \frac{4}{\Delta t}(m q_{n, m-1} - t_n q_{n,m}) & 1\leq m\leq n \label{Nrecursion_qnm4}\\
	& q_{n+1, n+1} = \frac{4}{\Delta t}(n+1)q_{n,n} \label{Nrecursion_qnn4}
\end{align}
and
\begin{align}
	& \tq_{n+1,0} = -\frac{4}{\Delta t}t_n \tq_{n,0} \label{Nrecursion_tqn04}\\
	& \tq_{n+1,m} = \frac{4}{\Delta t}(\tq_{n, m-1} - t_n \tq_{n,m}) & 1\leq m\leq n \label{Nrecursion_tqnm4}\\
	& \tq_{n+1, n+1} = \frac{4}{\Delta t}\tq_{n,n} \label{Nrecursion_tqnn4}
\end{align}

The $\vs_m$'s and the $\vts_m$'s are computed as in Eqs.~\eqref{q2s_dvd}, \eqref{tq2s_dvd}.

\subsection{Conversion scheme for a Chebyshev expansion}

The Chebyshev expansion is defined for the domain \text{$[-1, 1]$}. Suppose we want approximate $\vs(t)$ in an arbitrary domain \text{$t\in[t_{min},t_{max}]$} by a Chebyshev expansion. We define a variable \text{$y\in[-1, 1]$} such that
\begin{equation}\label{t2y}
	y \equiv \frac{2t - t_{min} - t_{max}}{\Delta t}
\end{equation}
where \text{$\Delta t = t_{max} - t_{min}$} (see Sec.~\ref{app:ChebApprox}). Then we expand $\vs(t)$ in a Chebyshev series:
\begin{equation}
	\vs(t) \approx \sum_{n=0}^{M-1} \vc_n T_n(y) = \sum_{n=0}^{M-1} \vc_n T_n\left(\frac{2t - t_{min} - t_{max}}{\Delta t} \right)
\end{equation}
Let us define the following set of polynomials:
\begin{equation}
	\phi_n(t) \equiv T_n\left(\frac{2t - t_{min} - t_{max}}{\Delta t} \right)
\end{equation}
Note that \eqref{t2y} is a linear transformation. Hence, $\phi_n(t)$ remains a polynomial of degree $n$, like $T_n(y)$. The Chebyshev expansion can be rewritten as a polynomial series in the terms of the $\phi_n(t)$'s:
\begin{equation}
	\vs(t) \approx \sum_{n=0}^{M-1} \vc_n \phi_n(t)
\end{equation}
Using this form, the coefficients of the Taylor like form, $\vs_m$, can be computed from the Chebyshev coefficients, $\vc_m$, via Eq.~\eqref{q2s}.

First, we expand the $\phi_n(t)$'s in a Taylor form (Cf. Eq.~\eqref{PnTaylor}):
\begin{equation}\label{phi2Taylor}
	\phi_n(t) = \sum_{m=0}^n q_{n,m} \frac{t^m}{m!}
\end{equation}
We can utilize the recurrence relation between the Chebyshev polynomials in order to find the $q_{n,m}$'s. The Chebyshev polynomials satisfy the following recursion formula:
\begin{equation}\label{Crecursion}
	T_{n+1}(y) = 2yT_n(y) - T_{n-1}(y)
\end{equation}
where
\begin{align}
	& T_0(y) = 1 \label{T0} \\
	& T_1(y) = y \label{T1}
\end{align}
The recursion formula can be rewritten in the terms of $t$ and the $\phi_n(t)$'s:
\begin{equation}\label{phirecursion}
	\phi_{n+1}(t) = \frac{4t - 2(t_{min} + t_{max})}{\Delta t}\phi_n(t) - \phi_{n-1}(t)
\end{equation}
where
\begin{align}
	& \phi_0(t) = 1 \label{phi0} \\
	& \phi_1(t) = \frac{2t - t_{min} - t_{max}}{\Delta t} \label{phi1}
\end{align}
It can be observed from Eqs.~\eqref{phi0}, \eqref{phi1}, that:
\begin{align}
	& q_{0,0} = 1 \label{q00C}\\
	& q_{1,0} = -\frac{t_{min}+t_{max}}{\Delta t} \label{q10}\\
	& q_{1,1} = \frac{2}{\Delta t} \label{q11}
\end{align}
Here, again, the rest of the $q_{n,m}$'s can be obtained from Eqs.~\eqref{q00C}-\eqref{q11} by the derivation of recurrence relations.

Plugging Eq.~\eqref{phi2Taylor} into \eqref{phirecursion} we obtain:
\begin{align}
	\phi_{n+1}(t) &= \frac{4}{\Delta t}\sum_{m=0}^n q_{n,m} \frac{t^{m+1}}{m!} - \frac{2(t_{min} + t_{max})}{\Delta t}\sum_{m=0}^n q_{n,m}\frac{t^m}{m!} - \sum_{m=0}^{n-1} q_{n-1,m} \frac{t^m}{m!} \nonumber \\
	&= \frac{4}{\Delta t}\sum_{m=1}^{n+1} q_{n,m-1} \frac{t^{m}}{(m-1)!} - \frac{2(t_{min} + t_{max})}{\Delta t}\sum_{m=0}^n q_{n,m}\frac{t^m}{m!} - \sum_{m=0}^{n-1} q_{n-1,m} \frac{t^m}{m!} \label{phinplus1a}
\end{align}
On the other hand, from \eqref{phi2Taylor} we have:
\begin{equation}\label{phinplus1b}
	\phi_{n+1}(t) = \sum_{m=0}^{n+1} q_{n+1,m} \frac{t^m}{m!}
\end{equation}
From Eqs.~\eqref{phinplus1a}, \eqref{phinplus1b}, we obtain:
\begin{equation}
	\sum_{m=0}^{n+1} q_{n+1,m} \frac{t^m}{m!} = \frac{4}{\Delta t}\sum_{m=1}^{n+1} q_{n,m-1} \frac{t^{m}}{(m-1)!} - \frac{2(t_{min} + t_{max})}{\Delta t}\sum_{m=0}^n q_{n,m}\frac{t^m}{m!} - \sum_{m=0}^{n-1} q_{n-1,m} \frac{t^m}{m!}
\end{equation}
The recurrence relations are obtained by equating the coefficients of similar powers of $t$:
\begin{align}
	& q_{n+1,0} = -\frac{2(t_{min} + t_{max})}{\Delta t}q_{n,0} - q_{n-1,0} \label{Crecursion_qn0} \\
	& q_{n+1,m} = \frac{4}{\Delta t} mq_{n, m-1} -\frac{2(t_{min} + t_{max})}{\Delta t}q_{n,m} - q_{n-1,m} & &1\leq m\leq n-1 \label{Crecursion_qnm}\\
	& q_{n+1, n} = \frac{4}{\Delta t} nq_{n, n-1} -\frac{2(t_{min} + t_{max})}{\Delta t}q_{n,n} \label{Crecursion_qnp1n}\\
	& q_{n+1, n+1} = \frac{4}{\Delta t}(n+1)q_{n,n} \label{Crecursion_qnp1np1}
\end{align}

The Taylor like coefficients are given by (Cf.~Eq.~\eqref{q2s}):
\begin{equation}\label{q2s_cheb}
	\vs_m = \sum_{n=m}^{M-1} q_{n,m}\vc_n
\end{equation}

The $\tq_{n,m}$'s can be computed in an analogous manner to the $q_{n,m}$'s. From Eqs.~\eqref{phi0}, \eqref{phi1}, we have:
\begin{align}
    & \tq_{0,0} = 1 \label{tq00C}\\
	& \tq_{1,0} = -\frac{t_{min}+t_{max}}{\Delta t} \label{tq10}\\
	& \tq_{1,1} = \frac{2}{\Delta t} \label{tq11}
\end{align}
Using the same technique as for the $q_{n,m}$'s, we obtain the following recurrence relations: 
\begin{align}
	& \tq_{n+1,0} = -\frac{2(t_{min} + t_{max})}{\Delta t}\tq_{n,0} - \tq_{n-1,0} \label{Crecursion_tqn0} \\
	& \tq_{n+1,m} = \frac{4}{\Delta t}\tq_{n, m-1} -\frac{2(t_{min} + t_{max})}{\Delta t}\tq_{n,m} - \tq_{n-1,m} & &1\leq m\leq n-1 \label{Crecursion_tqnm}\\
	& \tq_{n+1, n} = \frac{4}{\Delta t}\tq_{n, n-1} -\frac{2(t_{min} + t_{max})}{\Delta t}\tq_{n,n} \label{Crecursion_tqnp1n}\\
	& \tq_{n+1, n+1} = \frac{4}{\Delta t}\tq_{n,n} \label{Crecursion_tqnp1np1}
\end{align}

The $\vts_m$'s are given by
\begin{equation}\label{tq2s_cheb}
	\vts_m = \sum_{n=m}^{M-1} \tq_{n,m}\vc_n
\end{equation}

\section{Error estimation and control}\label{app:error}

One of the most important issues in any numerical method is the ability to estimate the magnitude of the error of the method. When the error can be estimated, it can usually be controlled by altering the parameters of the approximation. Thus, we should point out the possible sources of inaccuracy in the propagation procedure, and provide estimations of the magnitude of the error.

First, we focus on the error of the local solution \emph{in a given time-step and a given iteration}. Then, we discuss the relation between the local errors and the global error of the algorithm, \ie the error of the whole propagation process.

\subsection{Local error}

The local solution is computed in step~\ref{pr:uts} of the algorithm in Sec.~\ref{ssec:algorithm}. There are three sources of inaccuracy in this computation:
\begin{enumerate}
	\item \emph{Convergence error}: The computation is based on the previous $\vu(t_{k,l})$, \ie the guess solution or the solution from the previous iteration (step~\ref{pr:loop_beginning});
	\item \emph{Time-discretization error}: The time behaviour of $\sext(\vu(t), t)$ is approximated from sampling at the discrete Chebyshev points (steps~\ref{pr:loop_beginning}, \ref{pr:sext2b});
	\item \emph{Function of matrix computation error}: \text{$f_{M_k}(\tG, t_{k,l}-t_{k,0})\vv_{M_k}$} is approximated by a polynomial expansion, or by the Arnoldi approach. 
\end{enumerate}

The different sources of inaccuracy result in an inadequate representation of Eq.~\eqref{sol_tk} by the algorithm. The effects of the different inaccuracy sources on this representation vary. In any case, the algorithm does not represent Eq.~\eqref{sol_tk} accurately, but something else. It is important to understand what the algorithm does represent, for the understanding of the behaviour of the algorithm in each situation in which the algorithm fails to yield the required accuracy. The different situations can be classified as follows: 
\begin{enumerate}
	\item The algorithm represents an equation which differs from Eq.~\eqref{sol_tk}; we can distinguish between two situations, which correspond to different sources of inaccuracy:
	\begin{enumerate}
			\item \emph{Time-discretization error}: The time-sampling is insufficient to represent $G(t)$ or $\vs(t)$ properly. We can view this situation in the following way: We actually solve an equation of the general form of Eq.~\eqref{TDodes}, but for another problem, in which $G(t)$ or $\vs(t)$ are replaced by their truncated time-expansions; \label{sit:Gs}
		\item \emph{Function of matrix computation error}: The expansion of \text{$f_{M_k}(\tG, t_{k,l}-t_{k,0})\vv_{M_k}$} does not approximate the expression properly. In this case, the equation represented by the algorithm does not correspond to an equation of the form of Eq.~\eqref{TDodes}. \label{sit:fM}
	\end{enumerate}
	\item The algorithm does not represent a continuous \emph{equation} of time, but a discretized \emph{vector problem} in time. The algorithm is based on samplings of $\vu(t)$ at several discrete time-points, which constitute a time-vector. When the time-sampling is insufficient to represent the time-behaviour of $\vu(t)$ properly, a time-discretization error results. In such a case, the algorithm does represent the requirement that Eq.~\eqref{sol_tk} will be satisfied at the sampling Chebyshev time-points, but fails to represent the requirement that it will be satisfied at the intermediate points. One outcome is that the resulting $\vv_j$'s become inaccurate, and so is $\vu(t)$. Another outcome is that Eq.~\eqref{sol_tk} is not satisfied at the intermediate points, even with the resulting $\vv_j$'s. \label{sit:u}
	\item The algorithm does not represent an equality at all. Eq.~\eqref{sol_tk} is an implicit equation, because of the dependence of the $\vv_j$'s on $\vu(t)$. Step~\ref{pr:uts} would have represented it accurately only if the $\vu(t_{k,l})$'s in step~\ref{pr:loop_beginning} were exact. Because of the convergence error, step~\ref{pr:uts} may represent the equality only within the required extent of accuracy. If the convergence error is too large, the algorithm fails to represent the equality to the required accuracy. \label{sit:conv}
\end{enumerate}

In what follows, we give estimations to the magnitude of the error for the three inaccuracy sources, and discuss the ways to control the error for each source.

\subsubsection{Convergence error}
An estimation of the convergence error is already included in the algorithm in Sec.~\ref{ssec:algorithm}, as the convergence criterion in step~\ref{pr:conv}. The convergence rate of the iterative process can be assumed to be fast. Consequently, the error of $\vu_{old}$ is larger by orders of magnitude than the error of the new solution. Hence, the $\vu(t_{k, M_k - 1})$ obtained in step~\ref{pr:uts} can practically represent the accurate solution in this context. Thus, the convergence criterion yields an excellent approximation to the relative error of the old solution at the edge of the time-step,
\begin{equation*}
	\frac{\left\Vert\vu_{old} - \vu(t_{k, M_k - 1})\right\Vert}{\left\Vert \vu(t_{k, M_k - 1})\right\Vert}
\end{equation*}
where $\vu(t)$ here is the \emph{exact solution}. We assumed that $\Vert \vu_{old}\Vert$ in the denominator of step~\ref{pr:conv} is close enough to the converged solution, and can safely replace $\Vert \vu(t_{k, M_k - 1})\Vert$. Assuming that the iterative process converges, the error of the old solution yields an upper limit to the error of the new one. 

The problem with this estimation is that it greatly overestimates the convergence error, since the error of the old solution is assumed to be larger than the error of the new one by several orders of magnitude. Much better estimations to the convergence error of the new solution can be obtained with some extra numerical effort. This topic is left for a future publication.

In the algorithm in Sec.~\ref{ssec:algorithm}, the magnitude of the convergence error is controlled by the number of iterations. The convergence error can be controlled also by altering the other parameters of the algorithm. A decrement of the length of the time-step, $\Delta t_k$, is effective in the reduction of the convergence error. An increment of the number of expansion terms in the time-expansion \emph{in the previous time-step}, $M_{k-1}$, may be also helpful, since the guess solution becomes more accurate (unless $M_{k-1}$ becomes too large; see Sec.~\ref{ssec:parmeters}). The same is true for $K_{k-1}$, the number of expansion terms for the function of matrix in the previous time-step.

\subsubsection{Time-discretization error}
The estimation of the time-discretization error requires some additional insight into the origin of the error. The time-discretization error actually results from the replacement of the exact $\sext(\vu(t), t)$, by another time-dependent inhomogeneous term, which approximates it by interpolation at the Chebyshev points. Let us denote the approximated $\sext(\vu(t), t)$ by $\sext^{int}(t)$. The absolute time-discretization error is obtained by the difference between the solutions of two different problems---the  approximated problem and the actual one:
\begin{equation}
	E^{int}(t) = \left\Vert \vu^{int}(t) - \vu(t) \right\Vert
\end{equation}
where $\vu^{int}(t)$ denotes the solution of the problem with $\sext^{int}(t)$ as the inhomogeneous term. By the Duhamel principle (Eqs.~\eqref{solih}, \eqref{DuhamelTD}), we have:
\begin{align}
	& \vu(t) = \exp\left[\tG(t-t_{k,0})\right]\vu(t_{k,0}) + \int_{t_{k,0}}^t \exp\left[\tG(t-\tau)\right]\sext(\vu(\tau), \tau)\,d\tau \\
	& \vu^{int}(t) = \exp\left[\tG(t-t_{k,0})\right]\vu(t_{k,0}) + \int_{t_{k,0}}^t \exp\left[\tG(t-\tau)\right]\sext^{int}(\tau)\,d\tau	
\end{align}
and thus:
\begin{align}
	E^{int}(t) &= \left\Vert \int_{t_{k,0}}^t \exp\left[\tG(t-\tau)\right]\left[\sext^{int}(\tau) - \sext(\vu(\tau), \tau)\right]\,d\tau \right\Vert \nonumber\\
	&= \left\Vert \int_{t_{k,0}}^t \exp\left[\tG(t-\tau)\right]\Delta\sext^{int}(\tau)\,d\tau \right\Vert \label{er_t_Duhamel}
\end{align}
where we defined:
\begin{equation}\label{DeltaS}
	\Delta\sext^{int}(t) \equiv \sext^{int}(t) - \sext(\vu(t), t).
\end{equation}
The integral expression can be readily used to yield an upper limit for the error:
\begin{align}
	E^{int}(t) &\leq \max_{\tau\in[t_{k,0}, t]}\left\Vert \exp\left[\tG(t-\tau)\right]\Delta\sext^{int}(\tau) \right\Vert(t - t_{k,0}) \nonumber \\
	&\equiv \left\Vert \exp\left[\tG(t-t_{max})\right]\Delta\sext^{int}(t_{max}) \right\Vert(t - t_{k,0})
\end{align}
where $t_{max}$ denotes the time-point in the interval \text{$[t_{k,0}, t]$}, which maximizes the magnitude of the expression.

Next, we utilize the fact that the time-step is assumed to be small, due to the stability requirements of the algorithm. Hence, we can assume that
\begin{equation}\label{er_t_small_ts}
	\left\Vert \exp\left[\tG(t-t_{max})\right]\Delta\sext^{int}(t_{max}) \right\Vert \approx \left\Vert \Delta\sext^{int}(t_{max}) \right\Vert
\end{equation}
Note that for a Hermitian Hamiltonian, $\tG$ becomes anti-Hermitian, and Eq.~\eqref{er_t_small_ts} becomes exact.

At the edge of the time-step, the expression for the absolute time-discretization error yields the following estimation:
\begin{equation}\label{er_t_abs}
	E^{int}(t_{k, M_k - 1}) \approx \left\Vert \Delta\sext^{int}(t_{max}) \right\Vert \Delta t_k
\end{equation}
The relative error can be estimated by
\begin{equation}\label{er_t_rel}
	E^{int}_{rel}(t_{k, M_k - 1}) \approx \frac{\left\Vert \Delta\sext^{int}(t_{max}) \right\Vert\Delta t_k}{\left\Vert \vu^{int}(t_{k, M_k - 1}) \right\Vert}
\end{equation}
Observing Eqs.~\eqref{er_t_abs}, \eqref{er_t_rel}, one encounters the problem that $t_{max}$ is unknown. However, the precise knowledge of $t_{max}$ is unnecessary, since we need only an estimation for the order of magnitude of the error. It is reasonable to use the point which is furthest from neighbouring sampling points instead of the precise $t_{max}$. Hence, we can choose the middle point between $t_{mid}$ and the next Chebyshev point.

$\Delta\sext^{int}(t)$ at the estimated $t_{max}$ can be computed directly, via Eq.~\eqref{DeltaS}. $\sext(\vu(t), t)$ at the estimated $t_{max}$ is computed in the same way as the samplings of $\sext(\vu(t), t)$ at the Chebyshev points (step~\ref{pr:loop_beginning} of the algorithm in Sec.~\ref{ssec:algorithm}). $\sext^{int}(t)$ at the estimated $t_{max}$ is computed by the evaluation of the approximation polynomial of $\sext(\vu(t), t)$ at the point, using the coefficients computed in step~\ref{pr:sext2b} of the algorithm.

This error estimation tends to overestimate the time-discretization error, typically by one or two orders of magnitude at the time-step edge. The reason lies in the oscillatory nature of $\Delta\sext^{int}(t)$, which is responsible for cancellation of errors during the integration in Eq.~\eqref{er_t_Duhamel}. An accurate estimation of the time-discretization error requires a more detailed analysis. This topic is left for a future publication.

The magnitude of the time-discretization error can be primarily controlled by the number of Chebyshev time-points, $M_k$. However, $M_k$ cannot be increased indefinitely in order to reduce the error, because of reduction in the efficiency of the algorithm in higher orders (see Sec.~\ref{ssec:parmeters}). An alternative option for error reduction is the decrement of $\Delta t_k$.

\subsubsection{Function of matrix computation error}
The estimation of the function of matrix computation error is more direct. It can be readily observed from the computation in step~\ref{pr:uts} that the absolute error of the computation of \text{$f_{M_k}(\tG, t_{k,l}-t_{k,0})\vv_{M_k}$} is just the same as the resulting absolute error of $\vu(t_{k,l})$ itself. Let us denote the absolute error as $E^{fm}(t)$; the relative error at the edge of the time-step can be estimated by
\begin{equation}
	E^{fm}_{rel}(t_{k, M_k - 1}) = \frac{E^{fm}(t_{k, M_k - 1})}{\left\Vert \vu^{fm}(t_{k, M_k - 1}) \right\Vert}
\end{equation}
where $\vu^{fm}(t)$ denotes the solution resulting from the approximation of \text{$f_{M_k}(\tG, t-t_{k,0})\vv_{M_k}$}. An estimation to $E^{fm}(t)$ is given by an estimation of the truncation error of the expansion of \text{$f_{M_k}(\tG, t-t_{k,0})\vv_{M_k}$}. In the case that a polynomial expansion approximation is used, one simple way to estimate the truncation error is by computation of the error of the same approximation at several test points; the value of \text{$f_{M_k}(z, \Delta t_k)$} is interpolated in $z$ at several representative test points in the eigenvalue domain, and the relative error from the exact value is computed; the estimated $E^{fm}(t_{k, M_k - 1})$ is given by the multiplication of the obtained relative error by \text{$\Vert f_{M_k}(\tG, \Delta t_k)\vv_{M_k} \Vert$}. An estimation of the error for the Arnoldi approach is given in Appendix~\ref{app:Arnoldi}.

The magnitude of the function of matrix computation error can be primarily controlled by the number of expansion terms for the approximation, $K_k$. A decrement of $\Delta t_k$ also reduces the error.

\subsection{Global error}

Our main interest is in the estimation of the global error of the final solution obtained by the algorithm. One would have suggest that the global error of the algorithm is just an additive sum of the local errors in each time-step. This is indeed the case when the propagation process is numerically stable. However, it is important to be aware of the fact that the global behaviour of the algorithm might be different; we may observe three distinguished behaviours of error accumulation in the algorithm:
\begin{enumerate}
	\item \emph{Additive accumulation}: The final error is the sum of the errors of the solutions in the last iteration of each time-step;
	\item \emph{Divergence with propagation}: The error accumulates in an explosive manner during the propagation. The magnitude of the solution tends to infinity \emph{with the propagation};
	\item \emph{Divergence of the iterative process}: The iterative process in a specific time-step fails to converge. The magnitude of the solution tends to infinity \emph{with the number of iterations}.
\end{enumerate}

In the last two behaviours, the estimation of the local error is useless for the estimation of the global error. The divergence of the iterative process can be always detected, since the algorithm fails to continue the propagation process. Most frequently, the divergence with propagation is also easily detected, by the occurrence of an overflow, or an unreasonably large magnitude of the solution. Seldom, it may occur that the divergent process has stopped at an early stage at the end of the propagation, and the magnitude of the solution is not unreasonable. In this case, it might not be easy to detect the divergent behaviour of the error.

Of course, it is highly desirable to prevent an unstable behaviour of the algorithm. First, we should attribute the different unstable behaviours to the responsible causes.

The divergence of the iterative process clearly originates in a too large time-step, which is outside the convergence radius of the algorithm. If a divergence of the iterative process occurs, the length of the time-step should be decreased.

The origin of the divergence with propagation is less obvious. Let us recall the classification into the different situations in which the algorithm fails to represent Eq.~\eqref{sol_tk} adequately. When the algorithm represents an equation of the general form of Eq.~\eqref{TDodes} (situation~\ref{sit:Gs} above), the behaviour of the solution is expected to preserve the characteristic features of Eq.~\eqref{TDodes}. For instance, in the homogeneous Schr\"odinger equation (with a Hermitian Hamiltonian) the norm of the solution is expected to be conserved. Hence, a divergent behaviour with the propagation is not expected. In contrary, the behaviour of the algorithm in the other situations is unexpected.

In practice, a divergent behaviour was observed only in situation~\ref{sit:fM} above, \ie when there is a function of matrix computation error. Experience shows that only a low accuracy expansion of \text{$f_{M_k}(\tG, t_{k, l}-t_{k,0})\vv_{M_k}$} leads to a divergent behaviour. The instability disappears when more expansion terms are used. We can conclude that it is not recommended to expand \text{$f_{M_k}(z, t_{k,l}-t_{k,0})$} by a low accuracy expansion, even when a low accuracy solution is sufficient. Further research is required to estimate the maximal allowed inaccuracy in the expansion for stability of the propagation. In the problems that were tested so far, the following criterion was found to be sufficient for stability:
\begin{equation*}
	\frac{E^{fm}(t_{k, M_k - 1})}{\left\Vert f_{M_k}(\tG, \Delta t_k)\vv_{M_k} \right\Vert} < 10^{-5}
\end{equation*}

It is noteworthy that a divergent behaviour with the propagation might be observed also for the convergence error, in a different version of the algorithm than that presented in Sec.~\ref{ssec:algorithm}; if one restricts the allowed number of iterations in each time-step (like in the numerical example of Sec.~\ref{sec:example}), a divergence with propagation might occur. This phenomenon is typical for high order $M$ values. The use of high order $M$ is not recommended anyway, by efficiency considerations (see Sec.~\ref{ssec:parmeters}).

It should be noted that the phenomenon of divergence with propagation is common to other propagators, when the time-step is too large.

The inclusion of tests for the magnitude of the error in the algorithm increases the robustness of the algorithm. The tests may be also used for an adaptive choice of the parameters during the propagation. In the future, we plan to develop an improved version of the algorithm, based on this principle.

\bibliographystyle{amsplain}
\bibliography{SemiGlobal}

\end{document}